\documentclass[aps,prd,preprint,preprintnumbers,
superscriptaddress,showpacs,nofootinbib,%
tightenlines,
a4paper]{revtex4-1}
\usepackage{lipsum}
\usepackage[utf8]{inputenc}
\usepackage{textcomp}
\usepackage{epsfig}
\usepackage{color}
\usepackage{amssymb,amsmath}
\usepackage{array}
\usepackage{bm}
\usepackage{hyperref}
\newcommand{\Gevs}{\text{GeV}^2}
\newcommand{\res}{\langle r_\text{E}^2 \rangle}

\newcommand{\rms}{\langle r_\text{M}^2 \rangle}

\newcommand{\GE}{G_\text{E}}
\newcommand{\GM}{G_\text{M}}
\newcommand{\GX}[1]{G_{\rm #1}}
\newcommand{\rX}[1]{\langle r_{\rm #1}^2\rangle^{1/2}}

\newcommand{\mev}{\mathrm{MeV}}
\newcommand{\gev}{\mathrm{GeV}}
\newcommand{\fm}{\mathrm{fm}}

\begin{document}

\title{Isovector electromagnetic form factors of the nucleon from lattice QCD
and the proton radius puzzle}
\author{D.~Djukanovic}
 \affiliation{Helmholtz Institute Mainz, Staudingerweg 18, D-55128 Mainz, Germany}
 \affiliation{GSI Helmholtzzentrum für Schwerionenforschung, Darmstadt (Germany)}
\author{T.~Harris}
 \affiliation{School of Physics and Astronomy, University of Edinburgh, Edinburgh EH9
 3JZ, UK}
\author{G.~von~Hippel}
 \affiliation{PRISMA$^+$ Cluster of Excellence and Institute for Nuclear Physics, Johannes Gutenberg University of Mainz, Johann-Joachim-Becher-Weg 45, D-55128 Mainz, Germany}
\author{P.M.~Junnarkar}
\affiliation{Institut f\"ur Kernphysik, Technische Universit\"at Darmstadt,
Schlossgartenstra{\ss}e 2, 64289 Darmstadt}
\author{H.~B.~Meyer}
 \affiliation{Helmholtz Institute Mainz, Staudingerweg 18, D-55128 Mainz, Germany}
 \affiliation{GSI Helmholtzzentrum für Schwerionenforschung, Darmstadt (Germany)}
 \affiliation{PRISMA$^+$ Cluster of Excellence and Institute for Nuclear Physics, Johannes Gutenberg University of Mainz, Johann-Joachim-Becher-Weg 45, D-55128 Mainz, Germany}
 \author{D.~Mohler}
  \affiliation{Helmholtz Institute Mainz, Staudingerweg 18, D-55128 Mainz, Germany}
 \affiliation{GSI Helmholtzzentrum für Schwerionenforschung, Darmstadt (Germany)}
\author{K.~Ottnad}
 \affiliation{PRISMA$^+$ Cluster of Excellence and Institute for Nuclear Physics, Johannes Gutenberg University of Mainz, Johann-Joachim-Becher-Weg 45, D-55128 Mainz, Germany}
\author{T.~Schulz}
 \affiliation{PRISMA$^+$ Cluster of Excellence and Institute for Nuclear Physics, Johannes Gutenberg University of Mainz, Johann-Joachim-Becher-Weg 45, D-55128 Mainz, Germany}
\author{J.~Wilhelm}
 \affiliation{PRISMA$^+$ Cluster of Excellence and Institute for Nuclear Physics, Johannes Gutenberg University of Mainz, Johann-Joachim-Becher-Weg 45, D-55128 Mainz, Germany}
\author{H.~Wittig}
 \affiliation{Helmholtz Institute Mainz, Staudingerweg 18, D-55128 Mainz, Germany}
 \affiliation{GSI Helmholtzzentrum für Schwerionenforschung, Darmstadt (Germany)}
 \affiliation{PRISMA$^+$ Cluster of Excellence and Institute for Nuclear Physics, Johannes Gutenberg University of Mainz, Johann-Joachim-Becher-Weg 45, D-55128 Mainz, Germany}\date{\today}
\begin{abstract}
We present results for the isovector electromagnetic form
factors of the nucleon computed on the CLS ensembles with $N_f=2+1$ flavors of
$\mathcal{O}(a)$-improved Wilson fermions and an $\mathcal{O}(a)$-improved
vector current. The analysis includes ensembles with four lattice spacings 
and pion masses ranging from 130 MeV  up to 350 MeV and mainly targets the low-$Q^2$ region. 
In order to remove any bias from unsuppressed excited-state
contributions, we investigate several source-sink separations between
1.0 fm and 1.5 fm and apply the summation method as well as explicit
two-state fits. The chiral interpolation is performed by applying covariant chiral
perturbation theory including vector mesons directly to our form
factor data, thus avoiding an auxiliary parametrization of
the $Q^2$ dependence.
At the physical point, we obtain
$\mu=4.71(11)_{\mathrm{stat}}(13)_{\mathrm{sys}}$ for the
nucleon isovector magnetic moment, in good agreement with the
experimental value and $\langle
r_\mathrm{M}^2\rangle~=~0.661(30)_{\mathrm{stat}}(11)_{\mathrm{sys}}\,~\mathrm{fm}^2$
for the corresponding square-radius, again in good agreement
with the value inferred from the $ep$-scattering determination 
[Bernauer et~al., Phys. Rev. Lett., \textbf{105}, 242001 (2010)] of the proton radius.  Our estimate for the isovector electric charge
radius, $\langle r_\mathrm{E}^2\rangle =
0.800(25)_{\mathrm{stat}}(22)_{\mathrm{sys}}\,~\mathrm{fm}^2$, however, is in slight
tension with the larger value inferred from the aforementioned $ep$-scattering data, while being
in agreement with the value derived from the 2018 CODATA
average for the proton charge radius.
\end{abstract}

\maketitle

\section{Introduction}

The internal structure of the nucleon still poses many open questions.
Not only is the composition of its spin and momentum not completely understood
\cite{Ashman:1987hv,Aidala:2012mv,Chen:2009mr,Ji:2020ena},
but even its size is subject to significant uncertainty arising
from discrepancies between different determinations:
there is a decade-old inconsistency \cite{Pohl:2010zza,Karr:2020wgh}
between the electric charge radius
of the proton as obtained from $ep$-scattering
($\rX{p} = 0.879(8)\,\textrm{fm}$ \cite{Bernauer:2010wm})
in good agreement with the value $\rX{p}=0.8758(77)\,\textrm{fm}$
from hydrogen spectroscopy \cite{Mohr:2012tt} on the one hand, and 
the most accurate determination from the spectroscopy of muonic
hydrogen ($\rX{p} = 0.84087(39)\,\textrm{fm}$ \cite{Antognini:1900ns}) on the other.
This significant discrepancy, which has been dubbed the ``proton radius puzzle''
\cite{Carlson:2015jba}, has given rise to a variety of initiatives to better
determine the proton radius.
Recent measurements of $\rX{p}$ using (electronic) hydrogen spectroscopy 
\cite{Beyer:2013jla,Thomas:2019dad,Fleurbaey:2018fih} mostly tend to give
somewhat smaller values,
while the newest determinations from $ep$-scattering give different results,
albeit with still rather large uncertainties:
the A1 collaboration at MAMI uses an
Initial-State Radiation setup \cite{Mihovilovic:2016rkr}
to achieve very small momentum transfers
and finds a large value $\rX{p} = 0.870(28)\,\textrm{fm}$ \cite{Mihovilovic:2019jiz},
while the PRAD experiment at Jefferson Lab \cite{Gasparian:2017cgp}
has reported a small result of $\rX{p} = 0.831(14)\,\textrm{fm}$ \cite{Xiong:2019umf}.
An upgrade of PRAD has recently been proposed \cite{Gasparian:2020hog},
and a new $ep$-scattering experiment, MAGIX, is being prepared at Mainz
\cite{Grieser:2018qyq}.
To complement the result from muonic hydrogen spectroscopy with a result
from $\mu p$ scattering, the MUSE collaboration aims to measure the $\mu p$
scattering cross-section to sub-percent accuracy \cite{Gilman:2017hdr};
a similar experiment is in preparation at CERN by the COMPASS++/AMBER collaboration
\cite{Dreisbach:2019pkc}.
At present, a significant tension remains between the determinations
based on muonic hydrogen and $ep$-scattering, respectively.
A variety of possible theoretical explanations have been proposed,
but so far there has not been decisive
evidence in favour of any particular explanation
\cite{Carlson:2015jba,Krauth:2017ijq}. It remains notable that dispersive analyses
favour a smaller value of the proton charge radius \cite{Hoferichter:2016duk,
Hammer:2019uab}, and in fact have done so since before the muonic
measurements \cite{Belushkin:2006qa}.

In the context of scattering experiments, the proton charge radius is determined
from the derivative of the electric form factor $\GX{E}(Q^2)$ at $Q^2=0$. To
determine this derivative accurately, high-quality data at very low
momentum transfer $Q^2$ and/or a parameterization that remains trustworthy
over a large range of $Q^2$ are required.
In order to gain a proper understanding of the origin of the proton radius puzzle
and associated questions, theoretical
determinations of nucleon structure observables, and in particular of the
electromagnetic form factors of the nucleon, from first principles are required.
Lattice QCD calculations are therefore instrumental in predicting
the nucleon charge radii from QCD.
This has generated lively activity on this topic within the lattice QCD community
\cite{Gockeler:2003ay,Syritsyn:2009mx,Alexandrou:2013joa,Alexandrou:2017ypw,
Alexandrou:2018sjm,Shanahan:2014cga,Shanahan:2014uka,Bhattacharya:2013ehc,
Yamazaki:2009zq,Shintani:2018ozy,Ishikawa:2018rew,Green:2014xba,Chambers:2017tuf,
Capitani:2015sba}, although
at present, the precision of lattice results is not yet sufficient to rule out
either the electronic or the muonic result for the proton radius.

At large momentum transfers $Q^2$, the electromagnetic form factors are the
subject of another puzzle: while polarization-transfer experiments
\cite{Milbrath:1997de} find that the ratio of electric and magnetic form factor
of the proton, $\mu_{\rm p}\GX{E,p}(Q^2)/\GX{M,p}(Q^2)$,
decreases roughly linearly for large $Q^2$
\cite{Jones:1999rz,Gayou:2001qd,Punjabi:2005wq,Puckett:2010ac,Puckett:2011xg},
experiments based on the Rosenbluth separation formula \cite{Rosenbluth:1950yq}
find that it is roughly constant and of order one (albeit with rapidly
increasing errors at large $Q^2$, where $\GX{E,p}$ contributes little to the
total cross-section) \cite{Walker:1993vj,Andivahis:1994rq,Qattan:2004ht}.
This discrepancy has been explained theoretically as the result of two-photon
exchange contributions to the cross-section measurements
\cite{Guichon:2003qm,Arrington:2007ux,Afanasev:2017gsk},
but the situation is not yet completely clarified \cite{Bernauer:2020vue}.

In this paper, we present our lattice QCD-based determination of the isovector
electromagnetic
form factors of the nucleon from the $N_f=2+1$ CLS ensembles
\cite{Bruno:2014jqa}. We use two state-of-the-art methods, known as the summation method
and two-state fits to extract $\GX{E}$ and $\GX{M}$ from Euclidean correlation functions
for a range of momentum transfers
$Q^2\lesssim 1\,\textrm{GeV}^2$. Given the limited $Q^2$ range of our data, we focus
on extracting the electric and magnetic charge radii and the magnetic moment of the
nucleon using a variety of methods to check for consistency. Extrapolated to the
physical point, our results favour a small value of the electric charge radius, although
the present accuracy is not sufficient to make a decisive statement in this regard.

The paper is organized as follows: in section~\ref{sec:setup} we present the ensembles
and operators used in our simulations; section~\ref{sec:excited} describes the methods
we use to account for the presence of excited-state contaminations in our data,
and section~\ref{sec:Qsq} the methods we employ to parameterize the form factor data
on each lattice ensemble,
while section~\ref{sec:CCF} gives the results for the extrapolation of our results
to the continuum and infinite-volume limits at the physical pion mass.
Our conclusions and a discussion of the results are contained in
section~\ref{sec:summary}.
For completeness and ease of reference, we provide tables with the values of
all measured form factors in appendix~\ref{app:data}.
Appendix~\ref{app:priors} lists the priors we use to stabilize our two-state fits,
while the results of both dipole and $z$-expansion fits on each ensemble are given
in appendix~\ref{app:dipole}. Appendices~\ref{sec:app_hbchpt} and~\ref{sec:app_covchpt}
give the results of the extrapolations to the physical point using two variants of
chiral perturbation theory. We consider the ratio ${\GX{M}(Q^2)}/{\GX{E}(Q^2)}$ in appendix~\ref{app:gmoverge}.

\section{Lattice Setup}
\label{sec:setup}
We use the CLS $N_f=2+1$ ensembles \cite{Bruno:2014jqa} that have been generated 
with non-perturbatively $\mathcal{O}(a)$-improved Wilson
fermions \cite{Sheikholeslami:1985ij,Bulava:2013cta}
and the tree-level improved L\"uscher-Weisz gauge action \cite{Luscher:1984xn}. 
In order to prevent topological freezing \cite{Schaefer:2010hu},
the fields obey open boundary conditions in time \cite{Luscher:2011kk}, with the exception of
ensemble E250 which uses periodic boundary condition in the time
direction. The reweighting factors needed to correct for the treatment of the
strange quark determinant during the gauge field generation are
obtained using the method of Ref.~\cite{Mohler:2020txx}.
See Tab.~\ref{tab:ensembles} for a list of ensembles used in this work, which
cover the range of lattice spacings from $0.050$ fm to $0.086$ fm. Our setup in
this work is identical to that used in our paper on the isovector charges
and momentum fractions of the nucleon \cite{Harris:2019bih}, to which we refer
the reader for further details.

\begin{table}[!t]
 \centering
  \begin{tabular}{lcrccccrrrc}
   \hline\hline
   ID  & $\beta$ & $T/a$ & $L/a$ & $M_\pi~[\mev]$ & $M_\pi L$ & $M_N~[\gev]$ &
   $N_\mathrm{HP}$ & $N_\mathrm{LP}$ & $N_\mathrm{CFG}$ & $t_s~[\fm]$ \\
   \hline\hline
   H105 & 3.40 &  96 & 32 & 278(4) & 3.90 & 1.020(18) & 4076 & 48912 & 1019 & 1.0, 1.2, 1.4           \\
   C101 & 3.40 &  96 & 48 & 223(3) & 4.68 & 0.984(12) & 2000 & 64000 & 2000 & 1.0, 1.2, 1.4           \\
   \hline
   S400 & 3.46 & 128 & 32 & 350(4) & 4.34 & 1.123(15) & 1725 & 27600 & 1725 & 1.1, 1.2, 1.4, 1.5 \\ 
   \hline
   N203 & 3.55 & 128 & 48 & 347(4) & 5.42 & 1.105(13) & 1540 & 24640 & 1540 & 1.0, 1.2, 1.3, 1.4\\
   S201 & 3.55 & 128 & 32 & 293(4) & 3.05 & 1.097(21) & 2092 & 66944 & 2092 & 1.0, 1.2, 1.3, 1.4      \\
   N200 & 3.55 & 128 & 48 & 283(3) & 4.42 & 1.053(14) & 1697 & 20364 & 1697 & 1.0, 1.2, 1.3, 1.4      \\ 
   D200 & 3.55 & 128 & 64 & 203(3) & 4.23 & 0.960(13) & 1019 & 32608 & 1019 & 1.0, 1.2, 1.3, 1.4      \\
   E250 &  3.55 & 192  & 96 & 130(1) & 4.04 & 0.928(11) & 976 & 31232 & 244 & 1.0, 1.2, 1.3,  1.4\\
   \hline
   N302 & 3.70 & 128 & 48 & 353(4) & 4.28 & 1.117(15) & 1177 & 18832 & 1177 & 1.0, 1.1, 1.2, 1.3\\
   J303 & 3.70 & 192 & 64 & 262(3) & 4.24 & 1.052(17) &  531 &  8496 & 531 & 1.0, 1.1, 1.2, 1.3      \\ 
   \hline\hline
   \vspace*{0.1cm}
  \end{tabular}
  \caption{Overview of ensembles used in this study. The quoted errors on the
  pion and nucleon masses include the error from the scale setting
  \cite{Bruno:2016plf}.
  $N_\mathrm{HP}$ and $N_\mathrm{LP}$ denote the number of high-precision
  (HP) and low-precision (LP) measurements on all of the $N_\mathrm{CFG}$
  configurations for each value of the source-sink
  $t_s$, respectively. For E250 $N_\mathrm{HP}$ and $N_\mathrm{LP}$
  refer to the number of sources used for the two largest values of
  $t_s=1.3,1.4\,\rm{fm}$ while for each of the smaller values
  $t_s=1.2\,\rm{fm}$ and $t_s=1.0\, \rm{fm}$ the number of sources is
  reduced by a factor of two and four, respectively. }
 \label{tab:ensembles}
\end{table}

We obtain the matrix element of the vector current through the
ratio \cite{Alexandrou:2008rp} 
\begin{align}
\label{eq:matelement}
R^{J_\mu}(t,t_s;{\bf{q}})=\frac{C_3^{J_\mu}(t,t_s;{\bf{q}})}{C_2(t_s;{\bf{0}})} 
\sqrt{\frac{C_2(t_s-t;-{\bf{q}})\, C_2(t,{\bf{0}})\, C_2(t_s;{\bf{0}})}{C_2(t_s-t;{\bf{0}})\, 
C_2(t;-{\bf{q}}) \,C_2(t_s;-{\bf{q}})} } \, ,
\end{align}
where the nucleon two- and three-point-functions are given by
\begin{align}
C_2(t;{\bf{p}})&=\Gamma_{\alpha\beta} \sum\limits_{{\bf{x}}} e^{-i{\bf{px}}}
\Bigl\langle\Psi_\beta({\bf{x}},t) \overline{\Psi}_\alpha (0) \Bigr\rangle,\\
C_3^{J_\mu}(t,t_s;{\bf{q}})&=\Gamma_{\alpha\beta} \sum\limits_{{\bf{x,y}}} e^{i{\bf{qy}}}
\Bigl\langle\Psi_\beta({\bf{x}},t_s) J_\mu({\bf{y}},t) \overline{\Psi}_\alpha (0) \Bigr\rangle
\end{align}
in our setup, where the nucleon at the sink is at rest,
i.e. for a momentum transfer ${\bf{q}}$ the initial and final nucleon states have momenta
\begin{align}
{\bf{p'}}=0,\qquad {\bf{p}}=-{\bf{q}}.
\end{align}
Our interpolating operator
\begin{equation}
\Psi_\alpha(x) = \epsilon_{abc}
\left(\tilde{u}^T_a(x)C\gamma_5\tilde{d}_b(x)\right)\tilde{u}_{c,\alpha}(x)\,,
\end{equation}
for the proton is built using Gaussian-smeared \cite{Gusken:1989ad} quark fields
\begin{equation}
\tilde{q} = (1 + \kappa_{\rm G}\Delta)^{N_{\rm G}} q\,, \qquad q=u,d,
\end{equation}
using spatially APE-smeared \cite{Albanese:1987ds} gauge links in the covariant Laplacian $\Delta$
and tuning the parameters $\kappa_{\rm G}$ and $N_{\rm G}$ so that a smearing radius
$r_{\rm G}\sim 0.5\,\fm$ \cite{vonHippel:2013yfa} is realized. For the projection matrix $\Gamma$ we use
\begin{align}
\Gamma=\frac{1}{2}(1+\gamma_0) (1+i \gamma_5 \gamma_3).\label{eq:polarization}
\end{align}
Furthermore we use the improved conserved vector current 
\begin{align}
J_\mu(z)^{\text{c}}&=\frac{1}{2}\Bigl( \bar{q}(z+\hat{\mu}a) (1+\gamma_\mu) U_\mu(z)^\dagger q(z)
- \bar{q}(z) (1-\gamma_\mu) U_\mu(z) q(z+\hat{\mu}a)\Bigr) \, , \nonumber
\\
J_\mu(z)^{\rm Imp.} &= \frac{1}{2} \Bigl(J_\mu(z)^{\rm c }+ J_\mu(z-\hat\mu
a)^{\rm c}\Bigr)
- \frac{a\,  c^{cs}_{\rm V}}{2} \partial_\nu \Bigl(\bar q(z) [\gamma_\mu,\gamma_\nu]
q(z)\Bigr) \, ,
\label{eq:setup_improved_current}
\end{align}
with the improvement coefficient $c^{cs}_{\rm V}$ from \cite{Gerardin:2018kpy}. The
improvement term is implemented using the symmetric lattice derivative, i.e.
\begin{align}
	\partial_\nu \phi(x) &:=  \frac{\phi(x+ a\hat{\nu}) - \phi(x- a
	\hat{\nu} )}{2 a} \ .
		\label{eq:setup_sitesymm}
\end{align}
To compute the three-point functions, we employ extended propagators
in the ``fixed-sink'' method, requiring additional inversions for each value
of $t_s$ studied 
while allowing the momentum transfer to be varied via a phase factor
at the point of the current insertion \cite{Martinelli:1988rr}. As in
\cite{Harris:2019bih}, we apply the truncated solver method with bias correction
\cite{Bali:2009hu,Blum:2012uh,Shintani:2014vja} to reduce the cost of the inversions. The number of high-precision and
low-precision measurements carried out on each gauge configuration is indicated
in Table~\ref{tab:ensembles}.

\begin{figure}[!h]
	\includegraphics[scale=0.85]{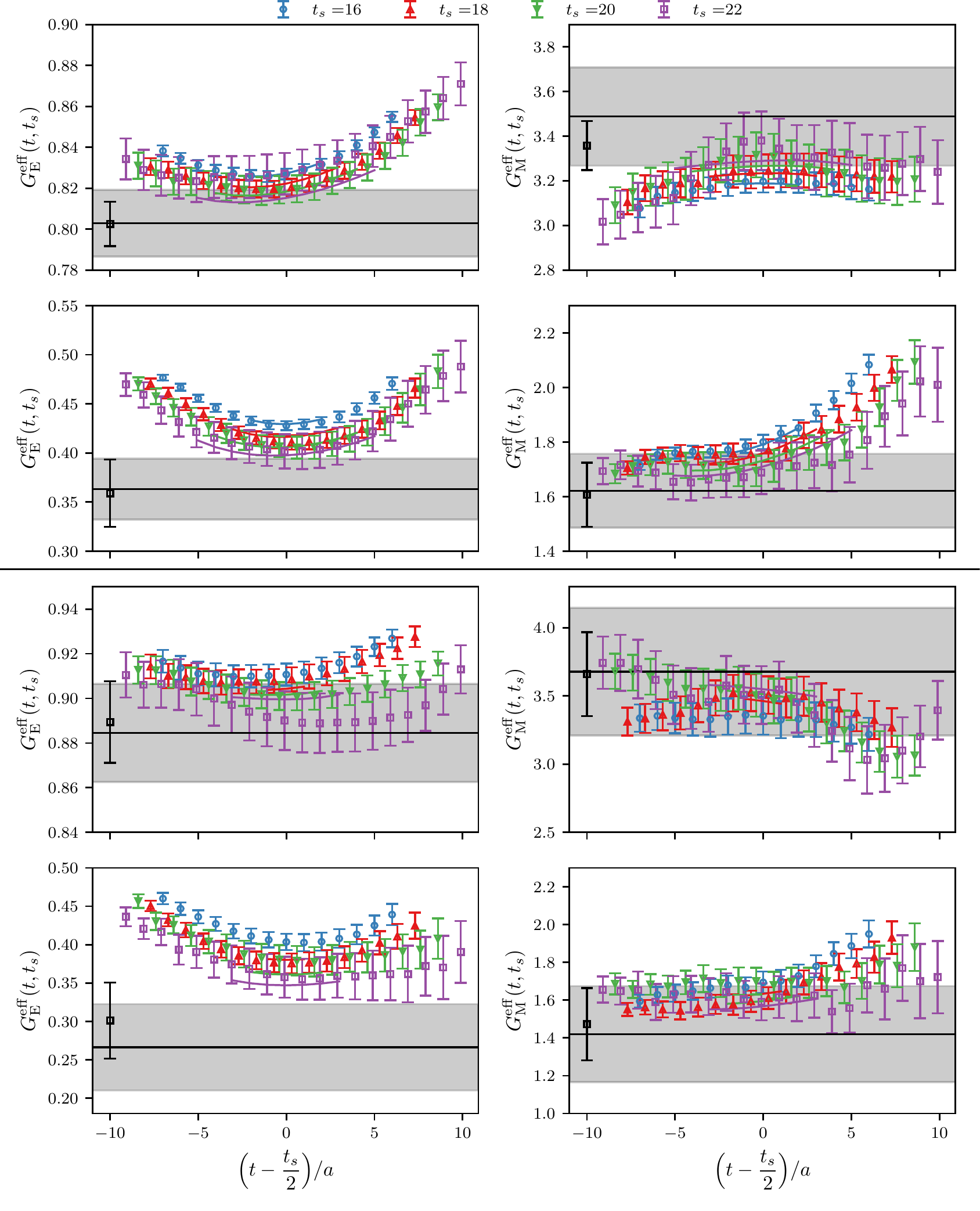}
	\caption{
		Effective form factors for ensemble D200 (upper panel) and E250 (lower
panel). In each panel, the first row corresponds to the smallest
non-vanishing momentum in the given ensemble, i.e. $Q^2=0.089,0.040 \Gevs$ for
D200 and E250, respectively, and the second row
corresponds to $Q^2 \sim 0.5\,$GeV$^2$.  For the four available
source-sink separations $t_s$, the effective form factors are
displayed as a function of the current insertion time $t$, offset
to the midpoint between nucleon source and sink.
The curves represent  the two-state fits in their respective fit intervals.
The gray band and black data point correspond to the estimate for the ground-state matrix
element for the summation and two-state method, respectively. The data
points are displaced for better visibility.
}
	\label{fig:eff_ff_e250_and_d200}
\end{figure}

For the polarization given in Eq.~(\ref{eq:polarization}) the asymptotic value for
the spectral decomposition of Eq.~(\ref{eq:matelement}) reads
\begin{subequations}
\begin{align}
R^{J_0}(t,t_s;{\bf{q}})& \equiv \sqrt{\frac{m_N+E_{{\bf{q}}}}{2E_{{\bf{q}}}}}\;
G^\mathrm{eff}_\mathrm{E}(Q^2,t,t_s)\,,\\
R^{J_0}(t,t_s;{\bf{q}})&\xrightarrow{t,(t_s-t) \gg 0}\sqrt{\frac{m_N+E_{{\bf{q}}}}{2E_{{\bf{q}}}}}\,
\GX{E}(Q^2)\,,
\label{eq:asymptotic_GEGM_groundstatea}
\\
\mathrm{Re} R^{J_i}(t,t_s;{\bf{q}})&\xrightarrow{t,(t_s-t) \gg 0}
\sqrt{\frac{1}{2E_{{\bf{q}}}(E_{{\bf{q}}}+M_N)}}\, \GX{M}(Q^2)
\epsilon_{ij3}q_j\,,
\label{eq:asymptotic_GEGM_groundstateb}
\end{align}
\end{subequations}
where $\GX{E}$ \footnote{In addition we have extracted the electric form factor
from the spatial components of the current matrix element, however the
extracted values are less accurate compared to
Eq.~(\ref{eq:asymptotic_GEGM_groundstatea}).} and $\GX{M}$ are the isovector electric and
magnetic Sachs form factors, respectively, with $\GE(0)=1$. The extraction of the magnetic form
factor via Eq.~(\ref{eq:asymptotic_GEGM_groundstateb}) amounts to solving a
system of linear equations, 
since, in general, several different choices of $\mathbf q$ produce the same value of $Q^2$.
Consequently there are several possibilities
for obtaining an estimate of $\GM(Q^2)$. We use the following
estimator,
\begin{align}
	\GM^\mathrm{eff}(Q^2,t,t_s)= \frac{\sqrt{2 E_{\bf{q}} (M_N +
	E_{\bf{q}})} }{q_1^2 + q_2^2 } \Bigl[q_2 \mathrm{Re} R^{J_1} - q_1
	\mathrm{Re} R^{J_2} \Bigr]  , \ q_1 \neq 0 \lor q_2 \neq 0 \, ,
	\label{eq:GM_extract_sol}
\end{align}
averaging over all momenta $\mathbf{q}$ contributing to the same $Q^2$.  The
resulting effective form factors for every source-sink separation for the first
non-zero momentum and a momentum close to $0.5\, \Gevs$ on the ensembles D200
and E250 are shown in Fig.~\ref{fig:eff_ff_e250_and_d200}.

Unless stated otherwise, errors are computed using the jackknife method on binned data with a bin
size of two for all ensembles except E250, where the spacing between two
analyzed configurations in terms of molecular dynamics time is twice as large
compared to e.g. D200 or C101 to begin with. For the conversion to physical units we use the lattice spacing
determination of \cite{Bruno:2016plf}.

\section{Excited-State Systematics}
\label{sec:excited}
Baryonic correlation functions suffer from a strong exponential growth
of the relative statistical noise when the distance in Euclidean time between
operators is increased \cite{Lepage:1989hd}. Therefore, for the typical
source-sink separations in current lattice calculations of baryon
structure observables, it cannot be guaranteed that contributions from
excited states are sufficiently suppressed. Evidently, special care is
required to avoid any bias from unwanted excited-state contributions
\cite{Green:2018vxw,Ottnad:2020qbw}. Predominantly, two approaches have been
widely adopted to address this problem: the summation method
\cite{Gusken:1989qx,Maiani:1987by,Doi:2009sq,Bulava:2011yz,Capitani:2012gj,Capitani:2015sba}
and multi-state fits
\cite{Yoon:2016jzj,Capitani:2015sba,Chang:2018uxx,Jang:2019jkn,Gupta:2018qil}.

While the former is (in its simplest incarnation) a straightforward method to
apply, the latter is more involved as one is forced to make specific assumptions
and/or parameter choices, regarding e.g. fit windows, at various steps of the
analysis. In this section we give details on our implementation of the two
respective methods and discuss how errors related to methodology are
incorporated in the final results. The form factor values obtained with both
methods are collected in Appendix~\ref{app:data} for all ensembles.

\begin{figure}[t]
	\includegraphics[scale=.95]{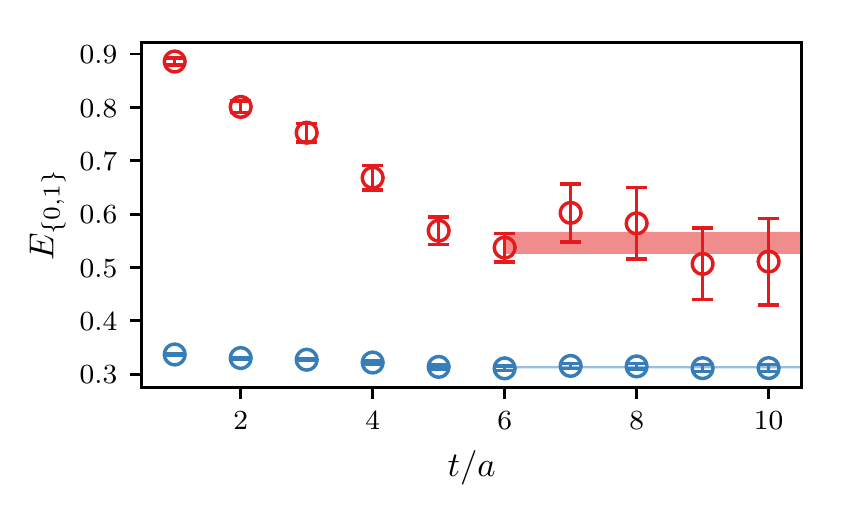}
	\includegraphics[scale=.95]{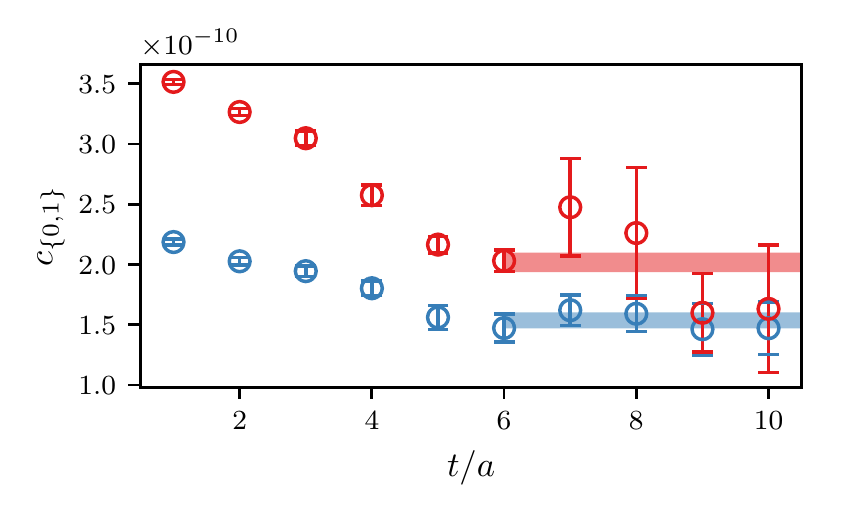}
	\caption{
Energy levels (left) and overlap factors (right) extracted from
the zero-momentum nucleon two-point function on ensemble D200,
for the ground state (blue) and the first excited state (red).
All quantities are given in lattice units.
}
\label{fig:excited_energy_levels_overlap_d200_0_mom}
\end{figure}

For the two-state fits of the effective form factors, we use priors obtained
from an analysis of the two point functions.  We fit the two-point function
with the ansatz 
\begin{align} 
	C_2(t, \mathbf{p}) &=  c_0(\mathbf{p}^2)
	e^{-E_0(\mathbf{p}^2) t}+c_1(\mathbf{p}^2) e^{-E_1 (\mathbf{p}^2) t},
\end{align} 
and extract the energy gap between
ground ($E_0$) and first excited state ($E_1$), as well as the ratio of the respective
overlaps, i.e. $\Delta(\mathbf{p}^2)= E_1(\mathbf{p}^2)-E_0(\mathbf{p}^2)$ and
$\rho(\mathbf{p}^2)=c_1(\mathbf{p}^2)/c_0(\mathbf{p}^2)$. In practice, the gaps and ratios
depend on the choice of fit ranges in time, especially the starting timeslice.
We therefore repeat the fits for different starting timeslices and obtain our
best estimate as a weighted average over the region where the results have
stabilized (see Fig.~\ref{fig:excited_energy_levels_overlap_d200_0_mom}).

For the nonlinear exponential fits we use the \textsc{VarPro} method
\cite{GolP73}, which only needs initial guesses for the energy levels.
Monitoring the ground state energy, we find that the extraction works well for
all ensembles for momenta up to 1 $\Gevs$, see Fig.
\ref{fig:excited_ground_state_disp}. The results, which are used in the
subsequent analysis, are given in Appendix \ref{app:priors} for all ensembles.

\begin{figure}[t]
	\includegraphics[scale=.95]{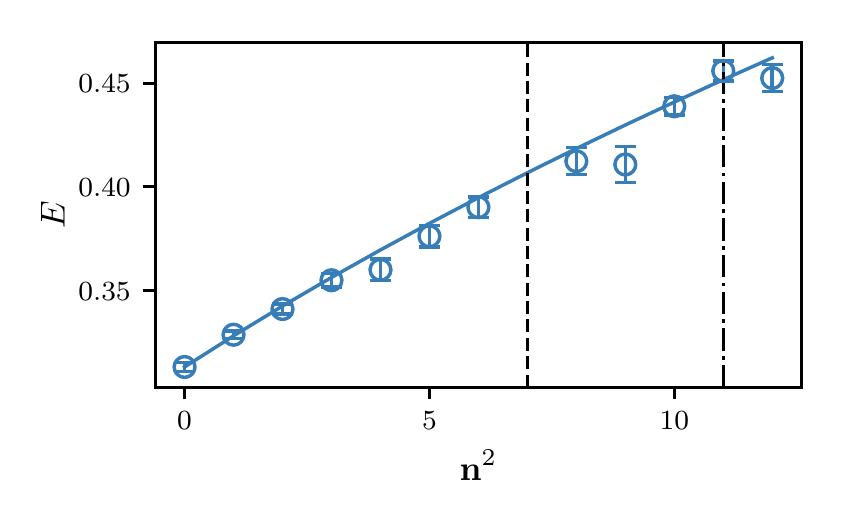}
	\includegraphics[scale=.95]{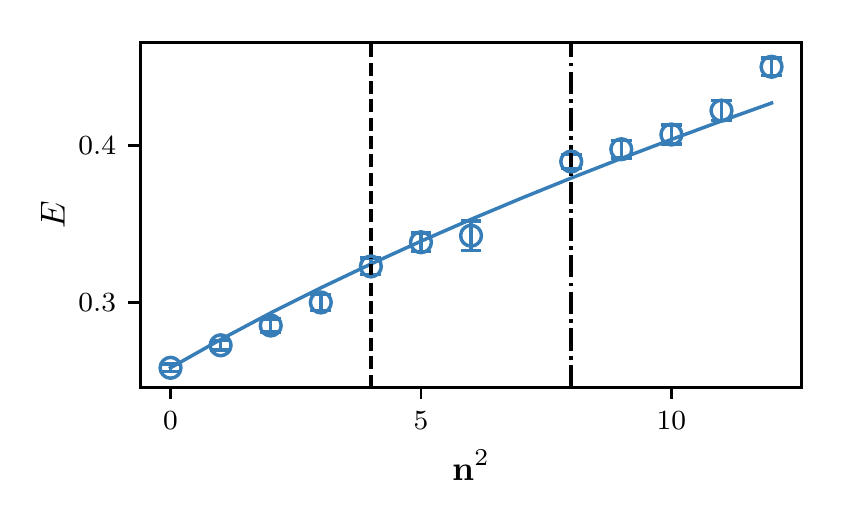}
\caption{
Ground state energy extracted in lattice units from two-state fits to the nucleon
two-point function on ensembles D200 (left) and J303 (right), where the blue line describes the relativistic
dispersion relation $({\bf p} = \frac{2\pi}{L}{\bf n})$.
The dashed, dashed-dotted lines indicate $Q^2 \leq 0.6$\,GeV$^2$, $Q^2 \leq 1.0$\,GeV$^2$, respectively.
}

\label{fig:excited_ground_state_disp}
\end{figure}

\begin{figure}[h]
	\includegraphics[scale=.95]{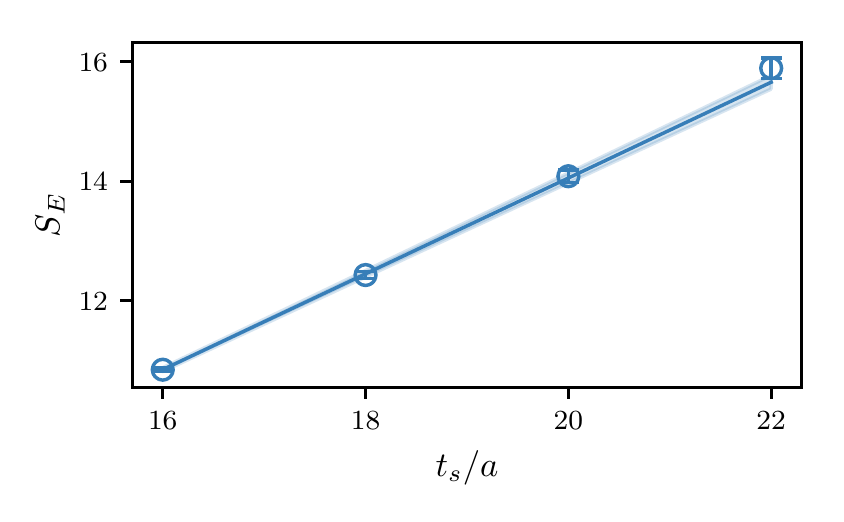}
	\includegraphics[scale=.95]{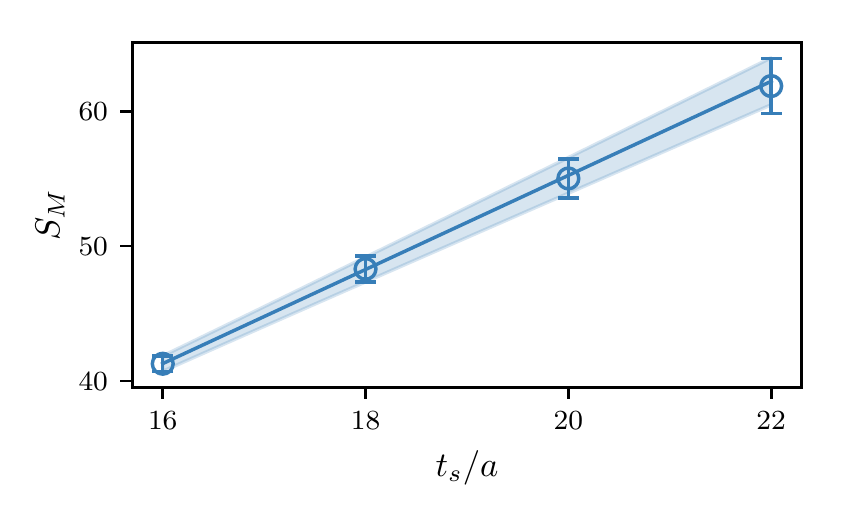}
\caption{Dependence of the summed ratios for the electric (left) and magnetic
	(right) effective form factors on the source-sink separation
	Eq.~(\ref{eq:summed_ratio_EM}). Data is shown for
	the first non-zero momentum on D200, i.e. $Q^2=0.089\,\Gevs$, together with a linear fit using
Eq.~(\ref{eq:sratio_gs}).}
\label{fig:summed_effective_ff_linear_fit}
\end{figure}

For the asymptotic limit of the ratio in Eq.~(\ref{eq:matelement}) we obtain
\begin{align}
	R^\text{as}(t,t_s,Q^2)&= r_{00} \Bigl\{
			1 
			+ \frac{\rho(\mathbf{q}^2)}{2}
			\bigl[e^{- \Delta(\mathbf q^2)  (t_s -t) }
			-e^{-\Delta(\mathbf q^2) t_s}\bigr] +
			\frac{\rho(\mathbf{0})}{2}\bigl[ e^{- \Delta(\mathbf 0 )
			\, t}-e^{- \Delta(\mathbf 0 ) \, t_s } \bigr]
		\Bigr\}
			\nonumber\\
			&+ r_{01} e^{-\Delta(\mathbf q ^2)  \, t} +  r_{10}
			e^{-\Delta(\mathbf 0 ) \, (t_s -t)} + r_{11}
			e^{-\Delta(\mathbf q ^2) \,  t} e^{-\Delta(\mathbf 0 )
			\, (t_s -t)}  + \dots \,, 
\label{eq:asymptotic_limit_ratio}
\end{align}
where $r_{00}$ is proportional to the ground state matrix element\footnote{Note
	that $r_{01},r_{10}$ and $r_{11}$, even though the indexing might
	suggest otherwise, are not directly proportional to the matrix
element of the current. The ellipsis denotes terms with at least one exponential
from the 2- and 3-point functions and further terms from the excitation spectrum.} $G_\text{E/M}(Q^2)$, and the last two terms in the first
line come from
the expansion of the two-point functions in Eq.(\ref{eq:matelement}).
For each value of the momentum $\mathbf q^2$, the gap $\Delta(\mathbf q^2)$ and
the terms proportional to $\rho$ are universal, and we therefore proceed by
fitting the electric and magnetic effective form factors simultaneously. The
fits are performed up to a maximum transfer momentum of about 1\,GeV$^2$. The
fits to the effective form factors are stabilized using Gaussian priors for
$\Delta$ and $\rho$ (see Appendix \ref{app:priors}), whose central values are
set to the results of the fits to the two-point function.  We monitor the
impact of our particular choice for the priors on the extracted form factor
values by varying the width of the priors in all fits. To this end we multiply
the errors of $\Delta$ and $\rho$ by a factor between 1 and 5. The associated
fit results are labeled 1x, 2x, $\dots$, 5x to
reflect their dependence on the prior width. In order for the prior to be
effective, we constrain the width to maximally half the mean value. The idea is
to strike a balance between the statistical accuracy of our extracted values and
any potential bias introduced by the priors. Therefore we choose the final
values for the two-state method to come from the analysis with the maximum prior
width that (a) gives values compatible within errors with all determinations
based on a smaller width, and (b) maintains an acceptable error increase.  We
made a rather conservative choice for the latter, allowing
the error to increase by factors between 2 and 4 for prior widths 2x to 5x.  For
the ensembles in this work, it turns out that, for $Q^2\leq 0.5\, \Gevs$ the
final values come from an analysis with prior width 5x.

In addition to the above analysis, we perform fits to the summed correlators.
The summation  method takes advantage of the fact that in the ratios of
Eq.~(\ref{eq:asymptotic_limit_ratio}),
when summed over timeslices in between source and sink, the contributions from
excited states are parametrically suppressed. 
Summing Eq.~(\ref{eq:asymptotic_limit_ratio}) over $t$, omitting $t_\mathrm{skip}$
timeslices at both ends, we obtain\footnote{Note that the ellipsis denotes
	terms after the expansion of the ratio in
	Eq.~(\ref{eq:asymptotic_limit_ratio}) to more than one exponential in the
nucleon energy.}
\begin{align}
	S (t_s)&=  \sum\limits_{t=t_{\mathrm{skip}}}^{t_s-t_{\mathrm{skip}}} R^{\text{as}}(t,t_s,Q^2)\nonumber\\
	&= r_{00}\Bigl[1 -\frac{\rho(\bm{q}^2)}{2}e^{-\Delta(\bm{q}^2) t_s}
	-\frac{ \rho(\bm{0})}{2} e^{-\Delta(\bm{0}) t_s} \Bigr]
	\frac{(t_s+a-2t_\mathrm{skip})}{a} \nonumber\\  
	&+\Bigl[r_{01}
	+r_{00}\frac{\rho(\bm{q}^2)}{2}\Bigr]\frac{(e^{\Delta(\bm{q}^2)
	(a-t_\mathrm{skip})}-e^{-\Delta(\bm{q}^2)
	(t_s-t_{\mathrm{skip}})})}{e^{a \Delta(\bm{q}^2)}-1}\nonumber\\
	&+\Bigl[r_{10} +r_{00}\frac{\rho(\bm{0})}{2}\Bigr]\frac{(e^{\Delta
	(\bm{0}) (a-t_\mathrm{skip})}-e^{-\Delta(\bm{0})
	(t_s-t_\mathrm{skip})})}{e^{a \Delta(\bm{0})}-1} + \dots .
\label{eq:asymptotic_limits_summed_ratio}
\end{align}
For the effective form factors we may write the summed ratios as
\begin{align}
        S_{\text{E}/\text{M}} (Q^2,t_s)&=  \sum\limits_{t=2 a}^{t_s-2a}
	G_{\text{E}/\text{M}}^{\text{eff}}(Q^2,t,t_s),
	\label{eq:summed_ratio_EM}
\end{align}
where in our analysis we use $t_\mathrm{skip}=2a$.
The summation method crucially depends on the computation of observables for
multiple source-sink separations. However, since the signal of the correlators
deteriorates with larger time separations, we are limited in the number of
available $t_s$. Our  current data does not deviate
significantly from linear behavior (see
Fig. \ref{fig:summed_effective_ff_linear_fit}) for any of the ensembles and we
therefore
fit Eq.~(\ref{eq:asymptotic_limits_summed_ratio}) in the asymptotic limit with only ground state contributions, i.e.
\begin{align}
	S_{\text{E}/\text{M}} (Q^2,t_s)&\xrightarrow{t_s\rightarrow \infty}
	C_{\text{E}/\text{M}}(Q^2) + \frac{t_s}{a} G_{\text{E}/\text{M}}(Q^2) + \dots \label{eq:sratio_gs}
\end{align}
where $C_{\text{E}/\text{M}}$ is  an irrelevant offset.

\begin{figure}[h]
	\includegraphics[scale=.95]{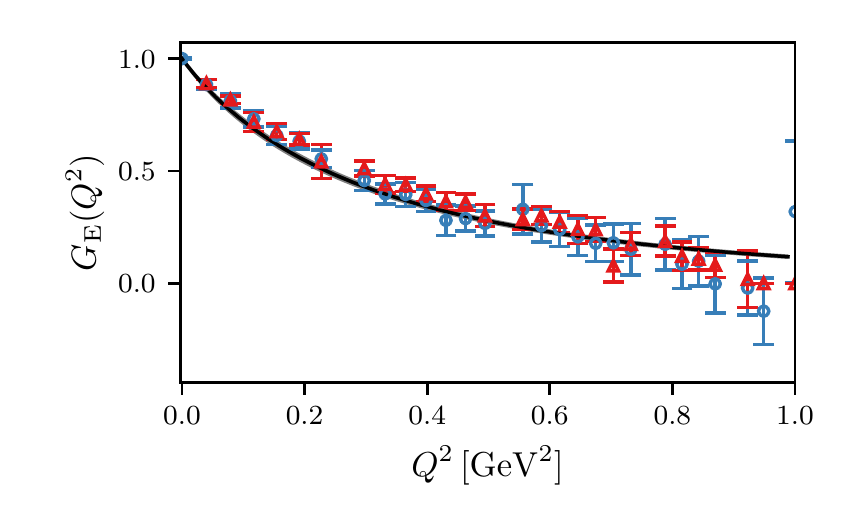}
	\includegraphics[scale=.95]{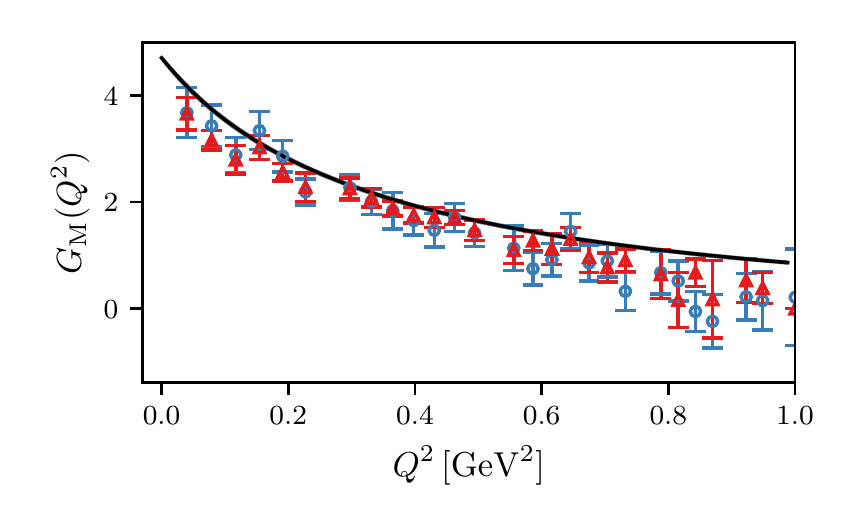}
	\caption{Comparison of the summation (blue circles) and two-state (red
	triangles) method for the priors in Appendix \ref{app:priors} on
ensemble E250. The black band, which corresponds to
the parameterization of \cite{Ye:2017gyb}, is displayed to enable a first comparison to phenomenology.
The continuum extrapolation of the lattice data is discussed in section~
\ref{sec:CCF}.}
	\label{fig:extracted_ff_summation_vs_twostate}
\end{figure}

The results of the two methods are shown together in
Fig.~\ref{fig:extracted_ff_summation_vs_twostate} for the near-physical ensemble
E250. In order to assess
systematics associated with excited-state effects, we apply all subsequent
analyses to the data obtained by both methods, hence we distinguish between
summation and explicit two-state fits in the following.

\section{Parameterization of the $Q^2$ Dependence}
\label{sec:Qsq}
Since the magnetic moment $\mu$ is defined as the intercept of $\GM$ and the
electric and magnetic radii are determined by the slope of the form
factors at zero momentum transfer, 
	\begin{subequations}
\begin{align}
	\mu &= G_\mathrm{M}(0) = \kappa+1 \, , \\
	\langle r_{\mathrm{E},\mathrm{M}}^2\rangle &= \left.
		-\frac{6}{G_{\mathrm{E},\mathrm{M}}(0)}
		\frac{\partial \, G_{\mathrm{E},\mathrm{M}}(Q^2)}{\partial  Q^2} \right|_{Q^2=0}\, , 
	\label{qs:def_res_rms_mu}
\end{align}
\end{subequations}
a description of the $Q^2$ dependence is
necessary. In analogy to \cite{Capitani:2015sba} we perform two
analyses. In the first analysis we parameterize the $Q^2$ dependence using
either a dipole or a
$z$-expansion ansatz \cite{Hill:2010yb} and subsequently perform chiral and continuum
extrapolations. In the second analysis we use covariant Baryon Chiral
Perturbation Theory (BChPT) \cite{Kubis:2000zd,Fuchs:2003ir,Schindler:2005ke,Bauer:2012pv} to fit
the available form factor data for all ensembles simultaneously. The latter approach combines the
chiral extrapolation and the fit to the $Q^2$ dependence, i.e. without
intermediary extraction of the radii using a separate ansatz of the $Q^2$ behavior for each
ensemble.  We use the expressions in  \cite{Bauer:2012pv}  as they include
vector meson degrees of freedom, e.g. $\rho$-mesons, in order to extend the
description of the form factors to larger values of $Q^2$. 

For the dipole fits we use the ansatz
\begin{align}
	G_{\text{E}/\text{M}}^\text{dipole} (Q^2)&=  \frac{a_{\text{E}/\text{M}}}{\Bigl( 1+ \frac{Q^2}{M^2} \Bigr)^2},
	\label{eq:dipole_ansatz}
\end{align}
where $a_{\text{E}}=1$. The dipole mass and $a_{\text{M}}$ are the fit parameters.  The
dipole fit is performed separately for the electric and the magnetic form
factor, thus the dipole mass is allowed to be different for $\GE$ and $\GM$. The
fits are performed with cuts in $Q^2$ of 0.6 $\Gevs$ and 0.9 $\Gevs$, and the
corresponding results
are collected in Appendix~\ref{app:dipole}.

\begin{figure}[h]
	\includegraphics[scale=.9]{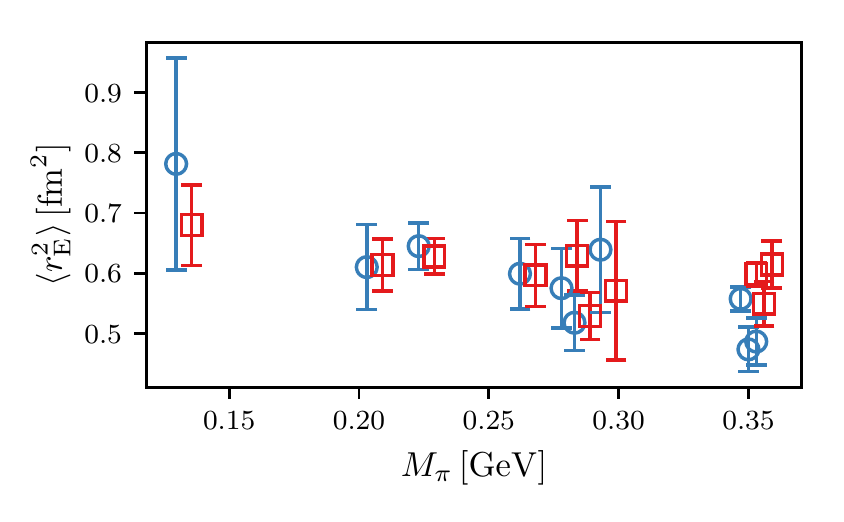}
	\includegraphics[scale=.9]{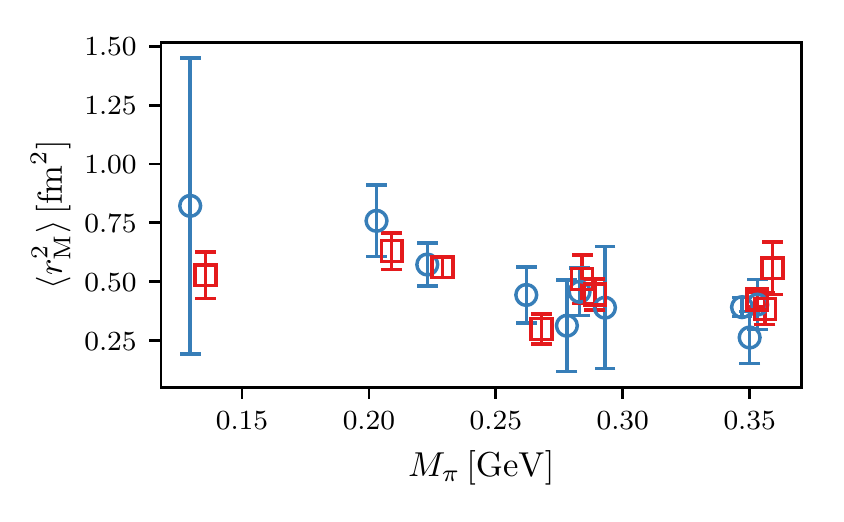}
	\includegraphics[scale=.9]{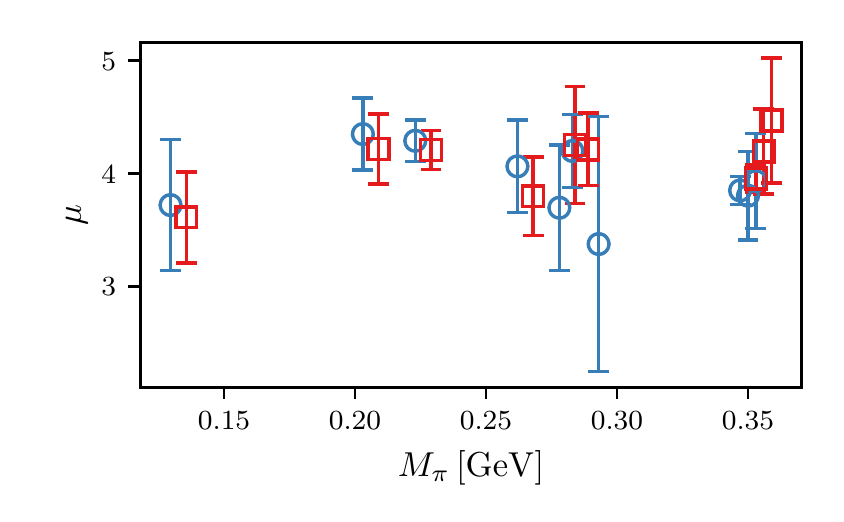}
	\caption{
		Dipole (red squares) and $z$-expansion results (blue circles) for the
radii and the magnetic moment, obtained with the summation method for
a momentum cut of 0.9\,GeV$^2$.  For better visibility, dipole and
$z$-expansion data points corresponding to the same ensemble have been
slightly separated from each other horizontally.
}
	\label{fig:dipole_qsq1}
\end{figure}

\begin{figure}[t]
	\centering
	\includegraphics[scale=0.9]{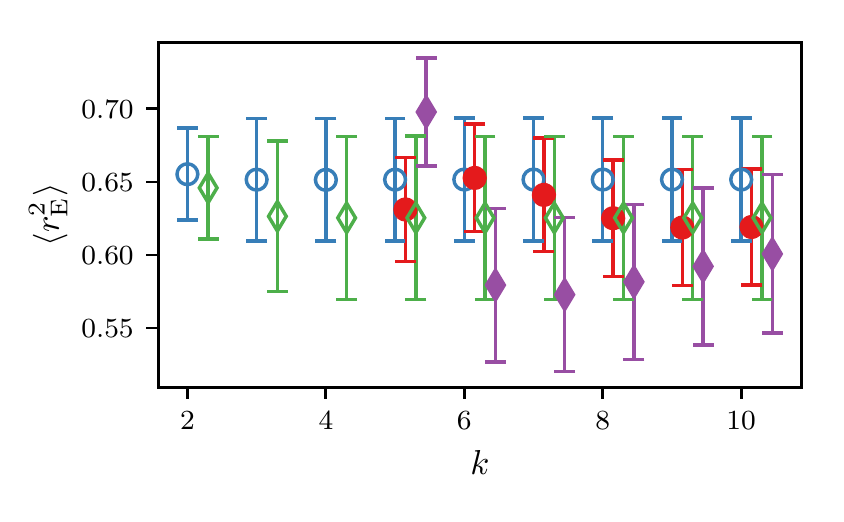}
	\includegraphics[scale=0.9]{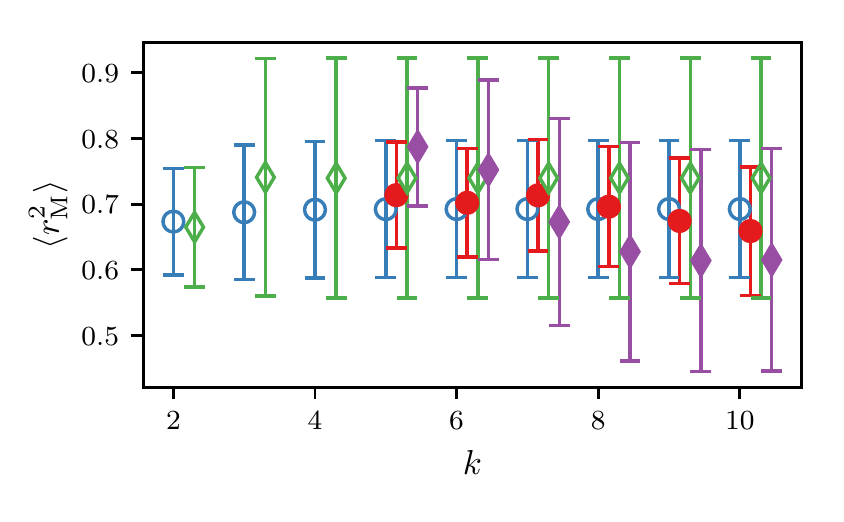}
	\caption{Dependence of the extracted electric and magnetic radii on the
	maximum order $k$ of the $z$-expansion for $t_0=0$ (circle) and
$t_0=-0.190 \,\Gevs$ (diamond) for the two-state method with (filled symbols)
and without (empty symbols) constraints of Eq.~(\ref{eq:largeqcond}) on ensemble D200.}
	\label{fig:z_exp_kord}
\end{figure}

A model-independent description of the $Q^2$ dependence of $G_{\rm{E},\rm{M}}$
can be obtained by employing the $z$-expansion \cite{Hill:2010yb}. The form factors may be decomposed as
\begin{subequations}
	\label{eq:zexp_GEGM}
\begin{align}
	G_\text{E}(Q^2) &= \sum \limits_{k=0}^\infty a_k\; z(Q^2)^k\, ,\\
	G_\text{M}(Q^2) &= \sum \limits_{k=0}^\infty b_k\; z(Q^2)^k\, ,
	\intertext{with}
	z(Q^2) &=  \frac{\sqrt{t_\text{cut}+Q^2} - \sqrt{t_\text{cut} -t_0}}{\sqrt{t_\text{cut}+Q^2}+\sqrt{t_\text{cut} -t_0}}.
\end{align}
\end{subequations}
The parameter $t_0$ is the value of $-Q^2$ which is mapped to $z=0$.
On each ensemble we set $t_{\mathrm{cut}}=4 M_\pi^2$, where $M_\pi$ denotes the
pion mass of the respective ensemble. 
In general we stabilize the fits using priors, where 
the coefficients are constrained using Gaussian distributions. The mean and
width of these distributions are chosen such that coefficients do not change
drastically with increasing  order of the expansion. 
To that end we first perform unconstrained fits to order $k=1$. For all subsequent
fits with orders $k>1$ in the $z$-expansion we add Gaussian priors for the first two
coefficients, centered around the means of the unconstrained fit, with widths of
5 times the corresponding error estimate. For  the remaining coefficients
we choose Gaussian priors
centered around zero with twice the maximum of the fitted coefficients
up to first order \footnote{We have checked that this procedure gives consistent
results with fits that use linearly extrapolated values for $\GM(0)$ and with
fits first removing a residual monopole dependence.}.
Throughout we always enforce $\GE(0)=
1$. Additionally we performed fits with stronger constraints on the large $k$ behavior
of the coefficients $a_k$ coming from the fall-off of the Sachs form factors
for large space like momentum transfer \cite{Lepage:1980fj}. These 
conditions may be implemented in the form of sum rules \cite{Lee:2015jqa}
\begin{align}
	\sum \limits_{k=n}^\infty k (k-1) \dots (k-n+1 ) a_k &= 0 , \qquad n=0,1,2,3.
	\label{eq:largeqcond}
\end{align}
For a given range in $Q^2$ the optimal value for $t_0$ is (see
Ref.~\cite{Hill:2010yb}) given by   
\begin{align}
	t_0^\mathrm{opt}(Q_{\text{max}}^2)&= t_{\text{cut}} \left( 1-
		\sqrt{1+Q_{\text{max}}^2/t_{\text{cut}}}
		\right).
\end{align}
We perform the fits with $t_0=0\, \Gevs$ and $t_0=t_0^\mathrm{opt}(0.6\, \Gevs)$ for momentum
cuts of $Q^2\leq 0.6 \, \Gevs $ and $Q^2\leq 0.9 \,\Gevs$; the results are
given in Appendix~\ref{app:dipole}. The dependence of the extracted electric
and magnetic radii on the maximum order of the $z$-expansion is shown in
Fig.~\ref{fig:z_exp_kord}. 
We find that the fits stabilize around  $k_\text{max}$ of 5 for the ansatz with weaker
assumptions about the bounds on the $z$-expansion parameters. Fits using the
large-$Q^2$ constraints of Eq.~(\ref{eq:largeqcond}) in general converge more
slowly, i.e. for larger values of
$k_\text{max}$; once plateaued, they however give compatible results. Since we are
interested in the low-$Q^2$ region, we choose the fits without imposing
Eq.~(\ref{eq:largeqcond}).
In Fig.~\ref{fig:dipole_qsq1} we show the extracted values for data with
$Q^2\leq 0.9\, \Gevs$ for the summation method. We see that the
values obtained from the different parameterizations of the $Q^2$ dependence are consistent, while
the dipole ansatz gives smaller errors.  We demand that at least four
non-vanishing values of $Q^2$ enter the dipole and $z$-expansion fits,
respectively. This implies, for instance, that ensemble H105 is excluded from
the analysis when $Q^2\leq 0.6 \ \Gevs$ is applied.

In the direct approach based on covariant BChPT, we fit the form factor data for
both $\GE$ and $\GM$ on all ensembles simultaneously, and thus the intermediate
parameterization of the $Q^2$
dependence is avoided.  While the ensembles are treated as statistically independent,
we do take the correlation among different $Q^2$ and between $\GE$ and
$\GM$  within an ensemble into account. 
Statistical errors are derived from the covariance matrix estimated in
the least-square fits that yield the results for charge radii, the magnetic
momentum and other fit parameters.
Since this method unifies the description of the $Q^2$ and the $M_\pi^2$ dependence,
it is presented in section~\ref{sec:CCF}.

\section{Chiral and Continuum Extrapolation}
\label{sec:CCF}

In this section we present our chiral and continuum extrapolation, in order to arrive
at results for the isovector electric and magnetic radii, as well as for the magnetic moment.
We adopt two strategies, the first of which follows up on the intermediate
results of section~\ref{sec:Qsq} and is presented in subsection~\ref{subsec:hbchpt_zexp}.
The second achieves a simultaneous description of the $Q^2$ and the $M_\pi^2$ dependence
of the results of section~\ref{sec:excited} and is presented in
subsection~\ref{subsec:direct_chpt}. The model-averaging procedure used to
arrive at our final results is explained in
subsection~\ref{subsec:model_average}.

\subsection{HBChPT extrapolation of the radii and the magnetic moment}
\label{subsec:hbchpt_zexp}

As detailed in Sec.~\ref{sec:Qsq}, we perform fits to the $Q^2$ dependence of
the Sachs form factors using a dipole and $z$-expansion ansatz. In this way we
obtain estimates for the magnetic moment and the electromagnetic radii on each
ensemble. 
For the chiral and continuum extrapolation of this data set we perform fits
based on Heavy
Baryon Chiral Perturbation Theory (HBChPT) \cite{Gockeler:2003ay}. We
apply momentum cuts to the data between $Q^2\leq 0.6 \, \Gevs$ and  $Q^2\leq 0.9 \,
\Gevs$  as well as a pion mass cut of $M_\pi\leq 0.29$ GeV. We fit the low-energy constants $\kappa_0,E_1,B_{10}$ and $B_{c2}$, while
setting the remaining constants to their phenomenological values\footnote{Note that when leaving $g_{\pi N\Delta}$ as a free parameter, the fit is in general not improved.}, i.e. 
\begin{subequations}
\begin{align}
		g_A&= 1.2724\,,  
		&F_\pi& =0.092 \,  \text{GeV}\, ,
		&c_v&= -2.26\, \text{GeV}^{-1}\,, \\
		\Delta &=  0.294 \, \text{GeV}\, ,&  g_{\pi N \Delta} &=  1.125\,. 
\end{align}
\end{subequations}

The results for
these fits extrapolated to the value of the pion mass in the isospin limit of
QCD (134.8 MeV \cite{Aoki:2016frl}) are given in 
Appendix~\ref{sec:app_hbchpt}. We have analyzed variations of the fit
function, amending the formulae with lattice spacing terms of $\mathcal{O}(a^2)$
and/or including finite volume effects\footnote{For the radii we choose a
simplified ansatz due to the subtleties in unambiguously defining finite volume
corrections, c.f. Ref.~\cite{Tiburzi:2007ep}.} \cite{Beane:2004tw}, i.e.

\begin{align}
	\begin{split}
	\kappa&= \kappa_{\text{HB}} + a^2 \kappa_a + \kappa_L M_\pi \Bigl(1-\frac{2}{M_\pi L}\Bigr) e^{-M_\pi L }\ ,  \\
	\res &= {\res}_\text{HB}  +  a^2 {\res}_a  + {\res}_L e^{-M_\pi L } \ ,\\
	\rms &= {\rms}_\text{HB}  +  a^2 {\rms}_a  + {\rms}_L e^{-M_\pi L } \ .
	\label{eq:hbchpt_ccf}
\end{split}
\end{align}
Here, the subscripts
HB, $a$ and L are used to distinguish the estimates in HBChPT from
the coefficients describing lattice artifacts and finite-volume
effects, respectively.
Fits to the HBChPT expression fail if they are applied to
the charge radii and the magnetic moment determined from the dipole ansatz.
Even excluding the results from ensemble E250, which are hardest to
accommodate in the fit, does not improve the HBChPT description
significantly. For the $z$-expansion extraction on the
other hand, a good
fit can be achieved for momentum cuts between $Q^2\leq 0.6\, \Gevs$ and
$Q^2\leq 0.9 \, \Gevs$ using HBChPT with and without lattice artifacts as
parameterized  in Eq.~(\ref{eq:hbchpt_ccf}). We find simultaneous fits of finite
lattice spacing and finite volume dependence to be unstable. Let us stress again
that we only include ensembles with at least four data points remaining after momentum cuts
are applied. This effectively limits the lower bound on the cut in $Q^2$, which
still allows for a chiral and continuum extrapolation, to roughly
$0.6$ $\Gevs$.
We also note that, even though for each ensemble a different
number of data points enters in the $z$-expansion, the relative weight in the
HBChPT fit does not reflect the density of available $Q^2$ points at low
momentum transfers. In this sense the two-step process, first performing $z$-expansion fits and
subsequently extrapolating using HBChPT, masks the relative paucity of data
points at small momentum transfer for some ensembles.

\begin{figure}[h]
	\includegraphics{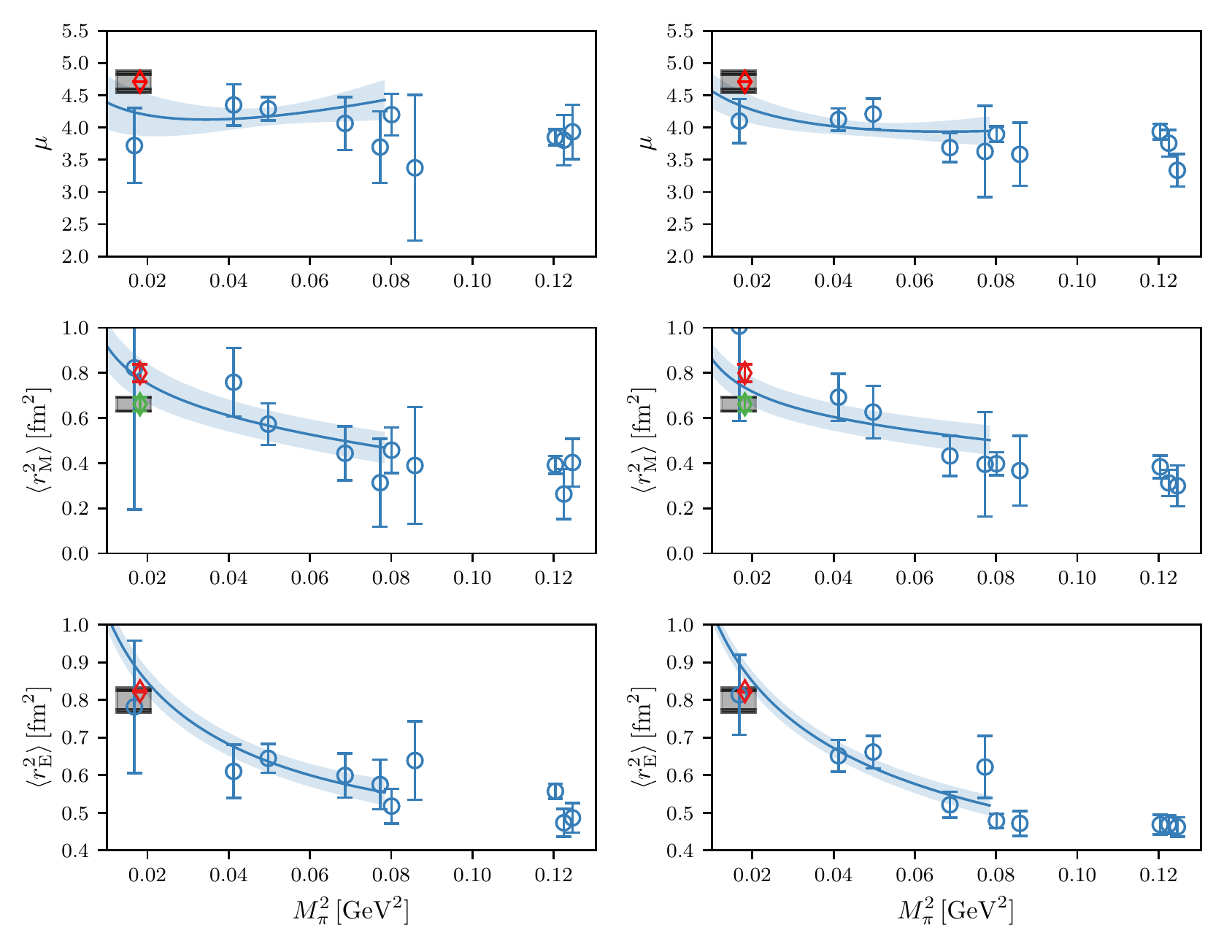}
	\caption{
		HBChPT fits to the radii and the magnetic moment, extracted via
		the $z$-expansion of the Sachs form factors determined 
		with the summation (left panel) and two-state method (right
		panel), with $Q^2 \leq 0.9$\,GeV$^2$ and $M_\pi \leq 0.28$\,GeV.
		Red points correspond to PDG values \cite{Zyla:2020zbs} for
		$\mu$ and $\res$. For $\rms$ we show the  result of a reanalysis
		of available world data from Ref.~\cite{Lee:2015jqa}, either
		based exclusively on the Mainz/A1 measurement \cite{Bernauer:2010wm}
		(green diamond) or excluding it from world data (red diamond).
		The gray bar depicts our final result of the model-averaged
		covariant BChPT analysis, where the width indicates statistical
		error, and the black bar includes systematic effects.}
\label{fig:ccf_comp_hbcpt_direct_fits}
\end{figure}

\subsection{Direct BChPT fits}
\label{subsec:direct_chpt}
As an alternative to the intermediate determination of the $Q^2$ dependence via
$z$-expansion or dipole fits, we perform direct fits of the covariant BChPT
expressions of \cite{Bauer:2012pv} to our form factor data. In this way we
obtain a combined description of the $Q^2$ and the $M_\pi$ dependence.  
The fit depends linearly on the four LECs $d_6,\tilde{c_6},d_x$ and
$G_\rho$ \cite{Bauer:2012pv}. It
turns out that an important advantage of this approach to extracting the
electromagnetic radii compared to the combined $z$-expansion and HBChPT analysis
is its stability against considerably lowering the momentum cuts applied.

For the direct fits we obtain results for various momentum cuts between $Q^2\leq
0.3 \, \Gevs$ and $Q^2\leq 0.6 \,  \Gevs$ for both the summation method and the
two-state method.
We perform the fits with and without terms
parameterizing the
lattice spacing and/or finite volume dependence, 
\begin{align}
	\GE(Q^2) &=  \GE(Q^2)^\chi + a^2 Q^2 \GE^a + Q^2 \GE^L e^{-M_\pi L} ,\nonumber\\
	\GM(Q^2) &=  \GM(Q^2)^\chi + a^2 \GM^a + \kappa_L M_\pi \Bigl(1-\frac{2}{M_\pi L}\Bigr) e^{-M_\pi L } + Q^2 \GM^L e^{-M_\pi L}.
	\label{fitmodel_cov_chpt}
\end{align}
Fits leaving  $\kappa_L$ as a free parameter are unstable, and we therefore fix
$\kappa_L$ to the value from HBChPT \cite{Beane:2004tw} \footnote{Note that in Ref.~\cite{Beane:2004tw} $f$ is
used instead of $F_\pi$ and that we are using the expression for the  isovector magnetic
moment.}, i.e.
\begin{align}
	\kappa_L&= - \frac{M_N g_A^2}{2 \pi F_\pi^2}.
\end{align}
As a cross-check, we perform fits where the lattice artifacts enter multiplicatively,
i.e.
\begin{align}
	\GE(Q^2) &=  \GE(Q^2)^\chi \left( 1 + a^2 Q^2 \GE^a + Q^2 \GE^L
	e^{-M_\pi L} \right),\nonumber\\
	\GM(Q^2) &=  \GM(Q^2)^\chi \left(1 + a^2 \GM^a + Q^2 \GM^L e^{-M_\pi
	L}\right)
	+ \kappa_L M_\pi \Bigl(1-\frac{2}{M_\pi L}\Bigr) e^{-M_\pi L } 
	\label{fitmodel_cov_chpt_method1} \, .
\end{align}

In total we have six models, i.e. without any lattice artifact and either
including discretization  or finite volume effects, for the additive
parameterization of Eq.~(\ref{fitmodel_cov_chpt}) or the multiplicative one of Eq.
(\ref{fitmodel_cov_chpt_method1}), respectively. 

Within our statistical errors, discretization and finite-volume
effects are hardly significant, implying that for most cuts the
corresponding coefficients are compatible with zero\footnote{Similar to the
	HBChPT fits we find simultaneous fits of finite
volume and lattice spacing dependence are not stable for all applied cuts and we do not include them
in our final estimate.}.
We perform simultaneous correlated fits of $G_\mathrm{E}(Q^2)$
and $G_\mathrm{M}(Q^2)$ for each model with the given data set and cuts applied
for every ensemble. The errors for the fit parameters are estimated using
derivatives of the $\chi^2$
function. 
The direct method leads to more stable results in comparison to the two-step
procedure in which fits to the  $z$-expansion are performed first, followed by
the  chiral
and continuum extrapolations using HBChPT. While the two methods give consistent
results, the direct
method has smaller errors, especially for the
magnetic form factor (see Fig.~\ref{fig:ccf_comp_hbcpt_direct_fits}). Moreover,
direct fits allow for more stringent cuts in the
momentum transfer, and the analysis is more driven by data in the low $Q^2$ region,
where the radii and magnetic moment are defined. Finally, the influence of priors
for the $z$-expansion is eliminated in the direct fits. For these reasons we
restrict the following presentation of the final results to the direct method, however noting
that the same procedure applied to the HBChPT extractions give consistent
results, albeit with larger errors.
\begin{figure}[!h]
	\includegraphics[scale=.92]{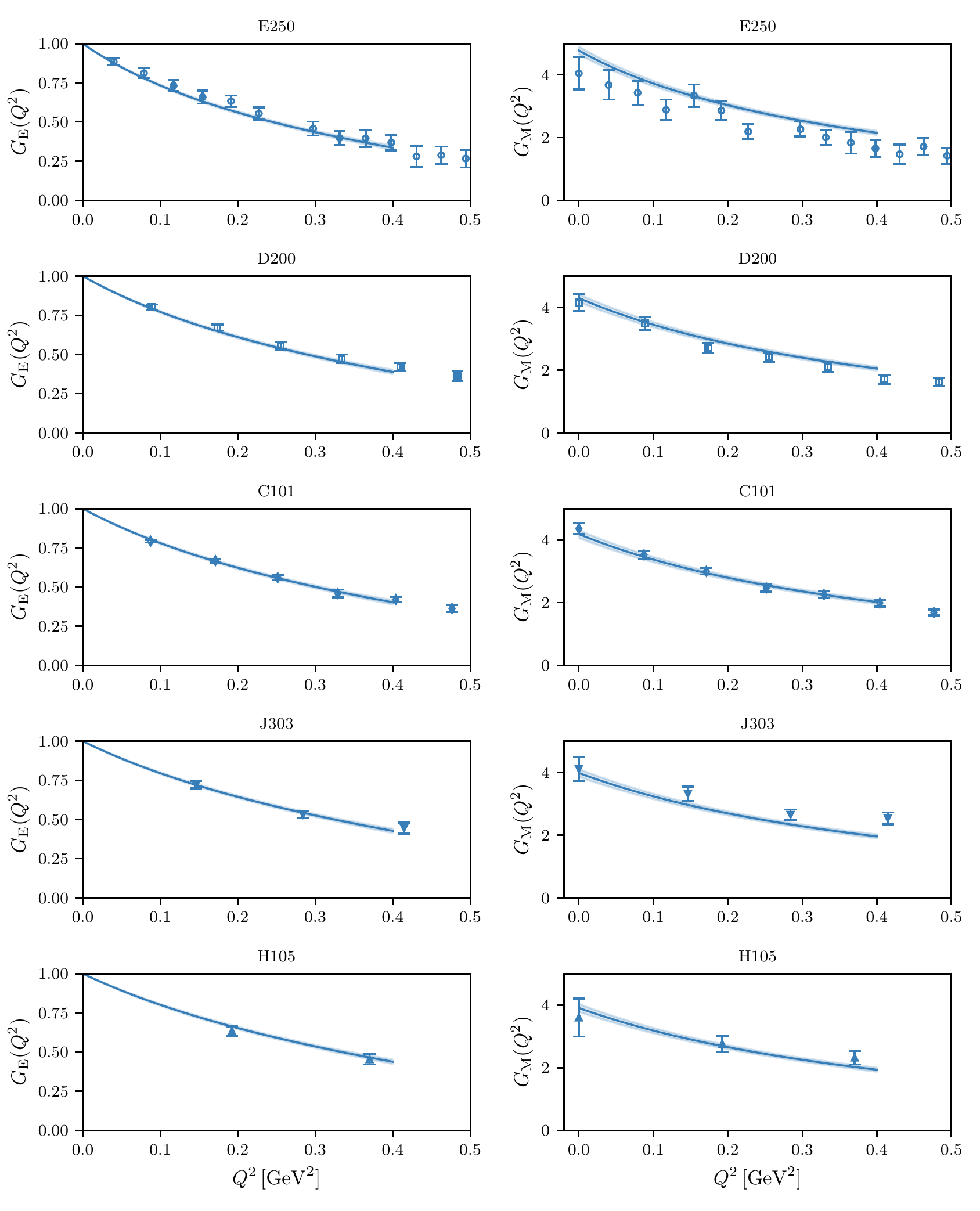}
	\caption{The summation-method data points for the Sachs form factors, and the blue band describing
the corresponding direct covariant BChPT fit with momentum cut 
	$Q^2\leq 0.4 \, \Gevs$, pion mass cut $0.28$ GeV and without lattice
artifacts. The data point for $\GM(0)$ is obtained from a linear fit to the
ratio of $\GM$ and $\GE$ and is not used in the direct covariant ChPT fit. The
fit depends linearly on the four LECs $d_6,\tilde{c_6},d_x$ and $G_\rho$ (c.f.
\cite{Capitani:2015sba}).}
	\label{fig:best_fit_direct_cov_sum}
\end{figure}

The quality of the direct covariant BChPT description is illustrated in
Fig.~\ref{fig:best_fit_direct_cov_sum}, where we present a typical fit to the
extracted form factors for summation data, corresponding to the model without
lattice artifacts or finite-volume corrections. The data is described rather well over the fit range in $Q^2$
for all ensembles, already suggesting that, within our statistical accuracy,
lattice artifacts are not discernible. Additionally,
Fig.~\ref{fig:best_fit_direct_cov_sum} illustrates the very different density of
low-$Q^2$ data points for each ensemble. For the magnetic moment, most recent
lattice determinations
\cite{Alexandrou:2018sjm,Alexandrou:2020aja,Jang:2019jkn,Shintani:2018ozy} lie
below the experimental value. For our most chiral ensemble, E250, we observe
(top right panel of Fig.~\ref{fig:best_fit_direct_cov_sum}) that the direct fit
lies somewhat above the data points for the magnetic form factor, while still
being compatible within the uncertainties. Thus the possible presence of a
non-negligible source of systematic error in calculations of the magnetic form
factor at the physical point merits further investigation.

\begin{figure}[!t]
	\centering
	\includegraphics[scale=.9]{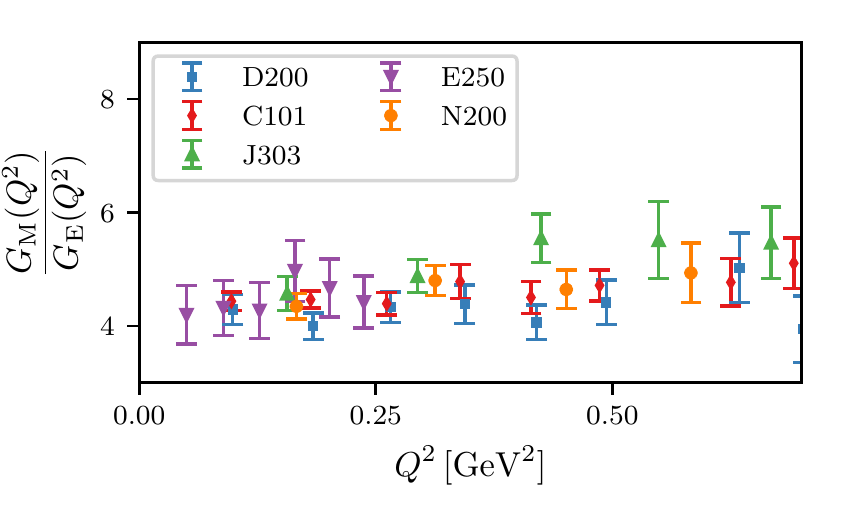}
	\includegraphics[scale=.9]{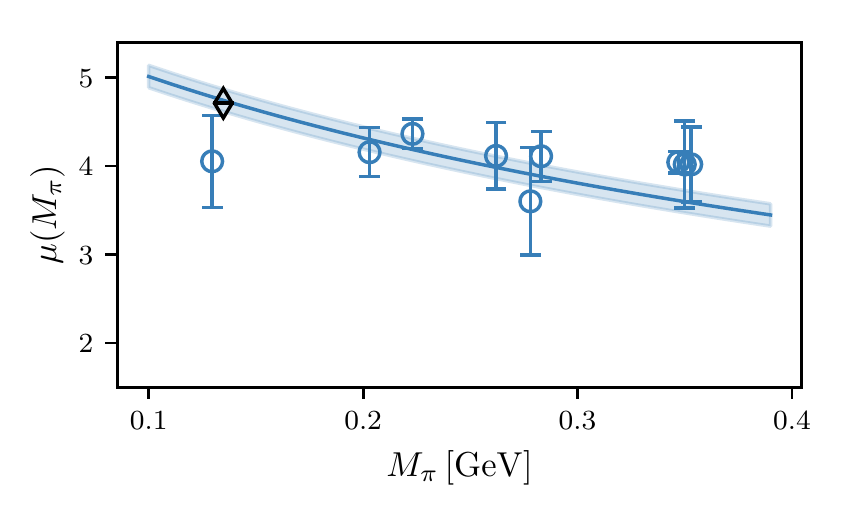}
	\caption{
The left panel shows the ratio of the magnetic and electric form factors for
summation method data. The right panel shows the direct covariant
BChPT result for the pion-mass dependence of the magnetic moment as a blue
shaded area, together with the data points obtained from a linear extrapolation
to $Q^2=0$ of the summation method data on the left panel. The black diamond corresponds
to the PDG value \cite{Zyla:2020zbs}. Note that the displayed
data points were not used in obtaining the covariant BChPT result.
	}

\label{fig:fit_cov_bchpt_vs_lineextrapol}
\end{figure}

From phenomenological dipole fits to experimental data, the ratio of the
electric and magnetic form factor is known to show a rather constant behavior
over a large range of $Q^2$. Indeed we observe similar behavior in our data
(see Fig.~\ref{fig:fit_cov_bchpt_vs_lineextrapol}), where the two-state and
summation data are rather flat with the exception of J303 showing signs of a
light upward slope. This pattern  motivates a linear extrapolation of the
ratio for  the magnetic moment up to $Q^2\leq 0.6 \, \Gevs$ ($Q^2\leq 0.29\,
\Gevs$ for E250), where the results are given in
Tab.~\ref{app:tab_ratio_intercept} and shown in
Fig.~\ref{fig:fit_cov_bchpt_vs_lineextrapol} (right). It is reassuring to see
that our fit, while not using these points, does reproduce the estimates
from a linear extrapolation of the ratio rather well.

\subsection{Model average and final result}
\label{subsec:model_average}
As we have no a priori preference for the results from the summation
method or two-state fits, we will treat both data sets on an equal
footing.  We obtain our final estimates and total errors from averages
over fit models and kinematic cuts using weights derived from the
Akaike Information Criterion (AIC) \cite{akaike1973second,Akaike:IEEE:1100705}.
In this context the momentum and pion mass cuts applied can be reinterpreted in
terms of a model selection problem
\cite{jay2020bayesian}. One may introduce for each would-be-cut data point an additional
fit parameter that matches the respective data point exactly, thus giving no contribution
to the $\chi^2$.  The corresponding data point is effectively excluded from the
least-square fit, while the model weight is decreased via the penalty
term for additional parameters in the AIC.
	\begin{figure}[b]
		\includegraphics[scale=0.9]{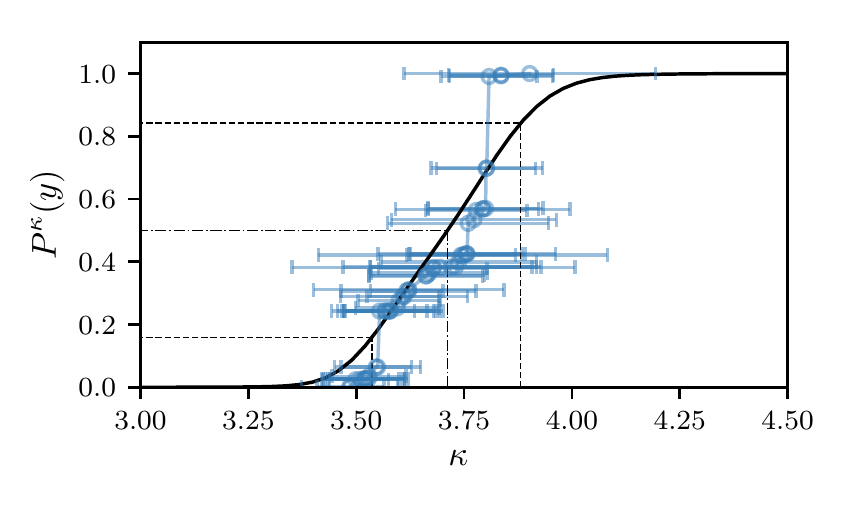}
		\includegraphics[scale=0.9]{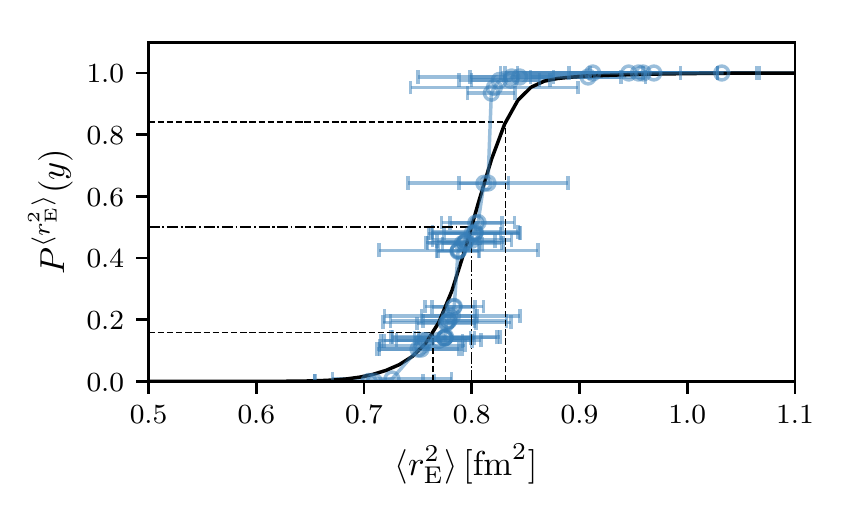}
		\includegraphics[scale=0.9]{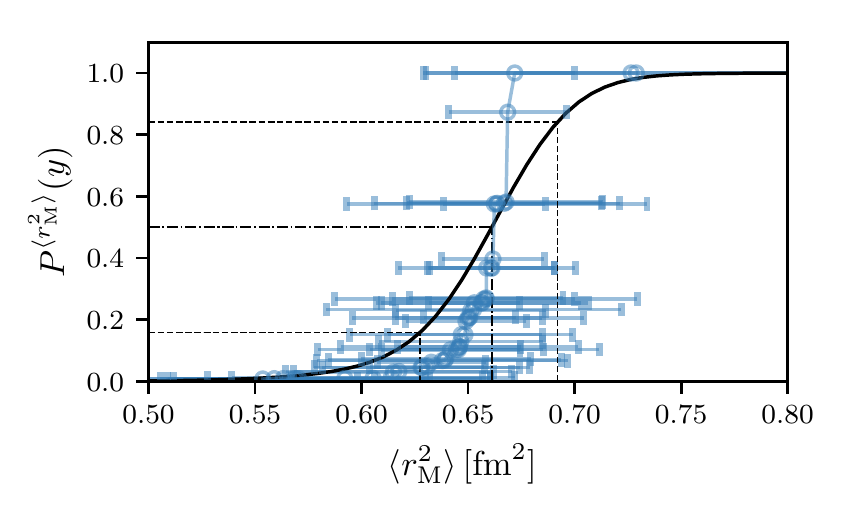}
		\caption{
			Cumulative distribution function of all fitted models, where dash-dotted
and short-dashed lines indicate median and 68\% percentiles, respectively.
		}
		\label{fig:ccf_cdf_model_average}
	\end{figure}
The AIC  reads
\begin{align}
	\text{AIC}_i = \chi_{\text{min},i}^2 + 2 n_f + 2 n_c,
	\label{eq:ccf_aic}
\end{align}
where $\chi_{\text{min},i}^2$ denotes the minimum of the weighted least
square, for the
i-th model, 
$n_f$ the number of fit parameters, and $n_c$ the number of cut data points.
In this way we obtain a criterion that takes the goodness of fit into account.
At the same time it penalizes increasing the number of fit parameters while it favors
including more actual data points \cite{Capitani:2015sba}.
For the weighting of different models on the same input data set we use
\begin{align}
	w^{\text{AIC}}_i &=  \frac{e^{-\frac{1}{2}\text{AIC}_i}}{
	\sum\limits_j e^{-\frac{1}{2}\text{AIC}_j}},
\end{align}
i.e. we normalize the AIC obtained for all models for summation and two-state
data separately. Finally, we apply a flat weight function to the estimates from
summation and two-state fits. We adopt the procedure from \cite{Borsanyi:2020mff}, which we briefly
sketch in the following,  for estimating the
systematic and statistical error of the model-averaged values. To that end we  treat the
model-averaged estimate as a random variable with the following cumulative
distribution function (CDF)
\begin{align}
	P^{x}(y) = \int\limits_{-\infty} ^y \sum \limits_i^n
	 w_i \mathcal{N}(y';x_i,\sigma_i^2) dy'
	 \label{eq:ccf_cdf_model_average}
\end{align}
i.e. the weighted sum of Gaussian distributions where the mean $x_i$ and
variance $\sigma_i^2$ is
given by the best estimate and fit error of each model, and the weight $w_i$ is obtained
as explained above. This effectively smoothens the otherwise rugged distribution
of model postdictions and allows for a more robust estimate of the distribution
parameters (see Fig.~\ref{fig:ccf_cdf_model_average}). The final value and total
error are easily  read off from the distribution
in Eq.~(\ref{eq:ccf_cdf_model_average}) as the median, and the 1-$\sigma$ percentiles, respectively.  Under the
assumptions that a
rescaling of all errors $\sigma_i$ only affects the statistical error but not the
systematic one, we can further separate the statistical and systematic errors,
c.f. \cite{Borsanyi:2020mff}.

In our previous work based on $N_f=2$ ensembles \cite{Capitani:2015sba}, we used
the spread in the central values as an estimate of systematic errors. While this
procedure is robust, it is also very conservative and susceptible to
overestimating the true error due to systematics. Therefore, in order to not be
overly conservative and still be able to incorporate systematic errors in a
robust way, we adopt the above model averaging procedure using AIC weights and
obtain as our final results
\begin{align}
	\begin{split}
		\kappa &=  3.71  \pm 0.11  \pm 0.13  , \\
		\res   &=  0.800 \pm 0.025 \pm 0.022 \, \mathrm{fm}^2,\\
		\rms   &=  0.661 \pm 0.030 \pm 0.011\, \mathrm{fm}^2\, ,
	\label{eq:ccf_final_values}
\end{split}
\end{align}
where the first and second errors refer to statistical
uncertainty and the total systematic error, respectively.

One may even proceed further and estimate the
individual contributions for every variation to the total systematic error. That
is achieved by building the CDF in Eq.~(\ref{eq:ccf_aic}) not over all variations but rather
first iterating over  a particular feature, e.g. a momentum cut, and performing the
analysis for every variant of that feature
separately. From this we then build a \emph{secondary} CDF like
Eq.~(\ref{eq:ccf_aic})  and extract the variation-specific systematic error.
Repeating this analysis for all variations we  obtain the following systematic
error budget,
\begin{align}
	\begin{split}
	\delta \kappa &= 0.11_{\text{exc}} \pm  0.03_{\text{artifacts}} \pm
	0.04_{{Q^2}} \pm 0.02_{m_\pi} \pm 0.02_{\text{method}}   (= 0.13),\\ 
	\delta \res &=  0.017_{\text{exc}} \pm  0.008_{\text{artifacts}} \pm
	0.007_{{Q^2}} \pm 0.001_{m_\pi} \pm 0.008_{\text{method}}\,\mathrm{fm}^2 (=0.022
	\, \mathrm{fm}^2),\\
	\delta \rms &= 0.006_{\text{exc}} \pm  0.007_{\text{artifacts}} \pm
	0.005_{{Q^2}} \pm 0.005_{m_\pi} \pm 0.005_{\text{method}}\,\mathrm{fm}^2 (=0.012
\, \mathrm{fm}^2).
\label{eq:ccf_final_systematic_error}
\end{split}
\end{align}
We note that, due to correlations, the individual terms added in quadrature (as
indicated on each line by the number in brackets) need
not exactly reproduce the total error of Eq.~(\ref{eq:ccf_final_values}). For the magnetic moment
and the electric radius, the dominant source of systematic error remains the excited state
contribution, while for the magnetic radius all systematic effects considered
here have comparable size. Moreover, in the current analysis the  magnetic radius is least affected by
systematic errors.

\begin{figure}[!t]
	\centering
	\includegraphics{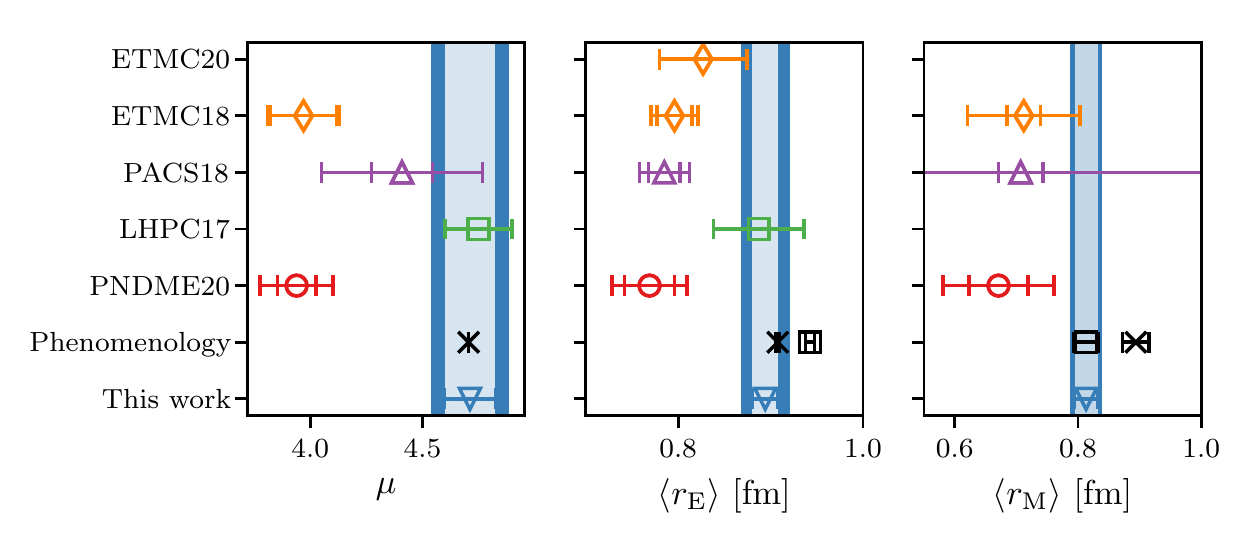}
	\caption{Comparison  of our best estimate (downward-pointing triangle)
		for the isovector quantities $\mu$, $\langle r_{\rm{E}}\rangle$,
		$\langle r_{\rm{M}}\rangle$ 
		to  other lattice calculations, i.e. PNDME \cite{Jang:2019jkn}
		(circle), ETMC \cite{Alexandrou:2020aja,Alexandrou:2018sjm}
		(diamond), PACS \cite{Shintani:2018ozy} (upward-pointing
		triangle), LHPC \cite{Hasan:2017wwt} (square).  The
		phenomenological value for $\mu$ is derived form the PDG values
		\cite{Zyla:2020zbs}. The two data points for $\langle
		r_\mathrm{E}\rangle$ are derived from CODATA 2018 (cross) or
	Mainz/A1 \cite{Bernauer:2010wm} (square) values for the proton electric
	charge radius, respectively, while the
		values for the neutron are taken from 
		\cite{Zyla:2020zbs} for both. The two data points for $\langle
		r_\mathrm{M}\rangle $ depict the values inferred from the proton
		results taken from the reanalysis of \cite{Lee:2015jqa} including only
		data from \cite{Bernauer:2010wm} (square) or excluding the
		Mainz/A1
		data set from the analysis (cross), while taking the values for
		the neutron magnetic radius from PDG \cite{Zyla:2020zbs} for
		both. For ease of comparison, the blue
		band represents our final result with the full uncertainty, with
	the light band indicating the statistical error.}
	\label{fig:comparison_result}
\end{figure}

In Fig.~\ref{fig:comparison_result} we compare our work to recent lattice
determinations and to the phenomenological values for the isovector  magnetic
moment and the isovector electromagnetic radii. While we postpone the comparison
of our results with phenomenological determinations of the radii to
section~\ref{sec:summary},
we remark that our value for the magnetic moment is in good agreement with the
experimentally precisely known difference of proton and neutron magnetic
moments.  As for the comparison with other lattice calculations, we note that
our estimate is compatible with the determinations from
\cite{Hasan:2017wwt,Alexandrou:2020aja}, while there is a sizeable difference to
the values from \cite{Jang:2019jkn,Shintani:2018ozy,Alexandrou:2018sjm}.  We
stress that the difference is not related to the issue of preferring direct fits
to the form factor data over the more conventional route via the $z$-expansion,
as the latter shows a trend to higher values for the radius for our data.  Our
error estimates for the statistical and systematic errors are comparable in size
with the other lattice determinations.  For the isovector magnetic moment we see
good agreement with phenomenology and \cite{Hasan:2017wwt,Shintani:2018ozy}. We
note that the  missing data point for $Q^2=0$ complicates the extraction of the
low-$Q^2$ observables in most recent lattice determinations.  Especially the
$z$-expansion fits, at least for orders $n\geq 2$, tend to overfit the dependence
of the form factor at low $Q^2$. In order to remedy this, either priors are
introduced or mock data points at $Q^2=0$, e.g. linear extrapolations of the
ratio of the isovector form factors, are used to stabilize the description.  We
note that the direct approach, in this sense, has less freedom and by itself
allows for considerably less variation in the form factors at low $Q^2$ (see
Fig.~\ref{fig:best_fit_direct_cov_sum}). We believe this to be responsible, in
large part, for the small errors we find in the isovector magnetic radius.

\section{Conclusions}
\label{sec:summary}

We have calculated the isovector electromagnetic form factors of the nucleon in lattice QCD
with dynamical up, down and strange quarks. The electromagnetic radii and the magnetic moment have
been extracted accounting for systematic effects due to excited states, finite
volume and non-zero lattice spacing. Our
final estimates are listed in Eq.~(\ref{eq:ccf_final_values}), with a detailed systematic
error budget given in Eq.~(\ref{eq:ccf_final_systematic_error}).

As an important benchmark, we reproduced the experimental value of the magnetic
moment with an overall precision of 3.6\%. 
The precision of the present calculation is significantly higher than
that of our earlier study in two-flavor QCD \cite{Capitani:2015sba}, especially
concerning the magnetic properties.
For the isovector electric charge radius, our result is in good agreement with the phenomenological estimate
inferred from the 2018 CODATA recommended value of the proton radius\footnote{The central value for the latter is very close, in comparison
to its uncertainty of 2.3 \textperthousand, to that extracted from muonic hydrogen
\cite{Antognini:1900ns},
which is yet 4.9 times more precise.}.
By contrast, after adding all errors in quadrature, we find a
2.4 $\sigma$ tension with the result from $ep$-scattering \cite{Bernauer:2010wm}.
For the isovector magnetic radius, on the other hand, our result agrees well with the
value inferred from the $ep$-scattering based determination
\cite{Bernauer:2010wm}, and exhibits a sizeable tension
with the other collected world data \cite{Lee:2015jqa}. For ease of comparison, we translate our estimate for the isovector
$\langle r_{\rm E}^2\rangle$ into a result for the proton radius with
the help of the experimental determination of the (squared) neutron
charge radius, $\langle r_{\rm n}^2\rangle=-0.1161(22)\,{\rm
fm}^2$ \cite{Zyla:2020zbs}. After combining all errors we obtain
$\langle r_{\rm p}^2\rangle^{1/2}=0.827(20)$\,fm, where the error is
completely dominated by the uncertainties of the lattice calculation.

Our analysis shows that in order to significantly improve on
the error for the radii and the magnetic moment, more points at smaller $Q^2$,
i.e. at large volumes and at physical pion mass are necessary. We plan to
extend our analysis to such ensembles as they become available.  
A promising strategy to further stabilize multi-state fits of the
three-point functions would be to perform a dedicated study of the
excitation spectrum.
Moreover, an analysis of the excited-state contributions in chiral effective
theory, as has been 
done 
for the axial form factor \cite{Bar:2019igf}, and the expression for the finite volume
dependence, would be highly desirable to improve the assessment of the related
systematic errors.

\subsection*{Acknowledgments}
This research is partly supported by the Deutsche Forschungsgemeinschaft (DFG,
German Research Foundation) through the Collaborative Research Center SFB 1044
``The low-energy frontier of the Standard Model'', under DFG grant HI 2048/1-2 (Project No. 399400745), and
in the Cluster of Excellence ``Precision Physics, Fundamental Interactions and
Structure of Matter'' (PRISMA+ EXC 2118/1) funded by the DFG within the German
Excellence strategy (Project ID 39083149). This work is supported by the European Research Council
(ERC) under the European Union’s Horizon 2020 research and innovation
programme through grant agreement 771971-SIMDAMA. Calculations for this project were
partly performed on the HPC clusters ``Clover'' and ``HIMster2'' at the Helmholtz 
Institute Mainz, and ``Mogon 2'' at Johannes Gutenberg-Universit{\"a}t Mainz.
Additional computer time has been allocated through projects HMZ21, HMZ23 and HMZ36 on
the supercomputer systems ``JUQUEEN'' and ``JUWELS'' at NIC, J\"ulich. 
The authors also gratefully acknowledge the Gauss Centre for Supercomputing e.V.
(www.gauss-centre.eu) for funding this project by providing computing time on
the GCS Supercomputer HAZEL HEN at H\"ochstleistungsrechenzentrum Stuttgart
(www.hlrs.de) under project GCS-HQCD.

Our programs use the QDP++ library \cite{Edwards:2004sx} and deflated SAP+GCR
solver from the openQCD package \cite{Luscher:2012av}, while the contractions
have been explicitly checked using \cite{Djukanovic:2016spv}. We are grateful to
our colleagues in the CLS initiative for sharing the gauge field configurations
on which this work is based.

\bibliography{refs}

\begin{thebibliography}{106}
\expandafter\ifx\csname natexlab\endcsname\relax\def\natexlab#1{#1}\fi
\expandafter\ifx\csname bibnamefont\endcsname\relax
  \def\bibnamefont#1{#1}\fi
\expandafter\ifx\csname bibfnamefont\endcsname\relax
  \def\bibfnamefont#1{#1}\fi
\expandafter\ifx\csname citenamefont\endcsname\relax
  \def\citenamefont#1{#1}\fi
\expandafter\ifx\csname url\endcsname\relax
  \def\url#1{\texttt{#1}}\fi
\expandafter\ifx\csname urlprefix\endcsname\relax\def\urlprefix{URL }\fi
\providecommand{\bibinfo}[2]{#2}
\providecommand{\eprint}[2][]{\url{#2}}

\bibitem[{\citenamefont{Ashman et~al.}(1988)}]{Ashman:1987hv}
\bibinfo{author}{\bibfnamefont{J.}~\bibnamefont{Ashman}} \bibnamefont{et~al.}
  (\bibinfo{collaboration}{European Muon Collaboration}),
  \bibinfo{journal}{Phys. Lett.} \textbf{\bibinfo{volume}{B206}},
  \bibinfo{pages}{364} (\bibinfo{year}{1988}), \bibinfo{note}{[,340(1987)]}.

\bibitem[{\citenamefont{Aidala et~al.}(2013)\citenamefont{Aidala, Bass, Hasch,
  and Mallot}}]{Aidala:2012mv}
\bibinfo{author}{\bibfnamefont{C.~A.} \bibnamefont{Aidala}},
  \bibinfo{author}{\bibfnamefont{S.~D.} \bibnamefont{Bass}},
  \bibinfo{author}{\bibfnamefont{D.}~\bibnamefont{Hasch}}, \bibnamefont{and}
  \bibinfo{author}{\bibfnamefont{G.~K.} \bibnamefont{Mallot}},
  \bibinfo{journal}{Rev. Mod. Phys.} \textbf{\bibinfo{volume}{85}},
  \bibinfo{pages}{655} (\bibinfo{year}{2013}), \eprint{1209.2803}.

\bibitem[{\citenamefont{Chen et~al.}(2009)\citenamefont{Chen, Sun, Lu, Wang,
  and Goldman}}]{Chen:2009mr}
\bibinfo{author}{\bibfnamefont{X.-S.} \bibnamefont{Chen}},
  \bibinfo{author}{\bibfnamefont{W.-M.} \bibnamefont{Sun}},
  \bibinfo{author}{\bibfnamefont{X.-F.} \bibnamefont{Lu}},
  \bibinfo{author}{\bibfnamefont{F.}~\bibnamefont{Wang}}, \bibnamefont{and}
  \bibinfo{author}{\bibfnamefont{T.}~\bibnamefont{Goldman}},
  \bibinfo{journal}{Phys. Rev. Lett.} \textbf{\bibinfo{volume}{103}},
  \bibinfo{pages}{062001} (\bibinfo{year}{2009}), \eprint{0904.0321}.

\bibitem[{\citenamefont{Ji et~al.}(2021)\citenamefont{Ji, Yuan, and
  Zhao}}]{Ji:2020ena}
\bibinfo{author}{\bibfnamefont{X.}~\bibnamefont{Ji}},
  \bibinfo{author}{\bibfnamefont{F.}~\bibnamefont{Yuan}}, \bibnamefont{and}
  \bibinfo{author}{\bibfnamefont{Y.}~\bibnamefont{Zhao}},
  \bibinfo{journal}{Nature Rev. Phys.} \textbf{\bibinfo{volume}{3}},
  \bibinfo{pages}{27} (\bibinfo{year}{2021}), \eprint{2009.01291}.

\bibitem[{\citenamefont{Pohl et~al.}(2010)}]{Pohl:2010zza}
\bibinfo{author}{\bibfnamefont{R.}~\bibnamefont{Pohl}} \bibnamefont{et~al.},
  \bibinfo{journal}{Nature} \textbf{\bibinfo{volume}{466}},
  \bibinfo{pages}{213} (\bibinfo{year}{2010}).

\bibitem[{\citenamefont{Karr et~al.}(2020)\citenamefont{Karr, Marchand, and
  Voutier}}]{Karr:2020wgh}
\bibinfo{author}{\bibfnamefont{J.-P.} \bibnamefont{Karr}},
  \bibinfo{author}{\bibfnamefont{D.}~\bibnamefont{Marchand}}, \bibnamefont{and}
  \bibinfo{author}{\bibfnamefont{E.}~\bibnamefont{Voutier}},
  \bibinfo{journal}{Nature Rev. Phys.} \textbf{\bibinfo{volume}{2}},
  \bibinfo{pages}{601} (\bibinfo{year}{2020}).

\bibitem[{\citenamefont{Bernauer et~al.}(2010)}]{Bernauer:2010wm}
\bibinfo{author}{\bibfnamefont{J.~C.} \bibnamefont{Bernauer}}
  \bibnamefont{et~al.} (\bibinfo{collaboration}{A1 Collaboration}),
  \bibinfo{journal}{Phys. Rev. Lett.} \textbf{\bibinfo{volume}{105}},
  \bibinfo{pages}{242001} (\bibinfo{year}{2010}), \eprint{1007.5076}.

\bibitem[{\citenamefont{Mohr et~al.}(2012)\citenamefont{Mohr, Taylor, and
  Newell}}]{Mohr:2012tt}
\bibinfo{author}{\bibfnamefont{P.~J.} \bibnamefont{Mohr}},
  \bibinfo{author}{\bibfnamefont{B.~N.} \bibnamefont{Taylor}},
  \bibnamefont{and} \bibinfo{author}{\bibfnamefont{D.~B.}
  \bibnamefont{Newell}}, \bibinfo{journal}{Rev. Mod. Phys.}
  \textbf{\bibinfo{volume}{84}}, \bibinfo{pages}{1527} (\bibinfo{year}{2012}),
  \eprint{1203.5425}.

\bibitem[{\citenamefont{Antognini et~al.}(2013)}]{Antognini:1900ns}
\bibinfo{author}{\bibfnamefont{A.}~\bibnamefont{Antognini}}
  \bibnamefont{et~al.}, \bibinfo{journal}{Science}
  \textbf{\bibinfo{volume}{339}}, \bibinfo{pages}{417} (\bibinfo{year}{2013}).

\bibitem[{\citenamefont{Carlson}(2015)}]{Carlson:2015jba}
\bibinfo{author}{\bibfnamefont{C.~E.} \bibnamefont{Carlson}},
  \bibinfo{journal}{Prog. Part. Nucl. Phys.} \textbf{\bibinfo{volume}{82}},
  \bibinfo{pages}{59} (\bibinfo{year}{2015}), \eprint{1502.05314}.

\bibitem[{\citenamefont{Beyer et~al.}(2013)\citenamefont{Beyer, Alnis,
  Khabarova, Matveev, Parthey, Yost, Pohl, Udem, Hänsch, and
  Kolachevsky}}]{Beyer:2013jla}
\bibinfo{author}{\bibfnamefont{A.}~\bibnamefont{Beyer}},
  \bibinfo{author}{\bibfnamefont{J.}~\bibnamefont{Alnis}},
  \bibinfo{author}{\bibfnamefont{K.}~\bibnamefont{Khabarova}},
  \bibinfo{author}{\bibfnamefont{A.}~\bibnamefont{Matveev}},
  \bibinfo{author}{\bibfnamefont{C.~G.} \bibnamefont{Parthey}},
  \bibinfo{author}{\bibfnamefont{D.~C.} \bibnamefont{Yost}},
  \bibinfo{author}{\bibfnamefont{R.}~\bibnamefont{Pohl}},
  \bibinfo{author}{\bibfnamefont{T.}~\bibnamefont{Udem}},
  \bibinfo{author}{\bibfnamefont{T.~W.} \bibnamefont{Hänsch}},
  \bibnamefont{and}
  \bibinfo{author}{\bibfnamefont{N.}~\bibnamefont{Kolachevsky}},
  \bibinfo{journal}{Annalen Phys.} \textbf{\bibinfo{volume}{525}},
  \bibinfo{pages}{671} (\bibinfo{year}{2013}).

\bibitem[{\citenamefont{Thomas et~al.}(2019)\citenamefont{Thomas, Fleurbaey,
  Galtier, Julien, Biraben, and Nez}}]{Thomas:2019dad}
\bibinfo{author}{\bibfnamefont{S.}~\bibnamefont{Thomas}},
  \bibinfo{author}{\bibfnamefont{H.}~\bibnamefont{Fleurbaey}},
  \bibinfo{author}{\bibfnamefont{S.}~\bibnamefont{Galtier}},
  \bibinfo{author}{\bibfnamefont{L.}~\bibnamefont{Julien}},
  \bibinfo{author}{\bibfnamefont{F.}~\bibnamefont{Biraben}}, \bibnamefont{and}
  \bibinfo{author}{\bibfnamefont{F.}~\bibnamefont{Nez}},
  \bibinfo{journal}{Annalen Phys.} \textbf{\bibinfo{volume}{2019}},
  \bibinfo{pages}{1800363} (\bibinfo{year}{2019}), \eprint{1903.04252}.

\bibitem[{\citenamefont{Fleurbaey et~al.}(2018)\citenamefont{Fleurbaey,
  Galtier, Thomas, Bonnaud, Julien, Biraben, Nez, Abgrall, and
  Gu\'ena}}]{Fleurbaey:2018fih}
\bibinfo{author}{\bibfnamefont{H.}~\bibnamefont{Fleurbaey}},
  \bibinfo{author}{\bibfnamefont{S.}~\bibnamefont{Galtier}},
  \bibinfo{author}{\bibfnamefont{S.}~\bibnamefont{Thomas}},
  \bibinfo{author}{\bibfnamefont{M.}~\bibnamefont{Bonnaud}},
  \bibinfo{author}{\bibfnamefont{L.}~\bibnamefont{Julien}},
  \bibinfo{author}{\bibfnamefont{F.}~\bibnamefont{Biraben}},
  \bibinfo{author}{\bibfnamefont{F.}~\bibnamefont{Nez}},
  \bibinfo{author}{\bibfnamefont{M.}~\bibnamefont{Abgrall}}, \bibnamefont{and}
  \bibinfo{author}{\bibfnamefont{J.}~\bibnamefont{Gu\'ena}},
  \bibinfo{journal}{Phys. Rev. Lett.} \textbf{\bibinfo{volume}{120}},
  \bibinfo{pages}{183001} (\bibinfo{year}{2018}), \eprint{1801.08816}.

\bibitem[{\citenamefont{Mihovilovič et~al.}(2017)}]{Mihovilovic:2016rkr}
\bibinfo{author}{\bibfnamefont{M.}~\bibnamefont{Mihovilovič}}
  \bibnamefont{et~al.}, \bibinfo{journal}{Phys. Lett.}
  \textbf{\bibinfo{volume}{B771}}, \bibinfo{pages}{194} (\bibinfo{year}{2017}),
  \eprint{1612.06707}.

\bibitem[{\citenamefont{Mihovilovi\v{c} et~al.}(2019)}]{Mihovilovic:2019jiz}
\bibinfo{author}{\bibfnamefont{M.}~\bibnamefont{Mihovilovi\v{c}}}
  \bibnamefont{et~al.} (\bibinfo{year}{2019}), \eprint{1905.11182}.

\bibitem[{\citenamefont{Gasparian}(2017)}]{Gasparian:2017cgp}
\bibinfo{author}{\bibfnamefont{A.~H.} \bibnamefont{Gasparian}}
  (\bibinfo{collaboration}{PRad Collaboration}), \bibinfo{journal}{JPS Conf.
  Proc.} \textbf{\bibinfo{volume}{13}}, \bibinfo{pages}{020052}
  (\bibinfo{year}{2017}).

\bibitem[{\citenamefont{Xiong et~al.}(2019)}]{Xiong:2019umf}
\bibinfo{author}{\bibfnamefont{W.}~\bibnamefont{Xiong}} \bibnamefont{et~al.},
  \bibinfo{journal}{Nature} \textbf{\bibinfo{volume}{575}},
  \bibinfo{pages}{147} (\bibinfo{year}{2019}).

\bibitem[{\citenamefont{Gasparian et~al.}(2020)}]{Gasparian:2020hog}
\bibinfo{author}{\bibfnamefont{A.}~\bibnamefont{Gasparian}}
  \bibnamefont{et~al.} (\bibinfo{collaboration}{PRad Collaboration})
  (\bibinfo{year}{2020}), \eprint{2009.10510}.

\bibitem[{\citenamefont{Grieser et~al.}(2018)\citenamefont{Grieser,
  Bonaventura, Brand, Hargens, Hetz, Le\ss{}mann, Westph{\"a}linger, and
  Khoukaz}}]{Grieser:2018qyq}
\bibinfo{author}{\bibfnamefont{S.}~\bibnamefont{Grieser}},
  \bibinfo{author}{\bibfnamefont{D.}~\bibnamefont{Bonaventura}},
  \bibinfo{author}{\bibfnamefont{P.}~\bibnamefont{Brand}},
  \bibinfo{author}{\bibfnamefont{C.}~\bibnamefont{Hargens}},
  \bibinfo{author}{\bibfnamefont{B.}~\bibnamefont{Hetz}},
  \bibinfo{author}{\bibfnamefont{L.}~\bibnamefont{Le\ss{}mann}},
  \bibinfo{author}{\bibfnamefont{C.}~\bibnamefont{Westph{\"a}linger}},
  \bibnamefont{and} \bibinfo{author}{\bibfnamefont{A.}~\bibnamefont{Khoukaz}},
  \bibinfo{journal}{Nucl. Instrum. Meth. A} \textbf{\bibinfo{volume}{906}},
  \bibinfo{pages}{120} (\bibinfo{year}{2018}), \eprint{1806.05409}.

\bibitem[{\citenamefont{Gilman et~al.}(2017)}]{Gilman:2017hdr}
\bibinfo{author}{\bibfnamefont{R.}~\bibnamefont{Gilman}} \bibnamefont{et~al.}
  (\bibinfo{collaboration}{MUSE Collaboration}) (\bibinfo{year}{2017}),
  \eprint{1709.09753}.

\bibitem[{\citenamefont{Dreisbach et~al.}(2019)}]{Dreisbach:2019pkc}
\bibinfo{author}{\bibfnamefont{C.}~\bibnamefont{Dreisbach}}
  \bibnamefont{et~al.} (\bibinfo{collaboration}{COMPASS++/AMBER working
  group}), \bibinfo{journal}{PoS} \textbf{\bibinfo{volume}{DIS2019}},
  \bibinfo{pages}{222} (\bibinfo{year}{2019}).

\bibitem[{\citenamefont{Krauth et~al.}(2017)}]{Krauth:2017ijq}
\bibinfo{author}{\bibfnamefont{J.}~\bibnamefont{Krauth}} \bibnamefont{et~al.},
  in \emph{\bibinfo{booktitle}{{52nd Rencontres de Moriond on EW Interactions
  and Unified Theories}}} (\bibinfo{year}{2017}), pp. \bibinfo{pages}{95--102},
  \eprint{1706.00696}.

\bibitem[{\citenamefont{Hoferichter et~al.}(2016)\citenamefont{Hoferichter,
  Kubis, Ruiz~de Elvira, Hammer, and Mei\ss{}ner}}]{Hoferichter:2016duk}
\bibinfo{author}{\bibfnamefont{M.}~\bibnamefont{Hoferichter}},
  \bibinfo{author}{\bibfnamefont{B.}~\bibnamefont{Kubis}},
  \bibinfo{author}{\bibfnamefont{J.}~\bibnamefont{Ruiz~de Elvira}},
  \bibinfo{author}{\bibfnamefont{H.~W.} \bibnamefont{Hammer}},
  \bibnamefont{and} \bibinfo{author}{\bibfnamefont{U.~G.}
  \bibnamefont{Mei\ss{}ner}}, \bibinfo{journal}{Eur. Phys. J. A}
  \textbf{\bibinfo{volume}{52}}, \bibinfo{pages}{331} (\bibinfo{year}{2016}),
  \eprint{1609.06722}.

\bibitem[{\citenamefont{Hammer and Mei\ss{}ner}(2020)}]{Hammer:2019uab}
\bibinfo{author}{\bibfnamefont{H.-W.} \bibnamefont{Hammer}} \bibnamefont{and}
  \bibinfo{author}{\bibfnamefont{U.-G.} \bibnamefont{Mei\ss{}ner}},
  \bibinfo{journal}{Sci. Bull.} \textbf{\bibinfo{volume}{65}},
  \bibinfo{pages}{257} (\bibinfo{year}{2020}), \eprint{1912.03881}.

\bibitem[{\citenamefont{Belushkin et~al.}(2007)\citenamefont{Belushkin, Hammer,
  and Mei{\ss}ner}}]{Belushkin:2006qa}
\bibinfo{author}{\bibfnamefont{M.~A.} \bibnamefont{Belushkin}},
  \bibinfo{author}{\bibfnamefont{H.~W.} \bibnamefont{Hammer}},
  \bibnamefont{and} \bibinfo{author}{\bibfnamefont{U.~G.}
  \bibnamefont{Mei{\ss}ner}}, \bibinfo{journal}{Phys. Rev. C}
  \textbf{\bibinfo{volume}{75}}, \bibinfo{pages}{035202}
  (\bibinfo{year}{2007}), \eprint{hep-ph/0608337}.

\bibitem[{\citenamefont{G{\"o}ckeler et~al.}(2005)\citenamefont{G{\"o}ckeler,
  Hemmert, Horsley, Pleiter, Rakow, Sch{\"a}fer, and
  Schierholz}}]{Gockeler:2003ay}
\bibinfo{author}{\bibfnamefont{M.}~\bibnamefont{G{\"o}ckeler}},
  \bibinfo{author}{\bibfnamefont{T.~R.} \bibnamefont{Hemmert}},
  \bibinfo{author}{\bibfnamefont{R.}~\bibnamefont{Horsley}},
  \bibinfo{author}{\bibfnamefont{D.}~\bibnamefont{Pleiter}},
  \bibinfo{author}{\bibfnamefont{P.~E.~L.} \bibnamefont{Rakow}},
  \bibinfo{author}{\bibfnamefont{A.}~\bibnamefont{Sch{\"a}fer}},
  \bibnamefont{and}
  \bibinfo{author}{\bibfnamefont{G.}~\bibnamefont{Schierholz}}
  (\bibinfo{collaboration}{QCDSF Collaboration}), \bibinfo{journal}{Phys. Rev.}
  \textbf{\bibinfo{volume}{D71}}, \bibinfo{pages}{034508}
  (\bibinfo{year}{2005}), \eprint{hep-lat/0303019}.

\bibitem[{\citenamefont{Syritsyn et~al.}(2010)}]{Syritsyn:2009mx}
\bibinfo{author}{\bibfnamefont{S.~N.} \bibnamefont{Syritsyn}}
  \bibnamefont{et~al.}, \bibinfo{journal}{Phys. Rev.}
  \textbf{\bibinfo{volume}{D81}}, \bibinfo{pages}{034507}
  (\bibinfo{year}{2010}), \eprint{0907.4194}.

\bibitem[{\citenamefont{Alexandrou et~al.}(2013)\citenamefont{Alexandrou,
  Constantinou, Dinter, Drach, Jansen, Kallidonis, and
  Koutsou}}]{Alexandrou:2013joa}
\bibinfo{author}{\bibfnamefont{C.}~\bibnamefont{Alexandrou}},
  \bibinfo{author}{\bibfnamefont{M.}~\bibnamefont{Constantinou}},
  \bibinfo{author}{\bibfnamefont{S.}~\bibnamefont{Dinter}},
  \bibinfo{author}{\bibfnamefont{V.}~\bibnamefont{Drach}},
  \bibinfo{author}{\bibfnamefont{K.}~\bibnamefont{Jansen}},
  \bibinfo{author}{\bibfnamefont{C.}~\bibnamefont{Kallidonis}},
  \bibnamefont{and} \bibinfo{author}{\bibfnamefont{G.}~\bibnamefont{Koutsou}},
  \bibinfo{journal}{Phys. Rev.} \textbf{\bibinfo{volume}{D88}},
  \bibinfo{pages}{014509} (\bibinfo{year}{2013}), \eprint{1303.5979}.

\bibitem[{\citenamefont{Alexandrou et~al.}(2017)\citenamefont{Alexandrou,
  Constantinou, Hadjiyiannakou, Jansen, Kallidonis, Koutsou, and Vaquero
  Aviles-Casco}}]{Alexandrou:2017ypw}
\bibinfo{author}{\bibfnamefont{C.}~\bibnamefont{Alexandrou}},
  \bibinfo{author}{\bibfnamefont{M.}~\bibnamefont{Constantinou}},
  \bibinfo{author}{\bibfnamefont{K.}~\bibnamefont{Hadjiyiannakou}},
  \bibinfo{author}{\bibfnamefont{K.}~\bibnamefont{Jansen}},
  \bibinfo{author}{\bibfnamefont{C.}~\bibnamefont{Kallidonis}},
  \bibinfo{author}{\bibfnamefont{G.}~\bibnamefont{Koutsou}}, \bibnamefont{and}
  \bibinfo{author}{\bibfnamefont{A.}~\bibnamefont{Vaquero Aviles-Casco}},
  \bibinfo{journal}{Phys. Rev.} \textbf{\bibinfo{volume}{D96}},
  \bibinfo{pages}{034503} (\bibinfo{year}{2017}), \eprint{1706.00469}.

\bibitem[{\citenamefont{Alexandrou et~al.}(2018)\citenamefont{Alexandrou,
  Bacchio, Constantinou, Finkenrath, Hadjiyiannakou, Jansen, Koutsou, and
  Casco}}]{Alexandrou:2018sjm}
\bibinfo{author}{\bibfnamefont{C.}~\bibnamefont{Alexandrou}},
  \bibinfo{author}{\bibfnamefont{S.}~\bibnamefont{Bacchio}},
  \bibinfo{author}{\bibfnamefont{M.}~\bibnamefont{Constantinou}},
  \bibinfo{author}{\bibfnamefont{J.}~\bibnamefont{Finkenrath}},
  \bibinfo{author}{\bibfnamefont{K.}~\bibnamefont{Hadjiyiannakou}},
  \bibinfo{author}{\bibfnamefont{K.}~\bibnamefont{Jansen}},
  \bibinfo{author}{\bibfnamefont{G.}~\bibnamefont{Koutsou}}, \bibnamefont{and}
  \bibinfo{author}{\bibfnamefont{A.~V.~A.} \bibnamefont{Casco}}
  (\bibinfo{year}{2018}), \eprint{1812.10311}.

\bibitem[{\citenamefont{Shanahan
  et~al.}(2014{\natexlab{a}})\citenamefont{Shanahan, Thomas, Young, Zanotti,
  Horsley, Nakamura, Pleiter, Rakow, Schierholz, and
  Stüben}}]{Shanahan:2014cga}
\bibinfo{author}{\bibfnamefont{P.~E.} \bibnamefont{Shanahan}},
  \bibinfo{author}{\bibfnamefont{A.~W.} \bibnamefont{Thomas}},
  \bibinfo{author}{\bibfnamefont{R.~D.} \bibnamefont{Young}},
  \bibinfo{author}{\bibfnamefont{J.~M.} \bibnamefont{Zanotti}},
  \bibinfo{author}{\bibfnamefont{R.}~\bibnamefont{Horsley}},
  \bibinfo{author}{\bibfnamefont{Y.}~\bibnamefont{Nakamura}},
  \bibinfo{author}{\bibfnamefont{D.}~\bibnamefont{Pleiter}},
  \bibinfo{author}{\bibfnamefont{P.~E.~L.} \bibnamefont{Rakow}},
  \bibinfo{author}{\bibfnamefont{G.}~\bibnamefont{Schierholz}},
  \bibnamefont{and} \bibinfo{author}{\bibfnamefont{H.}~\bibnamefont{Stüben}},
  \bibinfo{journal}{Phys. Rev.} \textbf{\bibinfo{volume}{D90}},
  \bibinfo{pages}{034502} (\bibinfo{year}{2014}{\natexlab{a}}),
  \eprint{1403.1965}.

\bibitem[{\citenamefont{Shanahan
  et~al.}(2014{\natexlab{b}})\citenamefont{Shanahan, Thomas, Young, Zanotti,
  Horsley, Nakamura, Pleiter, Rakow, Schierholz, and
  Stüben}}]{Shanahan:2014uka}
\bibinfo{author}{\bibfnamefont{P.~E.} \bibnamefont{Shanahan}},
  \bibinfo{author}{\bibfnamefont{A.~W.} \bibnamefont{Thomas}},
  \bibinfo{author}{\bibfnamefont{R.~D.} \bibnamefont{Young}},
  \bibinfo{author}{\bibfnamefont{J.~M.} \bibnamefont{Zanotti}},
  \bibinfo{author}{\bibfnamefont{R.}~\bibnamefont{Horsley}},
  \bibinfo{author}{\bibfnamefont{Y.}~\bibnamefont{Nakamura}},
  \bibinfo{author}{\bibfnamefont{D.}~\bibnamefont{Pleiter}},
  \bibinfo{author}{\bibfnamefont{P.~E.~L.} \bibnamefont{Rakow}},
  \bibinfo{author}{\bibfnamefont{G.}~\bibnamefont{Schierholz}},
  \bibnamefont{and} \bibinfo{author}{\bibfnamefont{H.}~\bibnamefont{Stüben}}
  (\bibinfo{collaboration}{CSSM, QCDSF/UKQCD Collaborations}),
  \bibinfo{journal}{Phys. Rev.} \textbf{\bibinfo{volume}{D89}},
  \bibinfo{pages}{074511} (\bibinfo{year}{2014}{\natexlab{b}}),
  \eprint{1401.5862}.

\bibitem[{\citenamefont{Bhattacharya et~al.}(2014)\citenamefont{Bhattacharya,
  Cohen, Gupta, Joseph, Lin, and Yoon}}]{Bhattacharya:2013ehc}
\bibinfo{author}{\bibfnamefont{T.}~\bibnamefont{Bhattacharya}},
  \bibinfo{author}{\bibfnamefont{S.~D.} \bibnamefont{Cohen}},
  \bibinfo{author}{\bibfnamefont{R.}~\bibnamefont{Gupta}},
  \bibinfo{author}{\bibfnamefont{A.}~\bibnamefont{Joseph}},
  \bibinfo{author}{\bibfnamefont{H.-W.} \bibnamefont{Lin}}, \bibnamefont{and}
  \bibinfo{author}{\bibfnamefont{B.}~\bibnamefont{Yoon}},
  \bibinfo{journal}{Phys. Rev.} \textbf{\bibinfo{volume}{D89}},
  \bibinfo{pages}{094502} (\bibinfo{year}{2014}), \eprint{1306.5435}.

\bibitem[{\citenamefont{Yamazaki et~al.}(2009)\citenamefont{Yamazaki, Aoki,
  Blum, Lin, Ohta, Sasaki, Tweedie, and Zanotti}}]{Yamazaki:2009zq}
\bibinfo{author}{\bibfnamefont{T.}~\bibnamefont{Yamazaki}},
  \bibinfo{author}{\bibfnamefont{Y.}~\bibnamefont{Aoki}},
  \bibinfo{author}{\bibfnamefont{T.}~\bibnamefont{Blum}},
  \bibinfo{author}{\bibfnamefont{H.-W.} \bibnamefont{Lin}},
  \bibinfo{author}{\bibfnamefont{S.}~\bibnamefont{Ohta}},
  \bibinfo{author}{\bibfnamefont{S.}~\bibnamefont{Sasaki}},
  \bibinfo{author}{\bibfnamefont{R.}~\bibnamefont{Tweedie}}, \bibnamefont{and}
  \bibinfo{author}{\bibfnamefont{J.}~\bibnamefont{Zanotti}},
  \bibinfo{journal}{Phys. Rev.} \textbf{\bibinfo{volume}{D79}},
  \bibinfo{pages}{114505} (\bibinfo{year}{2009}), \eprint{0904.2039}.

\bibitem[{\citenamefont{Shintani et~al.}(2019)\citenamefont{Shintani, Ishikawa,
  Kuramashi, Sasaki, and Yamazaki}}]{Shintani:2018ozy}
\bibinfo{author}{\bibfnamefont{E.}~\bibnamefont{Shintani}},
  \bibinfo{author}{\bibfnamefont{K.-I.} \bibnamefont{Ishikawa}},
  \bibinfo{author}{\bibfnamefont{Y.}~\bibnamefont{Kuramashi}},
  \bibinfo{author}{\bibfnamefont{S.}~\bibnamefont{Sasaki}}, \bibnamefont{and}
  \bibinfo{author}{\bibfnamefont{T.}~\bibnamefont{Yamazaki}},
  \bibinfo{journal}{Phys. Rev.} \textbf{\bibinfo{volume}{D99}},
  \bibinfo{pages}{014510} (\bibinfo{year}{2019}), \eprint{1811.07292}.

\bibitem[{\citenamefont{Ishikawa et~al.}(2018)\citenamefont{Ishikawa,
  Kuramashi, Sasaki, Tsukamoto, Ukawa, and Yamazaki}}]{Ishikawa:2018rew}
\bibinfo{author}{\bibfnamefont{K.-I.} \bibnamefont{Ishikawa}},
  \bibinfo{author}{\bibfnamefont{Y.}~\bibnamefont{Kuramashi}},
  \bibinfo{author}{\bibfnamefont{S.}~\bibnamefont{Sasaki}},
  \bibinfo{author}{\bibfnamefont{N.}~\bibnamefont{Tsukamoto}},
  \bibinfo{author}{\bibfnamefont{A.}~\bibnamefont{Ukawa}}, \bibnamefont{and}
  \bibinfo{author}{\bibfnamefont{T.}~\bibnamefont{Yamazaki}}
  (\bibinfo{collaboration}{PACS Collaboration}), \bibinfo{journal}{Phys. Rev.}
  \textbf{\bibinfo{volume}{D98}}, \bibinfo{pages}{074510}
  (\bibinfo{year}{2018}), \eprint{1807.03974}.

\bibitem[{\citenamefont{Green et~al.}(2014)\citenamefont{Green, Negele,
  Pochinsky, Syritsyn, Engelhardt, and Krieg}}]{Green:2014xba}
\bibinfo{author}{\bibfnamefont{J.~R.} \bibnamefont{Green}},
  \bibinfo{author}{\bibfnamefont{J.~W.} \bibnamefont{Negele}},
  \bibinfo{author}{\bibfnamefont{A.~V.} \bibnamefont{Pochinsky}},
  \bibinfo{author}{\bibfnamefont{S.~N.} \bibnamefont{Syritsyn}},
  \bibinfo{author}{\bibfnamefont{M.}~\bibnamefont{Engelhardt}},
  \bibnamefont{and} \bibinfo{author}{\bibfnamefont{S.}~\bibnamefont{Krieg}},
  \bibinfo{journal}{Phys. Rev.} \textbf{\bibinfo{volume}{D90}},
  \bibinfo{pages}{074507} (\bibinfo{year}{2014}), \eprint{1404.4029}.

\bibitem[{\citenamefont{Chambers et~al.}(2017)}]{Chambers:2017tuf}
\bibinfo{author}{\bibfnamefont{A.~J.} \bibnamefont{Chambers}}
  \bibnamefont{et~al.} (\bibinfo{collaboration}{QCDSF, UKQCD, CSSM
  Collaborations}), \bibinfo{journal}{Phys. Rev.}
  \textbf{\bibinfo{volume}{D96}}, \bibinfo{pages}{114509}
  (\bibinfo{year}{2017}), \eprint{1702.01513}.

\bibitem[{\citenamefont{Capitani et~al.}(2015)\citenamefont{Capitani,
  Della~Morte, Djukanovic, von Hippel, Hua, Jäger, Knippschild, Meyer, Rae,
  and Wittig}}]{Capitani:2015sba}
\bibinfo{author}{\bibfnamefont{S.}~\bibnamefont{Capitani}},
  \bibinfo{author}{\bibfnamefont{M.}~\bibnamefont{Della~Morte}},
  \bibinfo{author}{\bibfnamefont{D.}~\bibnamefont{Djukanovic}},
  \bibinfo{author}{\bibfnamefont{G.}~\bibnamefont{von Hippel}},
  \bibinfo{author}{\bibfnamefont{J.}~\bibnamefont{Hua}},
  \bibinfo{author}{\bibfnamefont{B.}~\bibnamefont{Jäger}},
  \bibinfo{author}{\bibfnamefont{B.}~\bibnamefont{Knippschild}},
  \bibinfo{author}{\bibfnamefont{H.~B.} \bibnamefont{Meyer}},
  \bibinfo{author}{\bibfnamefont{T.~D.} \bibnamefont{Rae}}, \bibnamefont{and}
  \bibinfo{author}{\bibfnamefont{H.}~\bibnamefont{Wittig}},
  \bibinfo{journal}{Phys. Rev.} \textbf{\bibinfo{volume}{D92}},
  \bibinfo{pages}{054511} (\bibinfo{year}{2015}), \eprint{1504.04628}.

\bibitem[{\citenamefont{Milbrath et~al.}(1998)}]{Milbrath:1997de}
\bibinfo{author}{\bibfnamefont{B.~D.} \bibnamefont{Milbrath}}
  \bibnamefont{et~al.} (\bibinfo{collaboration}{Bates FPP Collaboration}),
  \bibinfo{journal}{Phys. Rev. Lett.} \textbf{\bibinfo{volume}{80}},
  \bibinfo{pages}{452} (\bibinfo{year}{1998}), \bibinfo{note}{[Erratum: Phys.
  Rev. Lett.82,2221(1999)]}, \eprint{nucl-ex/9712006}.

\bibitem[{\citenamefont{Jones et~al.}(2000)}]{Jones:1999rz}
\bibinfo{author}{\bibfnamefont{M.~K.} \bibnamefont{Jones}} \bibnamefont{et~al.}
  (\bibinfo{collaboration}{Jefferson Lab Hall A Collaboration}),
  \bibinfo{journal}{Phys. Rev. Lett.} \textbf{\bibinfo{volume}{84}},
  \bibinfo{pages}{1398} (\bibinfo{year}{2000}), \eprint{nucl-ex/9910005}.

\bibitem[{\citenamefont{Gayou et~al.}(2002)}]{Gayou:2001qd}
\bibinfo{author}{\bibfnamefont{O.}~\bibnamefont{Gayou}} \bibnamefont{et~al.}
  (\bibinfo{collaboration}{Jefferson Lab Hall A Collaboration}),
  \bibinfo{journal}{Phys. Rev. Lett.} \textbf{\bibinfo{volume}{88}},
  \bibinfo{pages}{092301} (\bibinfo{year}{2002}), \eprint{nucl-ex/0111010}.

\bibitem[{\citenamefont{Punjabi et~al.}(2005)}]{Punjabi:2005wq}
\bibinfo{author}{\bibfnamefont{V.}~\bibnamefont{Punjabi}} \bibnamefont{et~al.},
  \bibinfo{journal}{Phys. Rev.} \textbf{\bibinfo{volume}{C71}},
  \bibinfo{pages}{055202} (\bibinfo{year}{2005}), \bibinfo{note}{[Erratum:
  Phys. Rev.C71,069902(2005)]}, \eprint{nucl-ex/0501018}.

\bibitem[{\citenamefont{Puckett et~al.}(2010)}]{Puckett:2010ac}
\bibinfo{author}{\bibfnamefont{A.~J.~R.} \bibnamefont{Puckett}}
  \bibnamefont{et~al.}, \bibinfo{journal}{Phys. Rev. Lett.}
  \textbf{\bibinfo{volume}{104}}, \bibinfo{pages}{242301}
  (\bibinfo{year}{2010}), \eprint{1005.3419}.

\bibitem[{\citenamefont{Puckett et~al.}(2012)}]{Puckett:2011xg}
\bibinfo{author}{\bibfnamefont{A.~J.~R.} \bibnamefont{Puckett}}
  \bibnamefont{et~al.}, \bibinfo{journal}{Phys. Rev.}
  \textbf{\bibinfo{volume}{C85}}, \bibinfo{pages}{045203}
  (\bibinfo{year}{2012}), \eprint{1102.5737}.

\bibitem[{\citenamefont{Rosenbluth}(1950)}]{Rosenbluth:1950yq}
\bibinfo{author}{\bibfnamefont{M.~N.} \bibnamefont{Rosenbluth}},
  \bibinfo{journal}{Phys. Rev.} \textbf{\bibinfo{volume}{79}},
  \bibinfo{pages}{615} (\bibinfo{year}{1950}).

\bibitem[{\citenamefont{Walker et~al.}(1994)}]{Walker:1993vj}
\bibinfo{author}{\bibfnamefont{R.~C.} \bibnamefont{Walker}}
  \bibnamefont{et~al.}, \bibinfo{journal}{Phys. Rev.}
  \textbf{\bibinfo{volume}{D49}}, \bibinfo{pages}{5671} (\bibinfo{year}{1994}).

\bibitem[{\citenamefont{Andivahis et~al.}(1994)}]{Andivahis:1994rq}
\bibinfo{author}{\bibfnamefont{L.}~\bibnamefont{Andivahis}}
  \bibnamefont{et~al.}, \bibinfo{journal}{Phys. Rev.}
  \textbf{\bibinfo{volume}{D50}}, \bibinfo{pages}{5491} (\bibinfo{year}{1994}).

\bibitem[{\citenamefont{Qattan et~al.}(2005)}]{Qattan:2004ht}
\bibinfo{author}{\bibfnamefont{I.~A.} \bibnamefont{Qattan}}
  \bibnamefont{et~al.}, \bibinfo{journal}{Phys. Rev. Lett.}
  \textbf{\bibinfo{volume}{94}}, \bibinfo{pages}{142301}
  (\bibinfo{year}{2005}), \eprint{nucl-ex/0410010}.

\bibitem[{\citenamefont{Guichon and Vanderhaeghen}(2003)}]{Guichon:2003qm}
\bibinfo{author}{\bibfnamefont{P.~A.~M.} \bibnamefont{Guichon}}
  \bibnamefont{and}
  \bibinfo{author}{\bibfnamefont{M.}~\bibnamefont{Vanderhaeghen}},
  \bibinfo{journal}{Phys. Rev. Lett.} \textbf{\bibinfo{volume}{91}},
  \bibinfo{pages}{142303} (\bibinfo{year}{2003}), \eprint{hep-ph/0306007}.

\bibitem[{\citenamefont{Arrington et~al.}(2007)\citenamefont{Arrington,
  Melnitchouk, and Tjon}}]{Arrington:2007ux}
\bibinfo{author}{\bibfnamefont{J.}~\bibnamefont{Arrington}},
  \bibinfo{author}{\bibfnamefont{W.}~\bibnamefont{Melnitchouk}},
  \bibnamefont{and} \bibinfo{author}{\bibfnamefont{J.~A.} \bibnamefont{Tjon}},
  \bibinfo{journal}{Phys. Rev.} \textbf{\bibinfo{volume}{C76}},
  \bibinfo{pages}{035205} (\bibinfo{year}{2007}), \eprint{0707.1861}.

\bibitem[{\citenamefont{Afanasev et~al.}(2017)\citenamefont{Afanasev, Blunden,
  Hasell, and Raue}}]{Afanasev:2017gsk}
\bibinfo{author}{\bibfnamefont{A.}~\bibnamefont{Afanasev}},
  \bibinfo{author}{\bibfnamefont{P.~G.} \bibnamefont{Blunden}},
  \bibinfo{author}{\bibfnamefont{D.}~\bibnamefont{Hasell}}, \bibnamefont{and}
  \bibinfo{author}{\bibfnamefont{B.~A.} \bibnamefont{Raue}},
  \bibinfo{journal}{Prog. Part. Nucl. Phys.} \textbf{\bibinfo{volume}{95}},
  \bibinfo{pages}{245} (\bibinfo{year}{2017}), \eprint{1703.03874}.

\bibitem[{\citenamefont{Bernauer et~al.}(2020)}]{Bernauer:2020vue}
\bibinfo{author}{\bibfnamefont{J.~C.} \bibnamefont{Bernauer}}
  \bibnamefont{et~al.} (\bibinfo{year}{2020}), \eprint{2008.05349}.

\bibitem[{\citenamefont{Bruno et~al.}(2015)}]{Bruno:2014jqa}
\bibinfo{author}{\bibfnamefont{M.}~\bibnamefont{Bruno}} \bibnamefont{et~al.},
  \bibinfo{journal}{JHEP} \textbf{\bibinfo{volume}{02}}, \bibinfo{pages}{043}
  (\bibinfo{year}{2015}), \eprint{1411.3982}.

\bibitem[{\citenamefont{Sheikholeslami and
  Wohlert}(1985)}]{Sheikholeslami:1985ij}
\bibinfo{author}{\bibfnamefont{B.}~\bibnamefont{Sheikholeslami}}
  \bibnamefont{and} \bibinfo{author}{\bibfnamefont{R.}~\bibnamefont{Wohlert}},
  \bibinfo{journal}{Nucl. Phys.} \textbf{\bibinfo{volume}{B259}},
  \bibinfo{pages}{572} (\bibinfo{year}{1985}).

\bibitem[{\citenamefont{Bulava and Schaefer}(2013)}]{Bulava:2013cta}
\bibinfo{author}{\bibfnamefont{J.}~\bibnamefont{Bulava}} \bibnamefont{and}
  \bibinfo{author}{\bibfnamefont{S.}~\bibnamefont{Schaefer}},
  \bibinfo{journal}{Nucl. Phys.} \textbf{\bibinfo{volume}{B874}},
  \bibinfo{pages}{188} (\bibinfo{year}{2013}), \eprint{1304.7093}.

\bibitem[{\citenamefont{L{\"u}scher and Weisz}(1985)}]{Luscher:1984xn}
\bibinfo{author}{\bibfnamefont{M.}~\bibnamefont{L{\"u}scher}} \bibnamefont{and}
  \bibinfo{author}{\bibfnamefont{P.}~\bibnamefont{Weisz}},
  \bibinfo{journal}{Commun. Math. Phys.} \textbf{\bibinfo{volume}{97}},
  \bibinfo{pages}{59} (\bibinfo{year}{1985}), \bibinfo{note}{[Erratum: Commun.
  Math. Phys.98,433(1985)]}.

\bibitem[{\citenamefont{Schaefer et~al.}(2011)\citenamefont{Schaefer, Sommer,
  and Virotta}}]{Schaefer:2010hu}
\bibinfo{author}{\bibfnamefont{S.}~\bibnamefont{Schaefer}},
  \bibinfo{author}{\bibfnamefont{R.}~\bibnamefont{Sommer}}, \bibnamefont{and}
  \bibinfo{author}{\bibfnamefont{F.}~\bibnamefont{Virotta}}
  (\bibinfo{collaboration}{ALPHA Collaboration}), \bibinfo{journal}{Nucl.
  Phys.} \textbf{\bibinfo{volume}{B845}}, \bibinfo{pages}{93}
  (\bibinfo{year}{2011}), \eprint{1009.5228}.

\bibitem[{\citenamefont{L{\"u}scher and Schaefer}(2011)}]{Luscher:2011kk}
\bibinfo{author}{\bibfnamefont{M.}~\bibnamefont{L{\"u}scher}} \bibnamefont{and}
  \bibinfo{author}{\bibfnamefont{S.}~\bibnamefont{Schaefer}},
  \bibinfo{journal}{JHEP} \textbf{\bibinfo{volume}{07}}, \bibinfo{pages}{036}
  (\bibinfo{year}{2011}), \eprint{1105.4749}.

\bibitem[{\citenamefont{Mohler and Schaefer}(2020)}]{Mohler:2020txx}
\bibinfo{author}{\bibfnamefont{D.}~\bibnamefont{Mohler}} \bibnamefont{and}
  \bibinfo{author}{\bibfnamefont{S.}~\bibnamefont{Schaefer}},
  \bibinfo{journal}{Phys. Rev. D} \textbf{\bibinfo{volume}{102}},
  \bibinfo{pages}{074506} (\bibinfo{year}{2020}), \eprint{2003.13359}.

\bibitem[{\citenamefont{Harris et~al.}(2019)\citenamefont{Harris, von Hippel,
  Junnarkar, Meyer, Ottnad, Wilhelm, Wittig, and Wrang}}]{Harris:2019bih}
\bibinfo{author}{\bibfnamefont{T.}~\bibnamefont{Harris}},
  \bibinfo{author}{\bibfnamefont{G.}~\bibnamefont{von Hippel}},
  \bibinfo{author}{\bibfnamefont{P.}~\bibnamefont{Junnarkar}},
  \bibinfo{author}{\bibfnamefont{H.~B.} \bibnamefont{Meyer}},
  \bibinfo{author}{\bibfnamefont{K.}~\bibnamefont{Ottnad}},
  \bibinfo{author}{\bibfnamefont{J.}~\bibnamefont{Wilhelm}},
  \bibinfo{author}{\bibfnamefont{H.}~\bibnamefont{Wittig}}, \bibnamefont{and}
  \bibinfo{author}{\bibfnamefont{L.}~\bibnamefont{Wrang}}
  (\bibinfo{year}{2019}), \eprint{1905.01291}.

\bibitem[{\citenamefont{Bruno et~al.}(2017)\citenamefont{Bruno, Korzec, and
  Schaefer}}]{Bruno:2016plf}
\bibinfo{author}{\bibfnamefont{M.}~\bibnamefont{Bruno}},
  \bibinfo{author}{\bibfnamefont{T.}~\bibnamefont{Korzec}}, \bibnamefont{and}
  \bibinfo{author}{\bibfnamefont{S.}~\bibnamefont{Schaefer}},
  \bibinfo{journal}{Phys. Rev. D} \textbf{\bibinfo{volume}{95}},
  \bibinfo{pages}{074504} (\bibinfo{year}{2017}), \eprint{1608.08900}.

\bibitem[{\citenamefont{Alexandrou et~al.}(2008)\citenamefont{Alexandrou,
  Korzec, Koutsou, Brinet, Carbonell, Drach, Harraud, and
  Baron}}]{Alexandrou:2008rp}
\bibinfo{author}{\bibfnamefont{C.}~\bibnamefont{Alexandrou}},
  \bibinfo{author}{\bibfnamefont{T.}~\bibnamefont{Korzec}},
  \bibinfo{author}{\bibfnamefont{G.}~\bibnamefont{Koutsou}},
  \bibinfo{author}{\bibfnamefont{M.}~\bibnamefont{Brinet}},
  \bibinfo{author}{\bibfnamefont{J.}~\bibnamefont{Carbonell}},
  \bibinfo{author}{\bibfnamefont{V.}~\bibnamefont{Drach}},
  \bibinfo{author}{\bibfnamefont{P.-A.} \bibnamefont{Harraud}},
  \bibnamefont{and} \bibinfo{author}{\bibfnamefont{R.}~\bibnamefont{Baron}}
  (\bibinfo{collaboration}{European Twisted Mass Collaboration}),
  \bibinfo{journal}{PoS} \textbf{\bibinfo{volume}{LATTICE2008}},
  \bibinfo{pages}{139} (\bibinfo{year}{2008}), \eprint{0811.0724}.

\bibitem[{\citenamefont{G{\"u}sken et~al.}(1989)\citenamefont{G{\"u}sken,
  L{\"o}w, Mutter, Sommer, Patel, and Schilling}}]{Gusken:1989ad}
\bibinfo{author}{\bibfnamefont{S.}~\bibnamefont{G{\"u}sken}},
  \bibinfo{author}{\bibfnamefont{U.}~\bibnamefont{L{\"o}w}},
  \bibinfo{author}{\bibfnamefont{K.~H.} \bibnamefont{Mutter}},
  \bibinfo{author}{\bibfnamefont{R.}~\bibnamefont{Sommer}},
  \bibinfo{author}{\bibfnamefont{A.}~\bibnamefont{Patel}}, \bibnamefont{and}
  \bibinfo{author}{\bibfnamefont{K.}~\bibnamefont{Schilling}},
  \bibinfo{journal}{Phys. Lett.} \textbf{\bibinfo{volume}{B227}},
  \bibinfo{pages}{266} (\bibinfo{year}{1989}).

\bibitem[{\citenamefont{Albanese et~al.}(1987)}]{Albanese:1987ds}
\bibinfo{author}{\bibfnamefont{M.}~\bibnamefont{Albanese}} \bibnamefont{et~al.}
  (\bibinfo{collaboration}{APE Collaboration}), \bibinfo{journal}{Phys. Lett.}
  \textbf{\bibinfo{volume}{B192}}, \bibinfo{pages}{163} (\bibinfo{year}{1987}).

\bibitem[{\citenamefont{von Hippel et~al.}(2013)\citenamefont{von Hippel,
  Jäger, Rae, and Wittig}}]{vonHippel:2013yfa}
\bibinfo{author}{\bibfnamefont{G.~M.} \bibnamefont{von Hippel}},
  \bibinfo{author}{\bibfnamefont{B.}~\bibnamefont{Jäger}},
  \bibinfo{author}{\bibfnamefont{T.~D.} \bibnamefont{Rae}}, \bibnamefont{and}
  \bibinfo{author}{\bibfnamefont{H.}~\bibnamefont{Wittig}},
  \bibinfo{journal}{JHEP} \textbf{\bibinfo{volume}{09}}, \bibinfo{pages}{014}
  (\bibinfo{year}{2013}), \eprint{1306.1440}.

\bibitem[{\citenamefont{Gérardin et~al.}(2019)\citenamefont{Gérardin, Harris,
  and Meyer}}]{Gerardin:2018kpy}
\bibinfo{author}{\bibfnamefont{A.}~\bibnamefont{Gérardin}},
  \bibinfo{author}{\bibfnamefont{T.}~\bibnamefont{Harris}}, \bibnamefont{and}
  \bibinfo{author}{\bibfnamefont{H.~B.} \bibnamefont{Meyer}},
  \bibinfo{journal}{Phys. Rev.} \textbf{\bibinfo{volume}{D99}},
  \bibinfo{pages}{014519} (\bibinfo{year}{2019}), \eprint{1811.08209}.

\bibitem[{\citenamefont{Martinelli and Sachrajda}(1989)}]{Martinelli:1988rr}
\bibinfo{author}{\bibfnamefont{G.}~\bibnamefont{Martinelli}} \bibnamefont{and}
  \bibinfo{author}{\bibfnamefont{C.~T.} \bibnamefont{Sachrajda}},
  \bibinfo{journal}{Nucl. Phys.} \textbf{\bibinfo{volume}{B316}},
  \bibinfo{pages}{355} (\bibinfo{year}{1989}).

\bibitem[{\citenamefont{Bali et~al.}(2010)\citenamefont{Bali, Collins, and
  Sch{\"a}fer}}]{Bali:2009hu}
\bibinfo{author}{\bibfnamefont{G.~S.} \bibnamefont{Bali}},
  \bibinfo{author}{\bibfnamefont{S.}~\bibnamefont{Collins}}, \bibnamefont{and}
  \bibinfo{author}{\bibfnamefont{A.}~\bibnamefont{Sch{\"a}fer}},
  \bibinfo{journal}{Comput. Phys. Commun.} \textbf{\bibinfo{volume}{181}},
  \bibinfo{pages}{1570} (\bibinfo{year}{2010}), \eprint{0910.3970}.

\bibitem[{\citenamefont{Blum et~al.}(2013)\citenamefont{Blum, Izubuchi, and
  Shintani}}]{Blum:2012uh}
\bibinfo{author}{\bibfnamefont{T.}~\bibnamefont{Blum}},
  \bibinfo{author}{\bibfnamefont{T.}~\bibnamefont{Izubuchi}}, \bibnamefont{and}
  \bibinfo{author}{\bibfnamefont{E.}~\bibnamefont{Shintani}},
  \bibinfo{journal}{Phys. Rev.} \textbf{\bibinfo{volume}{D88}},
  \bibinfo{pages}{094503} (\bibinfo{year}{2013}), \eprint{1208.4349}.

\bibitem[{\citenamefont{Shintani et~al.}(2015)\citenamefont{Shintani, Arthur,
  Blum, Izubuchi, Jung, and Lehner}}]{Shintani:2014vja}
\bibinfo{author}{\bibfnamefont{E.}~\bibnamefont{Shintani}},
  \bibinfo{author}{\bibfnamefont{R.}~\bibnamefont{Arthur}},
  \bibinfo{author}{\bibfnamefont{T.}~\bibnamefont{Blum}},
  \bibinfo{author}{\bibfnamefont{T.}~\bibnamefont{Izubuchi}},
  \bibinfo{author}{\bibfnamefont{C.}~\bibnamefont{Jung}}, \bibnamefont{and}
  \bibinfo{author}{\bibfnamefont{C.}~\bibnamefont{Lehner}},
  \bibinfo{journal}{Phys. Rev.} \textbf{\bibinfo{volume}{D91}},
  \bibinfo{pages}{114511} (\bibinfo{year}{2015}), \eprint{1402.0244}.

\bibitem[{\citenamefont{Lepage}(1989)}]{Lepage:1989hd}
\bibinfo{author}{\bibfnamefont{G.}~\bibnamefont{Lepage}}, in
  \emph{\bibinfo{booktitle}{{Theoretical Advanced Study Institute in Elementary
  Particle Physics}}} (\bibinfo{year}{1989}), pp. \bibinfo{pages}{97--120}.

\bibitem[{\citenamefont{Green}(2018)}]{Green:2018vxw}
\bibinfo{author}{\bibfnamefont{J.}~\bibnamefont{Green}}, \bibinfo{journal}{PoS}
  \textbf{\bibinfo{volume}{LATTICE2018}}, \bibinfo{pages}{016}
  (\bibinfo{year}{2018}), \eprint{1812.10574}.

\bibitem[{\citenamefont{Ottnad}(2020)}]{Ottnad:2020qbw}
\bibinfo{author}{\bibfnamefont{K.}~\bibnamefont{Ottnad}}, in
  \emph{\bibinfo{booktitle}{{38th International Symposium on Lattice Field
  Theory}}} (\bibinfo{year}{2020}), \eprint{2011.12471}.

\bibitem[{\citenamefont{Gusken}(1990)}]{Gusken:1989qx}
\bibinfo{author}{\bibfnamefont{S.}~\bibnamefont{Gusken}},
  \bibinfo{journal}{Nucl. Phys. Proc. Suppl.} \textbf{\bibinfo{volume}{17}},
  \bibinfo{pages}{361} (\bibinfo{year}{1990}).

\bibitem[{\citenamefont{Maiani et~al.}(1987)\citenamefont{Maiani, Martinelli,
  Paciello, and Taglienti}}]{Maiani:1987by}
\bibinfo{author}{\bibfnamefont{L.}~\bibnamefont{Maiani}},
  \bibinfo{author}{\bibfnamefont{G.}~\bibnamefont{Martinelli}},
  \bibinfo{author}{\bibfnamefont{M.}~\bibnamefont{Paciello}}, \bibnamefont{and}
  \bibinfo{author}{\bibfnamefont{B.}~\bibnamefont{Taglienti}},
  \bibinfo{journal}{Nucl. Phys. B} \textbf{\bibinfo{volume}{293}},
  \bibinfo{pages}{420} (\bibinfo{year}{1987}).

\bibitem[{\citenamefont{Doi et~al.}(2009)\citenamefont{Doi, Deka, Dong, Draper,
  Liu, Mankame, Mathur, and Streuer}}]{Doi:2009sq}
\bibinfo{author}{\bibfnamefont{T.}~\bibnamefont{Doi}},
  \bibinfo{author}{\bibfnamefont{M.}~\bibnamefont{Deka}},
  \bibinfo{author}{\bibfnamefont{S.-J.} \bibnamefont{Dong}},
  \bibinfo{author}{\bibfnamefont{T.}~\bibnamefont{Draper}},
  \bibinfo{author}{\bibfnamefont{K.-F.} \bibnamefont{Liu}},
  \bibinfo{author}{\bibfnamefont{D.}~\bibnamefont{Mankame}},
  \bibinfo{author}{\bibfnamefont{N.}~\bibnamefont{Mathur}}, \bibnamefont{and}
  \bibinfo{author}{\bibfnamefont{T.}~\bibnamefont{Streuer}},
  \bibinfo{journal}{Phys. Rev. D} \textbf{\bibinfo{volume}{80}},
  \bibinfo{pages}{094503} (\bibinfo{year}{2009}), \eprint{0903.3232}.

\bibitem[{\citenamefont{Bulava et~al.}(2012)\citenamefont{Bulava, Donnellan,
  and Sommer}}]{Bulava:2011yz}
\bibinfo{author}{\bibfnamefont{J.}~\bibnamefont{Bulava}},
  \bibinfo{author}{\bibfnamefont{M.}~\bibnamefont{Donnellan}},
  \bibnamefont{and} \bibinfo{author}{\bibfnamefont{R.}~\bibnamefont{Sommer}},
  \bibinfo{journal}{JHEP} \textbf{\bibinfo{volume}{01}}, \bibinfo{pages}{140}
  (\bibinfo{year}{2012}), \eprint{1108.3774}.

\bibitem[{\citenamefont{Capitani et~al.}(2012)\citenamefont{Capitani,
  Della~Morte, von Hippel, J{\"a}ger, J{\"u}ttner, Knippschild, Meyer, and
  Wittig}}]{Capitani:2012gj}
\bibinfo{author}{\bibfnamefont{S.}~\bibnamefont{Capitani}},
  \bibinfo{author}{\bibfnamefont{M.}~\bibnamefont{Della~Morte}},
  \bibinfo{author}{\bibfnamefont{G.}~\bibnamefont{von Hippel}},
  \bibinfo{author}{\bibfnamefont{B.}~\bibnamefont{J{\"a}ger}},
  \bibinfo{author}{\bibfnamefont{A.}~\bibnamefont{J{\"u}ttner}},
  \bibinfo{author}{\bibfnamefont{B.}~\bibnamefont{Knippschild}},
  \bibinfo{author}{\bibfnamefont{H.}~\bibnamefont{Meyer}}, \bibnamefont{and}
  \bibinfo{author}{\bibfnamefont{H.}~\bibnamefont{Wittig}},
  \bibinfo{journal}{Phys. Rev. D} \textbf{\bibinfo{volume}{86}},
  \bibinfo{pages}{074502} (\bibinfo{year}{2012}), \eprint{1205.0180}.

\bibitem[{\citenamefont{Yoon et~al.}(2017)}]{Yoon:2016jzj}
\bibinfo{author}{\bibfnamefont{B.}~\bibnamefont{Yoon}} \bibnamefont{et~al.},
  \bibinfo{journal}{Phys. Rev. D} \textbf{\bibinfo{volume}{95}},
  \bibinfo{pages}{074508} (\bibinfo{year}{2017}), \eprint{1611.07452}.

\bibitem[{\citenamefont{Chang et~al.}(2018)}]{Chang:2018uxx}
\bibinfo{author}{\bibfnamefont{C.}~\bibnamefont{Chang}} \bibnamefont{et~al.},
  \bibinfo{journal}{Nature} \textbf{\bibinfo{volume}{558}}, \bibinfo{pages}{91}
  (\bibinfo{year}{2018}), \eprint{1805.12130}.

\bibitem[{\citenamefont{Jang et~al.}(2020)\citenamefont{Jang, Gupta, Lin, Yoon,
  and Bhattacharya}}]{Jang:2019jkn}
\bibinfo{author}{\bibfnamefont{Y.-C.} \bibnamefont{Jang}},
  \bibinfo{author}{\bibfnamefont{R.}~\bibnamefont{Gupta}},
  \bibinfo{author}{\bibfnamefont{H.-W.} \bibnamefont{Lin}},
  \bibinfo{author}{\bibfnamefont{B.}~\bibnamefont{Yoon}}, \bibnamefont{and}
  \bibinfo{author}{\bibfnamefont{T.}~\bibnamefont{Bhattacharya}},
  \bibinfo{journal}{Phys. Rev. D} \textbf{\bibinfo{volume}{101}},
  \bibinfo{pages}{014507} (\bibinfo{year}{2020}), \eprint{1906.07217}.

\bibitem[{\citenamefont{Gupta et~al.}(2018)\citenamefont{Gupta, Jang, Yoon,
  Lin, Cirigliano, and Bhattacharya}}]{Gupta:2018qil}
\bibinfo{author}{\bibfnamefont{R.}~\bibnamefont{Gupta}},
  \bibinfo{author}{\bibfnamefont{Y.-C.} \bibnamefont{Jang}},
  \bibinfo{author}{\bibfnamefont{B.}~\bibnamefont{Yoon}},
  \bibinfo{author}{\bibfnamefont{H.-W.} \bibnamefont{Lin}},
  \bibinfo{author}{\bibfnamefont{V.}~\bibnamefont{Cirigliano}},
  \bibnamefont{and}
  \bibinfo{author}{\bibfnamefont{T.}~\bibnamefont{Bhattacharya}},
  \bibinfo{journal}{Phys. Rev. D} \textbf{\bibinfo{volume}{98}},
  \bibinfo{pages}{034503} (\bibinfo{year}{2018}), \eprint{1806.09006}.

\bibitem[{\citenamefont{Golub and Pereyra}(1973)}]{GolP73}
\bibinfo{author}{\bibfnamefont{G.~H.} \bibnamefont{Golub}} \bibnamefont{and}
  \bibinfo{author}{\bibfnamefont{V.}~\bibnamefont{Pereyra}},
  \bibinfo{journal}{SIAM Journal on Numerical Analysis}
  \textbf{\bibinfo{volume}{10}}, \bibinfo{pages}{413} (\bibinfo{year}{1973}).

\bibitem[{\citenamefont{Ye et~al.}(2018)\citenamefont{Ye, Arrington, Hill, and
  Lee}}]{Ye:2017gyb}
\bibinfo{author}{\bibfnamefont{Z.}~\bibnamefont{Ye}},
  \bibinfo{author}{\bibfnamefont{J.}~\bibnamefont{Arrington}},
  \bibinfo{author}{\bibfnamefont{R.~J.} \bibnamefont{Hill}}, \bibnamefont{and}
  \bibinfo{author}{\bibfnamefont{G.}~\bibnamefont{Lee}},
  \bibinfo{journal}{Phys. Lett. B} \textbf{\bibinfo{volume}{777}},
  \bibinfo{pages}{8} (\bibinfo{year}{2018}), \eprint{1707.09063}.

\bibitem[{\citenamefont{Hill and Paz}(2010)}]{Hill:2010yb}
\bibinfo{author}{\bibfnamefont{R.~J.} \bibnamefont{Hill}} \bibnamefont{and}
  \bibinfo{author}{\bibfnamefont{G.}~\bibnamefont{Paz}},
  \bibinfo{journal}{Phys. Rev.} \textbf{\bibinfo{volume}{D82}},
  \bibinfo{pages}{113005} (\bibinfo{year}{2010}), \eprint{1008.4619}.

\bibitem[{\citenamefont{Kubis and Mei{\ss}ner}(2001)}]{Kubis:2000zd}
\bibinfo{author}{\bibfnamefont{B.}~\bibnamefont{Kubis}} \bibnamefont{and}
  \bibinfo{author}{\bibfnamefont{U.-G.} \bibnamefont{Mei{\ss}ner}},
  \bibinfo{journal}{Nucl. Phys. A} \textbf{\bibinfo{volume}{679}},
  \bibinfo{pages}{698} (\bibinfo{year}{2001}), \eprint{hep-ph/0007056}.

\bibitem[{\citenamefont{Fuchs et~al.}(2004)\citenamefont{Fuchs, Gegelia, and
  Scherer}}]{Fuchs:2003ir}
\bibinfo{author}{\bibfnamefont{T.}~\bibnamefont{Fuchs}},
  \bibinfo{author}{\bibfnamefont{J.}~\bibnamefont{Gegelia}}, \bibnamefont{and}
  \bibinfo{author}{\bibfnamefont{S.}~\bibnamefont{Scherer}},
  \bibinfo{journal}{J. Phys. G} \textbf{\bibinfo{volume}{30}},
  \bibinfo{pages}{1407} (\bibinfo{year}{2004}), \eprint{nucl-th/0305070}.

\bibitem[{\citenamefont{Schindler et~al.}(2005)\citenamefont{Schindler,
  Gegelia, and Scherer}}]{Schindler:2005ke}
\bibinfo{author}{\bibfnamefont{M.~R.} \bibnamefont{Schindler}},
  \bibinfo{author}{\bibfnamefont{J.}~\bibnamefont{Gegelia}}, \bibnamefont{and}
  \bibinfo{author}{\bibfnamefont{S.}~\bibnamefont{Scherer}},
  \bibinfo{journal}{Eur. Phys. J. A} \textbf{\bibinfo{volume}{26}},
  \bibinfo{pages}{1} (\bibinfo{year}{2005}), \eprint{nucl-th/0509005}.

\bibitem[{\citenamefont{Bauer et~al.}(2012)\citenamefont{Bauer, Bernauer, and
  Scherer}}]{Bauer:2012pv}
\bibinfo{author}{\bibfnamefont{T.}~\bibnamefont{Bauer}},
  \bibinfo{author}{\bibfnamefont{J.~C.} \bibnamefont{Bernauer}},
  \bibnamefont{and} \bibinfo{author}{\bibfnamefont{S.}~\bibnamefont{Scherer}},
  \bibinfo{journal}{Phys. Rev.} \textbf{\bibinfo{volume}{C86}},
  \bibinfo{pages}{065206} (\bibinfo{year}{2012}), \eprint{1209.3872}.

\bibitem[{\citenamefont{Lepage and Brodsky}(1980)}]{Lepage:1980fj}
\bibinfo{author}{\bibfnamefont{G.~P.} \bibnamefont{Lepage}} \bibnamefont{and}
  \bibinfo{author}{\bibfnamefont{S.~J.} \bibnamefont{Brodsky}},
  \bibinfo{journal}{Phys. Rev.} \textbf{\bibinfo{volume}{D22}},
  \bibinfo{pages}{2157} (\bibinfo{year}{1980}).

\bibitem[{\citenamefont{Lee et~al.}(2015)\citenamefont{Lee, Arrington, and
  Hill}}]{Lee:2015jqa}
\bibinfo{author}{\bibfnamefont{G.}~\bibnamefont{Lee}},
  \bibinfo{author}{\bibfnamefont{J.~R.} \bibnamefont{Arrington}},
  \bibnamefont{and} \bibinfo{author}{\bibfnamefont{R.~J.} \bibnamefont{Hill}},
  \bibinfo{journal}{Phys. Rev.} \textbf{\bibinfo{volume}{D92}},
  \bibinfo{pages}{013013} (\bibinfo{year}{2015}), \eprint{1505.01489}.

\bibitem[{\citenamefont{Aoki et~al.}(2017)}]{Aoki:2016frl}
\bibinfo{author}{\bibfnamefont{S.}~\bibnamefont{Aoki}} \bibnamefont{et~al.},
  \bibinfo{journal}{Eur. Phys. J. C} \textbf{\bibinfo{volume}{77}},
  \bibinfo{pages}{112} (\bibinfo{year}{2017}), \eprint{1607.00299}.

\bibitem[{\citenamefont{Tiburzi}(2008)}]{Tiburzi:2007ep}
\bibinfo{author}{\bibfnamefont{B.~C.} \bibnamefont{Tiburzi}},
  \bibinfo{journal}{Phys. Rev. D} \textbf{\bibinfo{volume}{77}},
  \bibinfo{pages}{014510} (\bibinfo{year}{2008}), \eprint{0710.3577}.

\bibitem[{\citenamefont{Beane}(2004)}]{Beane:2004tw}
\bibinfo{author}{\bibfnamefont{S.~R.} \bibnamefont{Beane}},
  \bibinfo{journal}{Phys. Rev. D} \textbf{\bibinfo{volume}{70}},
  \bibinfo{pages}{034507} (\bibinfo{year}{2004}), \eprint{hep-lat/0403015}.

\bibitem[{\citenamefont{Zyla et~al.}(2020)}]{Zyla:2020zbs}
\bibinfo{author}{\bibfnamefont{P.}~\bibnamefont{Zyla}} \bibnamefont{et~al.}
  (\bibinfo{collaboration}{Particle Data Group}), \bibinfo{journal}{PTEP}
  \textbf{\bibinfo{volume}{2020}}, \bibinfo{pages}{083C01}
  (\bibinfo{year}{2020}).

\bibitem[{\citenamefont{Alexandrou et~al.}(2020)\citenamefont{Alexandrou,
  Hadjiyiannakou, Koutsou, Ottnad, and Petschlies}}]{Alexandrou:2020aja}
\bibinfo{author}{\bibfnamefont{C.}~\bibnamefont{Alexandrou}},
  \bibinfo{author}{\bibfnamefont{K.}~\bibnamefont{Hadjiyiannakou}},
  \bibinfo{author}{\bibfnamefont{G.}~\bibnamefont{Koutsou}},
  \bibinfo{author}{\bibfnamefont{K.}~\bibnamefont{Ottnad}}, \bibnamefont{and}
  \bibinfo{author}{\bibfnamefont{M.}~\bibnamefont{Petschlies}},
  \bibinfo{journal}{Phys. Rev. D} \textbf{\bibinfo{volume}{101}},
  \bibinfo{pages}{114504} (\bibinfo{year}{2020}), \eprint{2002.06984}.

\bibitem[{\citenamefont{Akaike et~al.}(1973)\citenamefont{Akaike, Petrov, and
  Csaki}}]{akaike1973second}
\bibinfo{author}{\bibfnamefont{H.}~\bibnamefont{Akaike}},
  \bibinfo{author}{\bibfnamefont{B.~N.} \bibnamefont{Petrov}},
  \bibnamefont{and} \bibinfo{author}{\bibfnamefont{F.}~\bibnamefont{Csaki}},
  \emph{\bibinfo{title}{Second international symposium on information theory}}
  (\bibinfo{year}{1973}).

\bibitem[{\citenamefont{{Akaike}}(1974)}]{Akaike:IEEE:1100705}
\bibinfo{author}{\bibfnamefont{H.}~\bibnamefont{{Akaike}}},
  \bibinfo{journal}{IEEE Transactions on Automatic Control}
  \textbf{\bibinfo{volume}{19}}, \bibinfo{pages}{716} (\bibinfo{year}{1974}).

\bibitem[{\citenamefont{Jay and Neil}(2020)}]{jay2020bayesian}
\bibinfo{author}{\bibfnamefont{W.~I.} \bibnamefont{Jay}} \bibnamefont{and}
  \bibinfo{author}{\bibfnamefont{E.~T.} \bibnamefont{Neil}},
  \emph{\bibinfo{title}{Bayesian model averaging for analysis of lattice field
  theory results}} (\bibinfo{year}{2020}), \eprint{2008.01069}.

\bibitem[{\citenamefont{Bors\'anyi et~al.}(2020)}]{Borsanyi:2020mff}
\bibinfo{author}{\bibfnamefont{S.}~\bibnamefont{Bors\'anyi}}
  \bibnamefont{et~al.} (\bibinfo{year}{2020}), \eprint{2002.12347}.

\bibitem[{\citenamefont{Hasan et~al.}(2018)\citenamefont{Hasan, Green, Meinel,
  Engelhardt, Krieg, Negele, Pochinsky, and Syritsyn}}]{Hasan:2017wwt}
\bibinfo{author}{\bibfnamefont{N.}~\bibnamefont{Hasan}},
  \bibinfo{author}{\bibfnamefont{J.}~\bibnamefont{Green}},
  \bibinfo{author}{\bibfnamefont{S.}~\bibnamefont{Meinel}},
  \bibinfo{author}{\bibfnamefont{M.}~\bibnamefont{Engelhardt}},
  \bibinfo{author}{\bibfnamefont{S.}~\bibnamefont{Krieg}},
  \bibinfo{author}{\bibfnamefont{J.}~\bibnamefont{Negele}},
  \bibinfo{author}{\bibfnamefont{A.}~\bibnamefont{Pochinsky}},
  \bibnamefont{and} \bibinfo{author}{\bibfnamefont{S.}~\bibnamefont{Syritsyn}},
  \bibinfo{journal}{Phys. Rev. D} \textbf{\bibinfo{volume}{97}},
  \bibinfo{pages}{034504} (\bibinfo{year}{2018}), \eprint{1711.11385}.

\bibitem[{\citenamefont{B{\"a}r}(2020)}]{Bar:2019igf}
\bibinfo{author}{\bibfnamefont{O.}~\bibnamefont{B{\"a}r}},
  \bibinfo{journal}{Phys. Rev. D} \textbf{\bibinfo{volume}{101}},
  \bibinfo{pages}{034515} (\bibinfo{year}{2020}), \eprint{1912.05873}.

\bibitem[{\citenamefont{Edwards and Joo}(2005)}]{Edwards:2004sx}
\bibinfo{author}{\bibfnamefont{R.~G.} \bibnamefont{Edwards}} \bibnamefont{and}
  \bibinfo{author}{\bibfnamefont{B.}~\bibnamefont{Joo}}
  (\bibinfo{collaboration}{SciDAC, LHPC, UKQCD Collaborations}),
  \bibinfo{journal}{Nucl. Phys. Proc. Suppl.} \textbf{\bibinfo{volume}{140}},
  \bibinfo{pages}{832} (\bibinfo{year}{2005}), \bibinfo{note}{[,832(2004)]},
  \eprint{hep-lat/0409003}.

\bibitem[{\citenamefont{L{\"u}scher and Schaefer}(2013)}]{Luscher:2012av}
\bibinfo{author}{\bibfnamefont{M.}~\bibnamefont{L{\"u}scher}} \bibnamefont{and}
  \bibinfo{author}{\bibfnamefont{S.}~\bibnamefont{Schaefer}},
  \bibinfo{journal}{Comput. Phys. Commun.} \textbf{\bibinfo{volume}{184}},
  \bibinfo{pages}{519} (\bibinfo{year}{2013}), \eprint{1206.2809}.

\bibitem[{\citenamefont{Djukanovic}(2020)}]{Djukanovic:2016spv}
\bibinfo{author}{\bibfnamefont{D.}~\bibnamefont{Djukanovic}},
  \bibinfo{journal}{Comput. Phys. Commun.} \textbf{\bibinfo{volume}{247}},
  \bibinfo{pages}{106950} (\bibinfo{year}{2020}), \eprint{1603.01576}.

\end{thebibliography}
\appendix
\section{Form factor data}
\label{app:data}

In this appendix, we present the results of extracting the isovector
electromagnetic form factors of the nucleon with either the summation method or
the two-state method, both described in section~\ref{sec:excited}, for every
gauge ensemble listed in Table~\ref{tab:ensembles}.

The summation-method results quoted below are obtained from a fit
using Eq.~(\ref{eq:sratio_gs}) as fit ansatz.  For the two-state-method, we perform
fits for different prior widths $\delta$, where we use integer factors
between one and five multiplying the initial error estimate. However,
for the prior to have an effect, we limit the width to at most 50\% of
the central value. We arrive at the numbers listed in the tables below
using the data for which (a) all values up to a given width are within
$2\sigma$, and (b) the error does not increase by more than a factor of
$\frac{2}{3}
(1+\delta)$, i.e. factors of $[2,2\frac{2}{3},3\frac{1}{3},4]$ for prior widths
$[2x,3x,4x,5x]$.

\begin{table}[h]
\begin{ruledtabular}
\begin{tabular}{lllll}
	$Q^2 \, [\Gevs]$ &         $\GE$ (sum) &  $\GE$ (two-state) &
	$\GM$ (sum) &  $\GM$ (two-state) \\
0.089 &  0.803(16) &  0.803(11) &  3.49(22) &   3.36(11) \\
0.174 &  0.671(20) &  0.663(12) &  2.71(16) &  2.688(77) \\
0.255 &  0.556(25) &  0.558(21) &  2.40(15) &  2.381(78) \\
0.334 &  0.473(28) &  0.465(29) &  2.09(15) &   2.04(10) \\
0.410 &  0.420(27) &  0.389(42) &  1.70(13) &   1.69(14) \\
0.484 &  0.363(31) &  0.359(34) &  1.62(14) &   1.61(12) \\
0.624 &  0.294(43) &  0.272(39) &  1.52(17) &   1.14(16) \\
0.692 &  0.301(43) &  0.256(27) &  1.16(16) &   1.11(11) \\
0.757 &  0.303(58) &  0.207(20) &  1.13(22) &  0.949(99) \\
0.821 &  0.319(60) &  0.197(20) &  1.21(23) &   0.93(10) \\
0.884 &  0.158(81) &  0.139(29) &  0.44(34) &   0.74(13) \\
\end{tabular}

\end{ruledtabular}
$\GE$ and $\GM$ for ensemble D200.
\end{table}
\begin{table}[h]
\begin{ruledtabular}
\begin{tabular}{lllll}
	$Q^2 \, [\Gevs]$ &         $\GE$ (sum) &  $\GE$ (two-state) &
	$\GM$ (sum) &  $\GM$ (two-state) \\
0.087 &  0.7921(91) &  0.755(22) &   3.53(13) &  3.35(21) \\
0.171 &   0.668(13) &  0.638(27) &   3.00(10) &  2.81(10) \\
0.252 &   0.559(15) &  0.501(53) &   2.47(11) &  2.35(15) \\
0.329 &   0.459(25) &  0.420(47) &   2.26(11) &  2.08(15) \\
0.404 &   0.420(18) &  0.381(35) &   1.99(11) &  1.79(14) \\
0.476 &   0.363(23) &  0.324(42) &  1.686(90) &  1.53(16) \\
0.615 &   0.323(26) &  0.276(13) &   1.58(14) &  0.72(31) \\
0.682 &   0.295(30) &  0.246(16) &   1.58(14) &  1.35(21) \\
0.747 &   0.266(37) &  0.245(34) &   1.42(15) &  1.11(18) \\
0.810 &   0.271(37) &  0.203(35) &   1.27(14) &  1.00(17) \\
0.872 &   0.320(77) &  0.155(48) &   1.12(32) &  0.80(20) \\
\end{tabular}

\end{ruledtabular}
$\GE$ and $\GM$ for ensemble C101.
\end{table}
\begin{table}[h]
\begin{ruledtabular}
\begin{tabular}{lllll}
	$Q^2 \, [\Gevs]$ &         $\GE$ (sum) &  $\GE$ (two-state) &
	$\GM$ (sum) &  $\GM$ (two-state) \\
0.193 &  0.632(32) &   0.590(58) &   2.75(26) &     2.58(31) \\
0.370 &  0.453(33) &   0.370(88) &   2.32(21) &     2.32(40) \\
0.536 &  0.275(49) &   0.233(69) &   1.73(22) &     1.58(27) \\
0.692 &  0.142(72) &   0.206(44) &   1.41(39) &     0.89(34) \\
0.840 &  0.209(62) &   0.166(72) &   0.82(24) &     0.82(27) \\
0.980 &  0.168(55) &   0.140(39) &   0.63(19) &     0.72(19) \\
1.244 &   0.20(12) &   0.111(45) &   0.10(47) &     0.54(21) \\
\end{tabular}

\end{ruledtabular}
$\GE$ and $\GM$ for ensemble H105.
\end{table}
\begin{table}[h]
\begin{ruledtabular}
\begin{tabular}{lllll}
	$Q^2 \, [\Gevs]$ &         $\GE$ (sum) &  $\GE$ (two-state) &
	$\GM$ (sum) &  $\GM$ (two-state) \\
0.156 &  0.723(19) &  0.7293(75) &  3.17(18) &  2.999(63) \\
0.303 &  0.561(22) &  0.5656(77) &  2.66(16) &  2.462(49) \\
0.441 &  0.470(31) &  0.4685(92) &  2.15(14) &  1.957(51) \\
0.573 &  0.315(44) &   0.381(12) &  1.57(18) &  1.695(52) \\
0.699 &  0.295(38) &   0.330(11) &  1.42(16) &  1.464(47) \\
0.820 &  0.331(46) &   0.315(12) &  1.56(20) &  1.329(55) \\
1.048 &   0.36(11) &   0.206(18) &  0.77(35) &  0.925(72) \\
1.156 &   0.22(10) &   0.172(21) &  0.60(41) &  0.825(85) \\
\end{tabular}

\end{ruledtabular}
$\GE$ and $\GM$ for ensemble N200.
\end{table}
\begin{table}[h]
\begin{ruledtabular}
\begin{tabular}{lllll}
	$Q^2 \, [\Gevs]$ &         $\GE$ (sum) &  $\GE$ (two-state) &
	$\GM$ (sum) &  $\GM$ (two-state) \\
0.156 &  0.6946(80) &   0.694(17) &  3.033(80) &  3.061(57) \\
0.304 &   0.512(11) &   0.515(23) &  2.428(69) &  2.408(59) \\
0.444 &   0.394(14) &   0.419(20) &  2.040(77) &  1.980(62) \\
0.577 &   0.322(19) &   0.329(21) &  1.679(89) &   1.54(13) \\
0.705 &   0.253(17) &   0.266(15) &  1.485(80) &   1.11(20) \\
0.827 &   0.190(22) &   0.208(18) &  1.272(94) &   0.91(22) \\
1.059 &   0.172(40) &   0.162(24) &   1.01(16) &  0.897(89) \\
1.170 &   0.074(41) &   0.059(54) &   0.66(16) &   0.50(17) \\
1.277 &   0.165(63) &   0.147(30) &   1.05(29) &   0.75(12) \\
1.381 &   0.144(69) &   0.144(15) &   0.80(30) &  0.723(88) \\
\end{tabular}

\end{ruledtabular}
$\GE$ and $\GM$ for ensemble N203.
\end{table}
\begin{table}[h]
\begin{ruledtabular}
\begin{tabular}{lllll}
	$Q^2 \, [\Gevs]$ &         $\GE$ (sum) &  $\GE$ (two-state) &
	$\GM$ (sum) &  $\GM$ (two-state) \\
0.256 &  0.575(21) &  0.613(12) &   2.62(15) &  2.435(80) \\
0.491 &  0.352(24) &  0.417(16) &   1.78(12) &  1.851(71) \\
0.708 &  0.149(36) &  0.276(29) &   1.32(16) &   1.39(13) \\
0.912 &  0.135(57) &  0.212(27) &   1.12(25) &   1.16(11) \\
1.105 &  0.123(55) &  0.165(36) &   1.30(25) &   0.95(12) \\
1.287 &  0.092(77) &  0.103(34) &   0.56(33) &   0.56(14) \\
\end{tabular}

\end{ruledtabular}
$\GE$ and $\GM$ for ensemble N302.
\end{table}
\begin{table}[h]
\begin{ruledtabular}
\begin{tabular}{lllll}
	$Q^2 \, [\Gevs]$ &         $\GE$ (sum) &  $\GE$ (two-state) &
	$\GM$ (sum) &  $\GM$ (two-state) \\
0.146 &  0.722(24) &  0.715(18) &  3.32(23) &   2.92(12) \\
0.284 &  0.532(25) &  0.544(25) &  2.65(17) &   2.36(10) \\
0.415 &  0.445(35) &  0.453(18) &  2.54(19) &   2.12(12) \\
0.539 &  0.348(49) &  0.373(18) &  2.17(23) &  1.753(84) \\
0.658 &  0.347(41) &  0.326(18) &  1.95(19) &  1.586(71) \\
0.772 &  0.360(52) &  0.288(14) &  1.86(23) &  1.381(64) \\
0.988 &  0.310(94) &  0.241(18) &  1.64(40) &  1.196(85) \\
1.090 &   0.40(12) &  0.181(39) &  2.09(50) &   0.93(14) \\
1.190 &   0.08(14) &  0.117(50) &  0.80(59) &   0.64(20) \\
1.287 &   0.10(16) &  0.112(32) &  0.53(63) &   0.46(16) \\
\end{tabular}

\end{ruledtabular}
$\GE$ and $\GM$ for ensemble J303.
\end{table}
\begin{table}[h]
\begin{ruledtabular}
\begin{tabular}{lllll}
	$Q^2 \, [\Gevs]$ &         $\GE$ (sum) &  $\GE$ (two-state) &
	$\GM$ (sum) &  $\GM$ (two-state) \\
0.340 &  0.448(54) &  0.522(17) &  2.06(32) &   2.27(10) \\
0.643 &  0.405(60) &  0.370(12) &  1.38(23) &  1.588(73) \\
0.919 &   0.59(23) &  0.238(29) &  2.31(69) &   1.11(12) \\
1.175 &   0.44(29) &  0.114(33) &  2.3(1.2) &   0.47(15) \\
\end{tabular}

\end{ruledtabular}
$\GE$ and $\GM$ for ensemble S201.
\end{table}
\begin{table}[h]
\begin{ruledtabular}
\begin{tabular}{lllll}
	$Q^2 \, [\Gevs]$ &         $\GE$ (sum) &  $\GE$ (two-state) &
	$\GM$ (sum) &  $\GM$ (two-state) \\
0.246 &  0.604(21) &  0.607(14) &  2.84(16) &  2.746(79) \\
0.471 &  0.422(23) &  0.412(18) &  2.26(13) &  2.090(58) \\
0.680 &  0.273(38) &  0.312(17) &  1.78(17) &  1.616(55) \\
0.877 &  0.138(63) &  0.204(29) &  0.94(28) &   1.20(11) \\
1.062 &  0.113(53) &  0.181(18) &  1.04(26) &  1.128(54) \\
1.239 &  0.106(85) &  0.112(35) &  0.99(37) &  0.772(99) \\
\end{tabular}

\end{ruledtabular}
$\GE$ and $\GM$ for ensemble S400.
\end{table}

\begin{table}[h]
\begin{ruledtabular}
\begin{tabular}{lllll}
	$Q^2 \, [\Gevs]$ &         $\GE$ (sum) &  $\GE$ (two-state) &
	$\GM$ (sum) &  $\GM$ (two-state) \\
0.040 &  0.885(22) &     0.889(18) &   3.68(47) &      3.66(31) \\
0.079 &  0.812(31) &     0.817(17) &   3.43(39) &      3.16(18) \\
0.117 &  0.732(36) &     0.718(42) &   2.88(33) &      2.80(27) \\
0.155 &  0.658(41) &     0.675(35) &   3.34(36) &      3.02(22) \\
0.191 &  0.633(36) &     0.641(26) &   2.86(30) &      2.56(16) \\
0.227 &  0.554(39) &     0.542(76) &   2.19(24) &      2.28(26) \\
0.297 &  0.457(44) &     0.511(34) &   2.27(24) &      2.26(20) \\
0.331 &  0.398(44) &     0.441(39) &   2.01(24) &      2.08(17) \\
0.365 &  0.395(54) &     0.438(29) &   1.84(34) &      1.88(14) \\
0.398 &  0.368(49) &     0.399(35) &   1.65(27) &      1.75(15) \\
0.431 &  0.280(68) &     0.364(40) &   1.47(31) &      1.71(19) \\
0.463 &  0.288(56) &     0.361(37) &   1.71(27) &      1.71(13) \\
0.494 &  0.266(56) &     0.301(49) &   1.42(25) &      1.47(19) \\
0.556 &   0.33(11) &     0.285(46) &   1.13(42) &      1.10(26) \\
0.586 &  0.256(72) &     0.301(40) &   0.75(31) &      1.27(18) \\
0.616 &  0.239(75) &     0.273(47) &   0.92(30) &      1.12(29) \\
0.646 &  0.207(84) &     0.240(62) &   1.45(33) &      1.31(22) \\
0.675 &  0.178(82) &     0.239(54) &   0.86(34) &      0.96(29) \\
0.704 &  0.180(83) &     0.079(73) &   0.90(31) &      0.77(27) \\
0.732 &   0.15(11) &     0.174(52) &   0.32(36) &      0.91(21) \\
0.788 &   0.17(12) &     0.188(67) &   0.68(40) &      0.64(46) \\
0.816 &   0.09(11) &     0.121(64) &   0.52(38) &      0.16(51) \\
0.843 &   0.10(11) &     0.109(50) &  -0.05(37) &      0.67(26) \\
0.870 &  -0.00(13) &     0.081(56) &  -0.24(50) &      0.17(73) \\
0.923 &  -0.02(12) &      0.02(13) &   0.22(44) &      0.53(41) \\
\end{tabular}

\end{ruledtabular}
$G_\mathrm{E}$ and $G_\mathrm{M}$ for ensemble E250.
\end{table}

\vfill
\clearpage
\section{Priors}
\label{app:priors}

The two-state method, as we apply it, requires a certain amount of
prior information in order to stabilize the fits.  In this appendix,
we summarize the priors, which are extracted from the nucleon two-point
functions, for the ratio of overlaps $\rho$ and for the energy
difference between ground and first excited state; for details see
Sec.~\ref{sec:excited}. The energy gap is given in lattice units.

\begin{table}[!h]
\begin{ruledtabular}
\begin{tabular}{llllllllll}
  $\mathbf{n}^2$ &      D200 &      C101 &      H105 &      N200 &      N203 &      N302 &      J303 &        S201 &      S400 \\
  0 &   1.35(8) &   1.17(7) &  1.16(13) &   1.37(3) &   1.08(5) &   1.47(5) &   1.74(9) &     1.23(4) &   1.08(4) \\
  1 &   1.33(7) &   1.14(5) &  1.24(13) &   1.42(3) &   1.07(5) &   1.55(5) &  1.89(11) &     1.30(4) &   1.17(4) \\
  2 &   1.40(8) &   1.13(5) &  1.31(11) &   1.48(3) &   1.04(6) &   1.52(5) &  2.14(18) &     1.40(4) &   1.22(4) \\
  3 &   1.39(9) &   1.16(6) &  1.43(14) &   1.54(4) &   1.03(7) &   1.51(5) &  2.28(24) &     1.41(5) &   1.37(6) \\
  4 &  1.68(13) &   1.19(7) &  1.26(14) &   1.54(4) &  1.48(14) &  1.83(20) &  2.09(20) &     1.57(7) &   1.40(7) \\
  5 &  1.59(12) &   1.20(9) &  1.39(18) &   1.62(5) &  1.37(15) &   1.50(9) &  2.12(22) &     1.69(8) &   1.44(8) \\
  6 &  1.66(15) &   1.19(7) &  1.83(50) &   1.66(6) &  1.33(17) &   1.42(9) &  2.52(42) &     1.51(9) &  1.66(14) \\
  8 &  1.76(23) &  1.32(18) &  1.69(51) &   1.76(8) &   1.62(4) &  1.40(10) &  1.99(17) &     1.74(7) &  1.70(13) \\
  9 &  2.41(45) &  1.29(13) &  1.55(35) &   1.82(7) &   1.64(5) &   1.69(7) &  2.23(24) &    1.89(10) &  2.30(21) \\
 10 &  2.25(34) &  1.31(10) &  1.65(36) &   1.77(7) &   1.65(5) &   1.78(9) &  2.50(33) &    1.93(11) &  2.20(29) \\
 11 &  2.18(51) &  1.35(11) &  1.48(27) &   1.94(9) &   1.74(6) &  1.86(10) &  2.40(33) &    2.14(17) &  3.04(44) \\
 12 &  2.07(34) &  1.42(26) &  1.57(21) &  2.00(14) &   1.72(8) &  1.81(13) &  2.04(19) &  3.28(1.28) &  3.50(91) \\
\end{tabular}
\caption{Overlap factors $\rho(\mathbf p^2)$ for all momenta on ensembles D200,
C101, H105, N200, N203, N302, J303, S201 and S400.}
\end{ruledtabular}
\end{table}
\begin{table}[!h]
\begin{ruledtabular}
\begin{tabular}{llllllllllll}
 $\mathbf{n}^2$ &      E250 &   $\mathbf{n}^2$ &      E250 &   $\mathbf{n}^2$ &     E250 &   $\mathbf{n}^2$ &      E250 &   $\mathbf{n}^2$ &      E250 &    $\mathbf{n}^2$ &      E250 \\
 0 &   1.24(9) &   6 &  1.47(12) &  13 &  1.57(6) &  20 &   1.65(8) &  27 &  1.75(12) &   35 &  1.74(17) \\
 1 &  1.22(10) &   8 &  1.49(14) &  14 &  1.54(6) &  21 &   1.59(8) &  29 &  1.65(12) &   36 &  1.73(15) \\
 2 &   1.28(9) &   9 &  1.50(18) &  16 &  1.70(8) &  22 &  1.63(12) &  30 &  1.76(15) &   &        \\
 3 &  1.48(11) &  10 &   1.61(5) &  17 &  1.58(7) &  24 &  1.68(11) &  32 &  1.74(14) &   &        \\
 4 &  1.28(11) &  11 &   1.58(5) &  18 &  1.64(7) &  25 &  1.74(11) &  33 &  1.73(14) &   &        \\
 5 &  1.35(11) &  12 &   1.66(6) &  19 &  1.57(8) &  26 &  1.71(10) &  34 &  1.72(16) &   &        \\
\end{tabular}
\caption{Overlap factors $\rho(\mathbf p^2)$ for all momenta on ensemble E250.}
\end{ruledtabular}
\end{table}
\begin{table}[!h]
\begin{ruledtabular}
\begin{tabular}{llllllllll}
  $\mathbf{n}^2$ &       D200 &       C101 &       H105 &       N200 &       N203 &       N302 &       J303 &       S201 &       S400 \\
  0 &  0.233(18) &  0.172(15) &  0.253(44) &  0.396(12) &  0.178(12) &  0.238(11) &  0.241(19) &  0.431(22) &  0.445(23) \\
  1 &  0.230(17) &   0.135(8) &  0.214(35) &  0.409(13) &  0.199(14) &   0.204(9) &  0.216(17) &  0.415(19) &  0.404(19) \\
  2 &  0.202(17) &  0.212(19) &  0.213(20) &  0.423(14) &  0.232(18) &  0.221(11) &  0.200(16) &  0.401(17) &  0.403(21) \\
  3 &  0.194(19) &  0.160(13) &  0.270(27) &  0.445(16) &  0.237(21) &  0.264(14) &  0.195(16) &  0.416(21) &  0.354(22) \\
  4 &  0.151(13) &  0.241(24) &  0.295(31) &  0.435(19) &  0.181(16) &  0.189(12) &  0.215(23) &  0.408(22) &  0.375(34) \\
  5 &  0.163(17) &  0.158(14) &  0.244(32) &  0.434(20) &  0.185(20) &  0.251(21) &  0.213(25) &  0.401(21) &  0.372(34) \\
  6 &  0.188(21) &  0.161(13) &  0.243(12) &  0.456(23) &  0.167(17) &  0.326(32) &  0.190(21) &  0.550(57) &  0.334(34) \\
  8 &  0.177(16) &  0.476(55) &  0.260(20) &  0.476(31) &   0.228(8) &  0.454(54) &  0.273(26) &  0.597(48) &  0.375(37) \\
  9 &   0.167(9) &  0.299(47) &  0.267(25) &  0.493(20) &  0.269(10) &  0.618(44) &  0.255(25) &  0.694(27) &  0.364(31) \\
 10 &  0.189(13) &  0.370(38) &  0.299(30) &  0.519(23) &  0.312(14) &  0.634(47) &  0.260(24) &  0.709(30) &  0.548(58) \\
 11 &  0.173(11) &  0.442(26) &  0.497(80) &  0.494(24) &  0.481(22) &  0.629(54) &  0.268(30) &  0.665(31) &  0.388(32) \\
 12 &  0.193(11) &  0.298(76) &  0.596(79) &  0.495(34) &  0.445(15) &  0.688(39) &  0.357(40) &  0.622(59) &  0.557(34) \\
\end{tabular}
\caption{Energy gap $\Delta(\mathbf{p}^2)$ for ensembles  D200,
C101, H105, N200, N203, N302, J303, S201 and S400 in lattice units.}

\end{ruledtabular}
\end{table}
\begin{table}[!h]
\begin{ruledtabular}
\begin{tabular}{llllllllllll}
 $\mathbf{n}^2$ &       E250 &   $\mathbf{n}^2$ &       E250 &   $\mathbf{n}^2$ &       E250 &   $\mathbf{n}^2$ &       E250 &   $\mathbf{n}^2$ &       E250 &    $\mathbf{n}^2$ &       E250 \\
 0 &  0.223(28) &   6 &  0.219(26) &  13 &  0.377(24) &  20 &  0.369(28) &  27 &  0.372(38) &   35 &  0.311(51) \\
 1 &  0.246(29) &   8 &  0.222(32) &  14 &  0.290(22) &  21 &  0.378(32) &  29 &  0.315(40) &   36 &  0.414(59) \\
 2 &  0.233(27) &   9 &  0.197(32) &  16 &  0.377(27) &  22 &  0.203(14) &  30 &  0.225(36) &   &         \\
 3 &  0.204(22) &  10 &  0.380(20) &  17 &  0.364(28) &  24 &  0.208(23) &  32 &  0.355(46) &   &         \\
 4 &  0.259(32) &  11 &  0.368(20) &  18 &  0.383(26) &  25 &  0.422(36) &  33 &  0.422(52) &   &         \\
 5 &  0.251(31) &  12 &  0.202(14) &  19 &  0.376(31) &  26 &  0.417(38) &  34 &  0.416(57) &   &         \\
\end{tabular}
\caption{Energy gap $\Delta(\mathbf{p}^2)$ for ensemble E250 in lattice units.}

\end{ruledtabular}
\end{table}
\vfill
\clearpage

\section{Dipole and $z$-expansion results}
\label{app:dipole}
This appendix presents the results for the magnetic moment and the electromagnetic radii
for the dipole and $z$-expansion fits for all
ensembles, both for the summation and two-state data given in
appendix~\ref{app:data}, for a momentum cut of $Q^2\leq 0.9 \, \Gevs$. The pion mass indicated in the first column of the
tables uniquely identifies the corresponding gauge ensemble via
Table~\ref{tab:ensembles}.

\begin{table}[h!]
\begin{ruledtabular}
\begin{tabular}{lllllll}
	$M_\pi\, [\rm{GeV}]$ &   $\mu$ (sum) & $\mu$ (two-state) &
	$\res$ (sum) &   $\res$ (two-state)  & $\rms$ (sum) & $\rms$ (two-state)\\
0.130 &  3.61(40) &  3.87(25) &   0.680(67) &   0.630(42) &   0.528(99) &   0.556(64) \\
0.203 &  4.22(31) &  4.12(15) &   0.614(43) &   0.644(24) &   0.630(78) &   0.624(47) \\
0.223 &  4.21(17) &  4.19(25) &   0.628(29) &   0.691(30) &   0.561(43) &   0.600(84) \\
0.262 &  3.80(35) &  3.63(19) &   0.597(52) &   0.541(24) &   0.299(63) &   0.374(31) \\
0.278 &  4.25(52) &  4.09(81) &   0.629(59) &  0.800(105) &  0.511(103) &  0.584(175) \\
0.283 &  4.21(32) &  3.93(11) &   0.529(39) &   0.501(13) &   0.446(65) &   0.423(23) \\
0.293 &  - &  - &  0.571(115) &   0.482(20) &    - &   - \\
0.347 &  3.96(13) &  4.12(12) &   0.599(20) &   0.606(25) &   0.425(30) &   0.464(34) \\
0.350 &  4.19(38) &  4.10(20) &   0.549(37) &   0.544(22) &   0.385(68) &   0.406(34) \\
0.353 &  4.47(56) &  3.52(23) &   0.614(39) &   0.515(20) &  0.557(112) &   0.364(53) \\
\end{tabular}

\caption{Dipole fits for $\mu$, $\res$  and $\rms$ (in $\mathrm{fm}^2$) on every
ensemble.}
\end{ruledtabular}
\end{table}

\begin{table}[h]
\begin{ruledtabular}
\begin{tabular}{rllllll}
	$M_\pi\, [\rm{GeV}]$ &   $\mu$ (sum) & $\mu$ (two-state) &
	$\res$ (sum) &   $\res$ (two-state)  & $\rms$ (sum) & $\rms$ (two-state)\\
0.130 &    3.72(58) &  4.10(34) &  0.782(176) &  0.814(106) &  0.822(629) &    1.01(42) \\
0.203 &    4.35(32) &  4.12(17) &   0.610(71) &   0.652(42) &  0.759(151) &  0.692(104) \\
0.223 &    4.29(18) &  4.21(24) &   0.645(39) &   0.662(43) &   0.573(92) &  0.626(116) \\
0.262 &    4.06(41) &  3.69(22) &   0.599(59) &   0.521(34) &  0.444(119) &   0.432(89) \\
0.278 &    3.70(56) &  3.63(71) &   0.575(66) &   0.622(82) &  0.313(195) &  0.395(232) \\
0.283 &    4.20(32) &  3.89(12) &   0.518(46) &   0.479(19) &  0.457(101) &   0.397(50) \\
0.293 &  3.37(1.13) &  3.58(49) &  0.639(104) &   0.472(33) &  0.390(259) &  0.367(155) \\
0.347 &    3.85(13) &  3.94(12) &   0.558(20) &   0.469(26) &   0.392(40) &   0.384(49) \\
0.350 &    3.80(39) &  3.76(21) &   0.474(37) &   0.469(25) &  0.263(111) &   0.313(58) \\
0.353 &    3.93(42) &  3.34(25) &   0.487(39) &   0.463(26) &  0.403(106) &   0.300(90) \\
\end{tabular}

\caption{$z$-expansion fits for $\mu$, $\res$  and $\rms$ (in $\mathrm{fm}^2$) on every
ensemble.}
\end{ruledtabular}
\end{table}

\clearpage
\section{HBChPT fits}
\label{sec:app_hbchpt}
Here we summarize the results for the physical values of the magnetic moment
$\mu=\kappa+1$ and the electromagnetic radii of the HBChPT fits to the
$z$-expansion data, applying a pion mass cut $M_\pi \leq 0.28\, \mathrm{GeV}$,
as explained in Section~\ref{sec:CCF}. Note that the value for AIC is not
corrected for the cut data points. The last column indicates which of the
corrections appearing in Eq.~(\ref{eq:hbchpt_ccf}) is included in the fit

\begin{table}[h]
\begin{ruledtabular}
\begin{tabular}{cccccccl}
	$\chi^2/\mathrm{DOF}$  &  $\kappa$  &  $\langle   r_\mathrm{E}^2\rangle
	\, [\rm{fm}^2] $          &  $\langle  r_\mathrm{M}^2
	\rangle\,[\rm{fm}^2]  $  &    p-value              &
	AIC   &   $Q^2$-cut  [$\Gevs$]  &  correction \\
0.45    &  3.08(40)  &  0.856(42)  &  0.88(12)   &  0.89  &  11.62  &  0.6  &                    -                       \\
0.52    &  2.62(62)  &  0.82(10)   &  0.94(28)   &  0.76  &  16.60  &  0.6  &                    $\mathcal{O}(a^2)$\\
0.53    &  3.56(75)  &  0.92(15)   &  0.86(38)   &  0.76  &  16.63  &  0.6  &                    $\mathcal{O}(e^{-m_\pi  L})$  \\
\hline
0.61    &  3.27(35)  &  0.871(37)  &  0.777(91)  &  0.82  &  14.71  &  0.7  &                    -                       \\
0.80    &  3.03(55)  &  0.858(94)  &  0.82(22)   &  0.61  &  20.36  &  0.7  &                    $\mathcal{O}(a^2)$      \\
0.45    &  3.84(50)  &  0.89(10)   &  0.91(26)   &  0.89  &  17.63  &  0.7  &                    $\mathcal{O}(e^{-m_\pi  L})$  \\
\hline
0.68    &  3.23(35)  &  0.869(37)  &  0.772(91)  &  0.76  &  15.44  &  0.8  &                    -                       \\
0.88    &  2.97(54)  &  0.856(93)  &  0.80(21)   &  0.53  &  21.05  &  0.8  &                    $\mathcal{O}(a^2)$      \\
0.54    &  3.78(49)  &  0.89(10)   &  0.91(26)   &  0.83  &  18.32  &  0.8  &                    $\mathcal{O}(e^{-m_\pi  L})$  \\
\hline
0.65    &  3.21(34)  &  0.872(36)  &  0.774(90)  &  0.78  &  15.18  &  0.9  &                    -                       \\
0.85    &  2.97(54)  &  0.851(92)  &  0.82(21)   &  0.56  &  20.76  &  0.9  &$\mathcal{O}(a^2)$  \\
0.53    &  3.75(48)  &  0.87(10)   &  0.93(26)   &  0.83  &  18.28  &  0.9  &                    $\mathcal{O}(e^{-m_\pi  L})$  \\
\end{tabular}
\caption{HBChPT fits for $z$-expansion extractions from summation data.}
\end{ruledtabular}
\end{table}

\begin{table}[h]
\begin{ruledtabular}
\begin{tabular}{cccccccl}
	$\chi^2/\mathrm{DOF}$ & $\kappa$ & $\langle r_\mathrm{E}^2\rangle\,
	[\rm{fm}^2] $ & $\langle r_\mathrm{M}^2 \rangle\, [\rm{fm}^2] $ &
	p-value  & AIC &    $Q^2$-cut  [$\Gevs$]  &  correction  \\
1.00    &  3.25(26)  &  0.875(36)  &  0.772(88)  &  0.43  &  16.03  &  0.6  &                    -                       \\
0.99    &  2.77(49)  &  0.781(84)  &  0.74(21)   &  0.42  &  18.97  &  0.6  &                    $\mathcal{O}(a^2)$\\
0.66    &  3.88(74)  &  1.24(21)   &  0.83(45)   &  0.66  &  17.29  &  0.6  &                    $\mathcal{O}(e^{-m_\pi  L})$  \\
\hline
0.79    &  3.26(24)  &  0.877(30)  &  0.748(74)  &  0.65  &  16.65  &  0.7  &                    -                       \\
0.70    &  2.88(42)  &  0.804(68)  &  0.72(16)   &  0.69  &  19.59  &  0.7  &                    $\mathcal{O}(a^2)$      \\
0.74    &  4.03(58)  &  0.92(12)   &  0.94(35)   &  0.66  &  19.90  &  0.7  &                    $\mathcal{O}(e^{-m_\pi  L})$  \\
\hline
0.95    &  3.34(23)  &  0.875(28)  &  0.732(68)  &  0.49  &  18.41  &  0.8  &                    -                       \\
0.63    &  2.86(40)  &  0.792(62)  &  0.68(15)   &  0.75  &  19.08  &  0.8  &                    $\mathcal{O}(a^2)$      \\
0.90    &  4.13(55)  &  0.92(12)   &  0.96(32)   &  0.52  &  21.16  &  0.8  &                    $\mathcal{O}(e^{-m_\pi  L})$  \\
\hline
0.89    &  3.31(22)  &  0.875(27)  &  0.735(66)  &  0.55  &  17.76  &  0.9  &                    -                       \\
0.59    &  2.87(38)  &  0.794(61)  &  0.69(15)   &  0.79  &  18.68  &  0.9  &$\mathcal{O}(a^2)$  \\
0.90    &  4.01(53)  &  0.90(11)   &  0.91(30)   &  0.52  &  21.18  &  0.9  &                    $\mathcal{O}(e^{-m_\pi  L})$  \\
\end{tabular}
\caption{HBChPT fits for $z$-expansion extractions from two-state data.}
\end{ruledtabular}
\end{table}

\clearpage
\section{Covariant B$\chi$PT fits}

\label{sec:app_covchpt}
Here we summarize the results for the physical values of the magnetic moment
$\mu=\kappa+1$ and the electromagnetic radii of the direct covariant ChPT fits
as discussed in Section~\ref{sec:CCF}, applying a pion mass cut of $M_\pi\leq
0.28 \mathrm{GeV}$. Note that the value for AIC is not corrected for the cut
data points. The entries with and without an asterisk in the last column
indicate which of the corrections appearing respectively in
Eq.~(\ref{fitmodel_cov_chpt_method1}) and Eq.~(\ref{fitmodel_cov_chpt}) is
included in the fit.

\begin{table}[h]
\begin{ruledtabular}
\begin{tabular}{cccccccccl}
	$\chi^2/\mathrm{DOF}$ & $\kappa$ & $\langle
	r_\mathrm{E}^2\rangle\,[\rm{fm}^2] $ & $\langle
	r_\mathrm{M}^2 \rangle \, [\rm{fm}^2] $ & $\langle r_\mathrm{E}
	\rangle\,[\rm{fm}]$ & $\langle
	r_\mathrm{M} \rangle \,[\rm{fm}]$ & p-value  & AIC & $Q^2$-cut &
	correction\\
1.63          &       3.75(11)  &  0.818(22)  &  0.663(27)  &  0.905(12)  &0.814(17)  &  0.00  &  112.11  &  0.6  &  -                         \\
1.68          &       3.83(19)  &  0.816(33)  &  0.652(35)  &  0.903(18)  &0.808(21)  &  0.00  &  115.87  &  0.6  &  $\mathcal{O}(a^2)$        \\
1.68          &       3.82(12)  &  0.790(31)  &  0.666(42)  &  0.889(18)  &0.816(26)  &  0.00  &  116.45  &  0.6  &  $\mathcal{O}(e^{-M_\pi    L})$  \\
1.67          &       3.77(32)  &  0.800(53)  &  0.662(63)  &  0.894(29)  &0.814(39)  &  0.00  &  115.84  &  0.6  &  $^*\mathcal{O}(a^2)$      \\
1.69          &       3.79(12)  &  0.770(45)  &  0.671(36)  &  0.878(26)  &0.819(22)  &  0.00  &  116.65  &  0.6  &  $^*\mathcal{O}(e^{-M_\pi  L})$  \\
\hline        \hline
1.49          &       3.81(11)  &  0.818(22)  &  0.669(28)  &  0.905(12)  &0.818(17)  &  0.01  &  91.44   &  0.5  &  -                         \\
1.55          &       3.80(20)  &  0.819(34)  &  0.670(36)  &  0.905(19)  &0.818(22)  &  0.01  &  95.44   &  0.5  &  $\mathcal{O}(a^2)$        \\
1.56          &       3.87(12)  &  0.797(33)  &  0.667(45)  &  0.893(19)  &0.817(28)  &  0.01  &  96.36   &  0.5  &  $\mathcal{O}(e^{-M_\pi    L})$  \\
1.54          &       3.82(32)  &  0.798(53)  &  0.668(64)  &  0.893(30)  &0.817(39)  &  0.01  &  95.12   &  0.5  &  $^*\mathcal{O}(a^2)$      \\
1.56          &       3.85(12)  &  0.783(47)  &  0.670(37)  &  0.885(27)  &0.819(23)  &  0.01  &  96.43   &  0.5  &  $^*\mathcal{O}(e^{-M_\pi  L})$  \\
\hline\hline
1.42          &       3.74(13)  &  0.800(26)  &  0.651(35)  &  0.895(15)  &0.807(22)  &  0.04  &  64.64   &  0.4  &  -                         \\
1.48          &       3.62(22)  &  0.803(40)  &  0.668(45)  &  0.896(22)  &0.817(28)  &  0.03  &  68.20   &  0.4  &  $\mathcal{O}(a^2)$        \\
1.52          &       3.80(13)  &  0.793(36)  &  0.664(58)  &  0.891(20)  &0.815(35)  &  0.02  &  69.60   &  0.4  &  $\mathcal{O}(e^{-M_\pi    L})$  \\
1.48          &       3.68(33)  &  0.777(59)  &  0.653(69)  &  0.881(34)  &0.808(43)  &  0.03  &  68.16   &  0.4  &  $^*\mathcal{O}(a^2)$      \\
1.52          &       3.79(13)  &  0.776(51)  &  0.639(55)  &  0.881(29)  &0.799(34)  &  0.02  &  69.61   &  0.4  &  $^*\mathcal{O}(e^{-M_\pi  L})$  \\
\hline\hline
1.75          &       3.76(13)  &  0.797(30)  &  0.660(42)  &  0.893(17)  &0.812(26)  &  0.01  &  56.95   &  0.3  &  -                         \\
1.88          &       3.76(25)  &  0.805(48)  &  0.661(53)  &  0.897(27)  &0.813(32)  &  0.00  &  60.87   &  0.3  &  $\mathcal{O}(a^2)$        \\
1.91          &       3.81(14)  &  0.781(45)  &  0.655(79)  &  0.883(25)  &0.810(49)  &  0.00  &  61.55   &  0.3  &  $\mathcal{O}(e^{-M_\pi    L})$  \\
1.87          &       3.80(35)  &  0.776(67)  &  0.658(77)  &  0.881(38)  &0.811(48)  &  0.00  &  60.71   &  0.3  &  $^*\mathcal{O}(a^2)$      \\
1.90          &       3.82(14)  &  0.769(60)  &  0.631(72)  &  0.877(34)  &0.794(46)  &  0.00  &  61.41   &  0.3  &  $^*\mathcal{O}(e^{-M_\pi  L})$  \\
\end{tabular}
\caption{Covariant BChPT fits for summation data.}
\end{ruledtabular}
\end{table}
\begin{table}[h]
\begin{ruledtabular}
\begin{tabular}{cccccccccl}
	$\chi^2/\mathrm{DOF}$ & $\kappa$ & $\langle r_\mathrm{E}^2\rangle\,
	[\rm{fm}^2] $ & $\langle
	r_\mathrm{M}^2 \rangle \, [\rm{fm}^2]  $ & $\langle r_\mathrm{E}
	\rangle\, [\rm{fm}] $& $\langle
	r_\mathrm{M} \rangle \, [\rm{fm}]$ & p-value  & AIC & $Q^2$-cut &
	correction \\
1.35          &       3.499(77)  &  0.791(18)  &  0.653(21)  &  0.889(10)  &0.808(13)  &  0.03  &  94.66   &  0.6  &  -                         \\
1.35          &       3.68(13)   &  0.779(26)  &  0.633(25)  &  0.883(15)  &0.795(16)  &  0.04  &  95.44   &  0.6  &  $\mathcal{O}(a^2)$        \\
1.42          &       3.509(90)  &  0.806(34)  &  0.614(47)  &  0.898(19)  &0.783(30)  &  0.02  &  100.06  &  0.6  &  $\mathcal{O}(e^{-M_\pi    L})$  \\
1.34          &       3.74(18)   &  0.755(36)  &  0.647(38)  &  0.869(21)  &0.804(24)  &  0.04  &  95.32   &  0.6  &  $^*\mathcal{O}(a^2)$      \\
1.37          &       3.530(86)  &  0.726(55)  &  0.554(45)  &  0.852(33)  &0.744(30)  &  0.03  &  96.70   &  0.6  &  $^*\mathcal{O}(e^{-M_\pi  L})$  \\
\hline        \hline
1.19          &       3.554(81)  &  0.787(20)  &  0.662(24)  &  0.887(11)  &0.814(15)  &  0.16  &  74.45   &  0.5  &  -                         \\
1.21          &       3.66(13)   &  0.776(27)  &  0.649(28)  &  0.881(16)  &0.806(18)  &  0.14  &  77.29   &  0.5  &  $\mathcal{O}(a^2)$        \\
1.23          &       3.599(95)  &  0.826(38)  &  0.647(52)  &  0.909(21)  &0.804(33)  &  0.12  &  78.55   &  0.5  &  $\mathcal{O}(e^{-M_\pi    L})$  \\
1.18          &       3.76(19)   &  0.750(38)  &  0.659(42)  &  0.866(22)  &0.812(26)  &  0.17  &  75.67   &  0.5  &  $^*\mathcal{O}(a^2)$      \\
1.24          &       3.617(84)  &  0.821(78)  &  0.615(47)  &  0.906(43)  &0.784(30)  &  0.11  &  79.02   &  0.5  &  $^*\mathcal{O}(e^{-M_\pi  L})$  \\
\hline\hline
1.46          &       3.529(91)  &  0.779(24)  &  0.661(29)  &  0.882(14)  &0.813(18)  &  0.03  &  66.33   &  0.4  &  -                         \\
1.51          &       3.61(15)   &  0.753(39)  &  0.650(34)  &  0.868(22)  &0.806(21)  &  0.02  &  69.20   &  0.4  &  $\mathcal{O}(a^2)$        \\
1.34          &       3.58(10)   &  0.908(53)  &  0.645(66)  &  0.953(28)  &0.803(41)  &  0.08  &  63.06   &  0.4  &  $\mathcal{O}(e^{-M_\pi    L})$  \\
1.42          &       3.73(20)   &  0.705(50)  &  0.656(47)  &  0.839(30)  &0.810(29)  &  0.04  &  66.10   &  0.4  &  $^*\mathcal{O}(a^2)$      \\
1.48          &       3.566(98)  &  0.844(94)  &  0.569(63)  &  0.919(51)  &0.754(42)  &  0.03  &  68.08   &  0.4  &  $^*\mathcal{O}(e^{-M_\pi  L})$  \\
\hline\hline
1.77          &       3.45(10)   &  0.772(29)  &  0.642(39)  &  0.879(16)  &0.801(24)  &  0.01  &  57.65   &  0.3  &  -                         \\
1.87          &       3.54(17)   &  0.740(50)  &  0.633(45)  &  0.860(29)  &0.796(28)  &  0.00  &  60.71   &  0.3  &  $\mathcal{O}(a^2)$        \\
1.65          &       3.58(12)   &  0.959(69)  &  0.727(96)  &  0.979(35)  &0.852(56)  &  0.02  &  54.87   &  0.3  &  $\mathcal{O}(e^{-M_\pi    L})$  \\
1.77          &       3.64(21)   &  0.690(60)  &  0.639(56)  &  0.830(36)  &0.800(35)  &  0.01  &  58.05   &  0.3  &  $^*\mathcal{O}(a^2)$      \\
1.88          &       3.50(11)   &  0.86(11)   &  0.584(85)  &  0.926(58)  &0.764(55)  &  0.00  &  60.97   &  0.3  &  $^*\mathcal{O}(e^{-M_\pi  L})$  \\
\end{tabular}
\caption{Covariant BChPT fits for two-state data.}
\end{ruledtabular}
\end{table}

\clearpage
\section{Ratio $G_\mathrm{M}/G_\mathrm{E}$}
\label{app:gmoverge}
A common observation among lattice determinations of the isovector form factors
is that the ratio of the magnetic and electric form factor exhibits a rather
flat behavior. One may therefore hope to extract the magnetic moment as the intercept of a linear
fit to that ratio over a restricted range in $Q^2$. We perform linear fits with
an upper limit of $Q^2\leq 0.29 \,  \Gevs$ for E250 and $Q^2\leq 0.6 \,  \Gevs$ for the
remaining ensembles. Note that S201, S400 and N302 do not allow for a linear
extrapolation with a cut off $Q^2\leq 0.6 \, \Gevs$. We stress that these points
do not enter in our final fits, but rather serve as a consistency check (see
Fig.~\ref{fig:best_fit_direct_cov_sum} and
Fig.~\ref{fig:fit_cov_bchpt_vs_lineextrapol}).
\begin{table}[h]
	\begin{tabular}{ccc}
		\hline\hline
		Ensemble  & $\mu$ (summation)  & $\mu$ (two-state)\\\hline
D200  & 4.16(28) & 3.73(18)\\
C101  & 4.36(17) & 4.55(70)\\
H105  & 3.60(61) & 3.61(73)\\
N200  & 4.11(28) & 3.91(7)\\
N203  & 4.04(12) & 4.02(28)\\
J303  & 4.11(38) & 3.78(16)\\
E250  & 4.05(52) & 4.06(39)\\
\hline
	\end{tabular}
	\caption{Extrapolated values of the magnetic moment using the ratio
	$\frac{G_\mathrm{M}(Q^2)}{G_\mathrm{E}(Q^2)}$. }
	\label{app:tab_ratio_intercept}
\end{table}

\end{document}